\documentclass[a4paper,11pt]{article}
\pdfoutput=1 

\usepackage{jheppub} 

\usepackage{inputenc}

\usepackage{subfigure}
\usepackage{array,multirow}
\usepackage{etoolbox}
\usepackage{booktabs}
\usepackage{float}
\usepackage{scalerel}
\usepackage{epsfig,color,wrapfig}
\usepackage{amsmath}
\usepackage{graphicx}
\usepackage{amssymb}
\usepackage{slashbox}
\usepackage{comment}
\usepackage{bm}
\usepackage[toc]{appendix}
\usepackage{slashed}
\usepackage{caption}
\usepackage{dcolumn,multirow}
\usepackage[normalem]{ulem}
\usepackage{hyperref}

{\newcommand{\lsim}{\mbox{\raisebox{-.6ex}{~$\stackrel{<}{\sim}$~}}}
	{\newcommand{\gsim}{\mbox{\raisebox{-.6ex}{~$\stackrel{>}{\sim}$~}}}

		\newcommand{\bmt}{\begin{pmatrix}}
			\newcommand{\emt}{\end{pmatrix}}
		\newcommand{\ba}{\begin{array}{c}}
			\newcommand{\ea}{\end{array}}
		\newcommand{\be}{\begin{equation}}
		\newcommand{\ee}{\end{equation}}
		\newcommand{\bea}{\begin{eqnarray}}
		\newcommand{\eea}{\end{eqnarray}}
		\newcommand{\nn}{\nonumber}
		\newcommand{\bi}{\begin{itemize}}
			\newcommand{\ei}{\end{itemize}}
		
		\newcommand{\baz}{\begin{array}{cc}}
			
			\newcommand{\mathsym}[1]{{}}

			\newcommand{\bt}{\begin{tabular}}
				\newcommand{\et}{\end{tabular}}

			\newcommand{\benu}{\begin{enumerate}}
				\newcommand{\eenu}{\end{enumerate}}
			
			\newcommand{\bav}{\begin{array}{cccc}}

\title{\boldmath Low Scale $U(1)_X$ Gauge Symmetry as an Origin of Dark Matter, Neutrino Mass and Flavour Anomalies}

\author{Debasish Borah,}
\emailAdd{dborah@iitg.ac.in}
\author{Lopamudra Mukherjee,}
\emailAdd{mukherjeelopa@iitg.ac.in}
\author{and Soumitra Nandi}
\emailAdd{soumitra.nandi@iitg.ac.in}
\affiliation{Department of Physics, Indian Institute of Technology Guwahati, Assam 781039, India}

\abstract{We study a generic leptophilic $U(1)_X$ extension of the standard model with a light gauge boson. The $U(1)_X$ charge assignments for the leptons are guided by lepton universality violating (LUV) observables in semileptonic $b \to s\ell\ell$ decays, muon anomalous magnetic moment and the origin of leptonic masses and mixing. Anomaly cancellation conditions require the addition of new chiral fermions in the model, one of which acts as a dark matter (DM) candidate when it is stabilised by an additional $\mathcal{Z}_2$ symmetry. From our analysis, we show two different possible models with similar particle content that lead to quite contrasting neutrino mass origin and other phenomenology. The proposed models also have the potential to address the anomalous results in $b\to c\ell\nu_{\ell}$ decays like $R(D), R(D^*)$, electron anomalous magnetic moment and the very recent KOTO anomaly in the kaon sector. We also discuss different possible collider signatures of our models which can be tested in future.}

\arxivnumber{2007.13778}

\begin{document} 
\maketitle

\section{Introduction}
The large hadron collider (LHC) at CERN is operational for more than last ten years and so far apart from the discovery of Higgs boson, no new particles or interactions have been found. No evidence for the theoretically well-motivated models like supersymmetry, extra dimension etc have been found. Yet there is a list of unsolved puzzles in particle physics. In the standard model (SM) of particle physics, we do not have explanations for neutrino masses, the existence of dark matter (DM) and the domination of matter over antimatter in the Universe \cite{pdg2018}. Nature may still be supersymmetric, or there may be an extra dimension; however, these extensions of SM have failed to show up at the LHC. Even if they are absent, there are a lot of things to learn, at a fundamental level. In principle, one could write down models which are consistent with the present observations at the collider and will show distinct features only at the high luminosity, as for example see \cite{Barman:2018jhz}. There are indications that we have not yet fully understood the working rules of our Universe at the fundamental level, and that is motivating enough for the particle physics community to keep looking for it. In addition to the ongoing LHC experiment, we already have a few experimental facilities which are operational or will start functioning very soon, and very quickly data will be collected in unprecedented amounts. We can hope that the upcoming data will guide us to establish the more fundamental theory of elementary particles and their interactions.
 
Apart from the direct searches at the collider, the low energy observables play an essential role for indirect detection of a new particle(s) or interaction(s). In this regard, B-factories have played a significant role in the last couple of decades \cite{Bevan:2014iga} and will remain productive in near future \cite{Kou:2018nap}. In the last couple of years, LHCb has also produced significant results, for a brief review see \cite{Bediaga:2012py,lhcb}. In the low energy data, new physics (NP) contributions in an observable can be pinpointed through the deviation of its measured value from the respective SM prediction. At the moment there are a few measurements in $b \to c$ and $b \to s$ decays which show some degree of discrepancies with their respective SM predictions, for very recent updates see \cite{hflav,Aoki:2019cca}. Apart from these long standing anomalies, more recently an excess of events have been observed in the rare $K_L \to \pi^0 \nu \bar{\nu}$ decay ($d \to s$ FCNC process) at the KOTO experiment at J-PARC \cite{KOTOTalk}.

The measurements of various angular observables in $B\to K^* \mu^+\mu^-$ \cite{Aaij:2015oid, Wehle:2016yoi} and $B_s\to \phi\mu^+\mu^-$ \cite{Aaij:2015esa} decays are available, and in a few of them there are discrepancies between the theory and experiment. Very recently, with the data collected by the LHCb experiment during the years 2011, 2012 and 2016, a complete set of CP-averaged angular observables has been measured in $B \to K^{(*)} \mu^+\mu^-$ decay \cite{Aaij:2020nrf}. To date, this is the most precise measurement, and the data still shows discrepancies between the theoretical predictions and the measured value in a couple of those angular observables. Note that these angular observables are not free from hadronic uncertainties. However, there are theoretically clean observables like  $R(K^{(*)}) = \frac{\mathcal{B}(B \to K^{(*)} \mu^+\mu^-)}{\mathcal{B}(B \to K^{(*)} e^+ e^-)}$ the measured values of which \cite{Aaij:2019wad, Abdesselam:2019wac} are not in good agreement with the corresponding SM expectations. There are new physics explanations of these observations, for a recent update on the model-independent new physics explanation of these data see \cite{Bhattacharya:2019dot,Biswas:2020uaq,Hurth:2020rzx} and the references therein. Similar to the observables $R(K^{(*)})$, we define $R(D^{(*)}) = \frac{\mathcal{B}(B \to D^{(*)} \tau\nu_{\tau})}{\mathcal{B}(B \to D^{(*)} \ell \nu_{\ell})}$ (with $\ell = \mu, e$) which is associated with the $b\to c$ decays. The measured values of these observables \cite{hflav} have also shown some degree of discrepancies with the respective SM predictions, for details see \cite{Gambino:2019sif, Bordone:2019vic,Jaiswal:2020wer}. The most recent predictions (in SM) differ from the one obtained using the old Belle data \cite{Bigi:2017jbd,Jaiswal:2017rve}. The bounds on the model-independent new physics Wilson-coefficients (WC) can be seen from \cite{Jaiswal:2020wer}. It is found that the data still allows sizeable NP contributions in these decays.               
 
Apart from the above mentioned results, the muon anomalous magnetic moment $(g-2)_{\mu}$ is another longstanding puzzle. It has been measured very precisely while it has also been predicted in the SM to a great accuracy. At present the difference between the predicted and the measured value is given by 
\begin{equation}
\Delta a_{\mu} = a_{\mu}^{exp} - a_{\mu}^{SM} = 26.1 (7.9)\times 10^{-10},
\label{eq:delamu_old} 
\end{equation}
which shows there is still room for NP beyond the SM (for details see~\cite{pdg2018}). In a recent article, the status of the SM calculation of muon magnetic moment has been updated \cite{Aoyama:2020ynm}. According to this study, the difference is given by 
\begin{equation}
 \Delta a_{\mu} = 27.9 (7.6)\times 10^{-10},
 \label{eq:delamu_new}
 \end{equation}
which is a 3.7$\sigma$ discrepancy. Analogous to muon magnetic moment, measurements are also available for electron magnetic moment $(g-2)_{e}$. The most recent result obtained from measurement of the fine structure constant of QED \cite{Parker_2018}, shows a deviation from the SM. The excess is given by $\Delta a_{e} = - 8.7 (3.6)\times 10^{-13}$.   

In this study, we look for a NP model which is capable of addressing all the above-mentioned results. At first, we consider a simple model by extending the SM gauge group with an additional $U(1)_X$ gauge symmetry\footnote{For a review of such Abelian gauge extension of SM, please see \cite{Langacker:2008yv}.}. The resulting complete gauge group of the model will be $SU(3)_c\times SU(2)_L\times U(1)_Y\times U(1)_X$ which is an extension of SM by an abelian factor. The advantage of such an extension is that it introduces a minimal set of free parameters. The other most important feature of the new gauge symmetry we adopt here is that it is  leptophilic in nature i.e. only the leptons will be charged under $U(1)_X$, not the quarks. For an explanation of the above mentioned anomalous results, the lepton generations must have different charges under $U(1)_X$. The degree of fermion non-universality should explain the observed discrepancies in $R(K^{(*)})$ and muon anomalous magnetic moment. In this minimal model with GeV scale mass of $U(1)_X$ gauge boson, we can not explain $R(D^{*})$ and the data on electron anomalous magnetic moment. 

However, charging the fermions under this new gauge group in the absence of additional chiral fermions generally leads to triangle anomalies which must be cancelled in order to validate the gauge theory at the quantum level. Hence, in order to cancel the gauge anomalies, we need to introduce additional degrees of freedom into our model, in terms of chiral fermions. Here, following the constraints from gauge anomaly cancellation, we discuss only two different possible scenarios in which we can explain the existing data on DM and neutrino oscillation. In extended version of such minimal model with more particle and interactions, there will be additional Feynman diagrams which will contribute to $R(D^{*})$ and $\Delta a_{e}$ that help us to explain the observed data. 

In a similar direction, studies are available in the literature with a heavy $U(1)_X$ gauge boson \cite{Sierra:2015fma, Crivellin:2015mga, Crivellin:2015lwa, Altmannshofer:2016brv,Altmannshofer:2016jzy,Bhatia:2017tgo,Baek:2017sew,Ballett:2019xoj,Han:2019diw}. While such models with heavy $U(1)_X$ gauge boson have been extensively studied, there have been very few studies on low mass regions \cite{Datta:2017pfz, Sala:2017ihs, Correia:2019pnn,Correia:2019woz, Darme:2020hpo}. However, our working model is very much different compared to the one discussed in the references mentioned above and we also correlate the flavour anomalies with origin of neutrino mass and dark matter. Both the scenarios we discuss here consider the viability of a leptophilic $U(1)_X$ gauge symmetry in a way that it is anomaly free, predicts lepton flavour non-universality and the origin of light neutrino masses while the stability of DM candidate is ensured by an additional $\mathcal{Z}_2$ symmetry which also plays a non-trivial role in neutrino mass generation for one of the models.
 
 This paper is organised as follows. In section \ref{sec2} we briefly discuss our overall framework followed by the corresponding analysis of flavour anomalies in section \ref{sec3} by considering only the SM particle spectrum along with a massive leptophilic and family non-universal $U(1)_X$ gauge boson. We then move onto the discussions of the complete models in sections \ref{sec4}, \ref{sec5} covering the details of flavour anomalies, dark matter and neutrino mass. In section \ref{sec6}, we discuss the possibility of explaining KOTO anomaly within our toy models. In section \ref{sec7} we discuss about different Higgs invisible and charged lepton flavour violating decays and also comment on other possible ways to probe our model at the LHC and finally summarise our findings in section \ref{sec9}.

\section{Our Framework}
\label{sec2}
As mentioned before, our goal is to extend the SM by an Abelian $U(1)_X$ symmetry with a corresponding massive gauge boson $X$. We restrict our study to only low mass regime (GeV scale) of this additional gauge boson and allow only the leptons to couple to it. The charge assignments of the different SM particles under the different gauge groups are listed in Table.~\ref{table_anomaly} and the NP interaction Lagrangian is given by
\begin{equation}
\mathcal{L}^{\rm NP}_{\text{int}} = i \sum \limits_{i=1}^3 n_i g_{X} ( \bar{\ell}^L_i\gamma^\mu \ell^L_i + \bar{e}^R_{i} \gamma^\mu e^R_i )  X_\mu - \frac{1}{4} X_{\mu \nu} X^{\mu \nu} + \frac{\epsilon}{4}B_{\mu \nu} X^{ \mu \nu},
\label{eq:zpint}
\end{equation}
where $g_X$ is the gauge coupling of the $U(1)_X$ group, $i$ represents the lepton generation and $n_i$ are the charges of the lepton families under $U(1)_X$ which we want to constrain from anomaly cancellation requirements as well as flavour phenomenology. Here, in the above Lagrangian, $\ell^L_i$ is the left-handed lepton doublet while $e^R_i$ is the right-handed singlet with same gauge charge $n_i$. While writing the above Lagrangian, we have assumed that the $U(1)_X$ charges for the right and left-handed leptons are same, leading to a vector type interaction. In eq.~\eqref{eq:zpint}, $B_{\mu \nu}$ and $X_{\mu \nu}$ are the standard $U(1)_Y$ and $U(1)_X$ field stress tensors, respectively, and the factor $\epsilon$ represents the kinetic mixing between them. We assume that the leptophilic $X$ mixes kinetically with the SM $Z$ boson with a strength $\epsilon$. This mixing will be helpful to get contributions in various low energy observables like $R(K)$, $R(K^*)$ through penguin diagrams with the lepton vertex dominated by the above interaction and the one-loop quark vertex modified by the mixing parameter $\epsilon$. In muon or electron anomalous magnetic moments or in other lepton flavour violating (LFV) decays, at leading order, this mixing parameter does not have any specific role. 

\begin{table}[htbp!]
\begin{center}
\renewcommand{\arraystretch}{0.8}
\begin{tabular}{|c|c|c|}
\hline
Particles & $SU(3)_c \times SU(2)_L \times U(1)_Y$ & $U(1)_X $   \\
\hline
$Q_L=\begin{pmatrix}u_{L}\\
d_{L}\end{pmatrix}$ & $(3, 2, \frac{1}{6})$ & 0  \\
$u_R$ & $(3, 1, \frac{2}{3})$ & 0  \\
$d_R$ & $(3, 1, -\frac{1}{3})$ & 0  \\
$\begin{pmatrix}\nu_{e}\\
e\end{pmatrix}_L$ & $(1, 2, -\frac{1}{2})$ & $n_1$  \\
$\begin{pmatrix}\nu_{\mu}\\
\mu \end{pmatrix}_L$ & $(1, 2, -\frac{1}{2})$ & $n_2$  \\
$\begin{pmatrix}\nu_{\tau}\\
\tau \end{pmatrix}_L$ & $(1, 2, -\frac{1}{2})$ & $n_3$  \\
$e_R$ & $(1, 1, -1)$ & $n_1$ \\
$\mu_R$ & $(1, 1, -1)$ & $n_2$ \\
$\tau_R$ & $(1, 1, -1)$ & $n_3$ \\
\hline
\end{tabular}
\end{center}
\caption{$U(1)_X$ charges of the SM fermions.}
\label{table_anomaly}
\end{table}

As mentioned before, assigning charges to the SM fermions under a generic $U(1)_X$ symmetry leads to non-zero contributions to the one-loop triangle diagrams and makes the model anomalous. Therefore in order to realise a anomaly-free renormalisable model, one needs to put additional chiral fermions into the model which may also provide a natural candidate for DM. At the same time the additional chiral fermions required for anomaly cancelation could be made useful for neutrino mass generation as well. For similar construction of Abelian gauge extended models in the context of DM and neutrino mass generation, see \cite{Davidson:1978pm, PhysRevLett.44.1316, Borah:2012qr, Adhikari:2015woo, Nanda:2017bmi, Borah:2018smz, Barman:2019aku, Biswas:2019ygr, Nanda:2019nqy} and references therein.

The equations that govern the anomaly cancellation requirements in our setup are given by :
\begin{enumerate}
\item[(A)] $\mathbf{[SU(2)]^2 [U(1)_X]}$ :
\begin{equation}
\begin{split}
& \bigg(\frac{1}{2}\bigg)\times 2 \times n_1 + \bigg(\frac{1}{2}\bigg)\times 2 \times n_2 + \bigg(\frac{1}{2}\bigg)\times 2 \times n_3 - \bigg(\frac{1}{2}\bigg)\times 1 \times n_1 - \bigg(\frac{1}{2}\bigg)\times 1 \times n_2 \\ & - \bigg(\frac{1}{2}\bigg)\times 1 \times n_3 = \mathbf{n_1 + n_2 + n_3}
\end{split}
\label{condA}
\end{equation}
\item[(B)] $\mathbf{[U(1)_Y]^2 [U(1)_X]}$ :
\begin{equation}
\begin{split}
& \bigg(-\frac{1}{2}\bigg)^2 \times 2 \times n_1 + \bigg(-\frac{1}{2}\bigg)^2 \times 2 \times n_2 + \bigg(-\frac{1}{2}\bigg)^2 \times 2 \times n_3 - (-1)^2 \times n_1 - (-1)^2 \times n_2  \\ & - (-1)^2  \times n_3 = \mathbf{-\frac{1}{2}(n_1 + n_2 + n_3)}
\end{split}
\label{condB}
\end{equation}

\item[(C)] $\mathbf{[U(1)_Y] [U(1)_X]^2}$ :
\begin{equation}
\bigg(-\frac{1}{2}\bigg) \times 2 \times n_1^2 + \bigg(-\frac{1}{2}\bigg) \times 2 \times n_2^2 + \bigg(-\frac{1}{2}\bigg) \times 2 \times n_3^2 - (-1) \times n_1^2 - (-1) \times n_2^2 - (-1) \times n_3^2 = \mathbf{0}
\label{condC}
\end{equation}

\item[(D)] $\mathbf{[U(1)_X]^3}$ :
\begin{equation}
n_1^3 \times 2 + n_2^3 \times 2 + n_3^3 \times 2 - n_1^3 -n_2^3 -n_3^3 = \mathbf{n_1^3 + n_2^3 + n_3^3}
\label{condD}
\end{equation}

\item[(E)] $\mathbf{[U(1)_X]}$ :
\begin{equation}
\mathbf{n_1 + n_2 + n_3}
\label{condE}
\end{equation}
\end{enumerate}

From the above set of conditions (A-E) one can infer that :
\begin{itemize}
\item $n_1 + n_2 + n_3  = 0$ ensures anomaly cancellation of all the anomalies except eq.~\eqref{condD}.
\item In order to ensure eq.~\eqref{condD} is also zero, we can add $N$ extra fermions with $U(1)_X$ charges ($m_1, m_2,...$ etc.) such that {\large{$\mathbf{\sum \limits_{i=1}^{N} m_i= 0}$}} and {\large{$\mathbf{\sum \limits_{i=1}^{N} m_i^3 = -\bigg(\sum \limits_{i=1}^{3} n_i^3 \bigg)}$}}.
\end{itemize}
The one way of cancelling the anomaly without adding more fermions is to consider equal and opposite charges for any two generations of leptons and let the charge of the third generation be zero. These are the symmetries like $U(1)_{L_e - L_\mu}$, $U(1)_{L_\mu - L_\tau}$ which has been discussed earlier in the references \cite{Altmannshofer:2016jzy,Baek:2017sew,Han:2019diw,Biswas:2019twf}. However, if we want to consider non-zero charges for all the three lepton generations, then we need to have additional chiral fermions in our model for anomaly cancellation. So without choosing random charges and adding fermions in an ad-hoc manner, we can try to constrain the possible values of $n_1$, $n_2$ and $n_3$ from the available low-energy data. Note that $n_1$ and $n_2$ will be sensitive to the observables like $R(K^{(*)})$ as well as electron and muon anomalous magnetic moments. There will not be any contributions to the lepton flavour violating decays and the rare decays like $B_s \to \mu\mu$ or $B_s\to e e$. Also, depending on the lepton in the final state, the $b\to c\ell\nu_{\ell}$ decays (with $\ell = e, \mu, \tau$) will be sensitive to the charges as mentioned above. However, due to the low mass of $X$, the new contributions in $B \to D^{(*)} \ell\nu_{\ell}$ decays are much smaller as compared to the corresponding SM counterpart. Therefore, effectively we can get constraints on $n_1$ and $n_2$ using the available data on $b\to s \ell\ell$ decays (for $\ell = \mu, e$) and anomalous magnetic moments; however, due to unavailability of sufficient data, $n_3$ can not be constrained. We then look for possible solutions for the charges ($n_1, n_2, n_3$) such that $n_1 + n_2 + n_3 = 0$. Such a prescription also allows us to constrain the mass of $X$ and the kinetic mixing parameter effectively. The detailed analysis is described in the next section.

Note that for general $U(1)_X$ charges of leptons, one can have a general structure of charged lepton mass matrices. One can have non-diagonal terms in charged lepton mass matrix even with vector-like couplings of leptons to $X$ (equal charge of left-handed doublet and right-handed singlets). This is because of the equality of $U(1)_X$ charges across multiple fermion generations or in case we extend our model with additional Higgs doublets with appropriate $U(1)_X$ charges. In such a case, the charged lepton mass matrix has to be diagonalised using a bi-unitary transformation as follows 
$$ M_l = V_L M^{\rm diag}_l V^{\dagger}_R $$
where $M^{\rm diag}_l = {\rm diag}(m_e, m_{\mu}, m_{\tau})$. As will be discussed later, the PMNS mixing matrix will get additional contribution from charged lepton sector via $U_{\rm PMNS} =V^{\dagger}_L U_{\nu}$ where $U_{\nu}$ diagonalises the complex symmetric Majorana light neutrino mass matrix $M_{\nu} = U_{\nu} M^{\rm diag}_{\nu} U^T_{\nu}$ with $M^{\rm diag}_{\nu} = {\rm diag}(m_1, m_2, m_3)$.

\section{Analysis}
\label{sec3}
In the following subsection, we will discuss different observables which will be useful to constrain various model parameters like $U(1)_X$ charges of leptons $n_i$, new gauge coupling $g_X$, new gauge boson mass $M_X$, and the kinetic mixing parameter $\epsilon$. 

\subsection{Exclusive $b\to s\ell\ell$ (with $\ell = e, \mu$) decays}

As mentioned earlier, the measured values of $R(K^{(*)})$ in the semi-leptonic $B$-meson decay reported by the experimental collaborations provide an indication of lepton flavour universality violation (LFUV). The measured value of $R(K)$ by LHCb is given by \cite{Aaij:2019wad}
\begin{equation}
R(K)  = 0.846^{+0.060 \> + 0.016}_{-0.054 \> -0.014},
\ \ \text{for $q^2 \in [1.1,6]$ {\rm {GeV}$^2$}}, 
\label{eq:rklhcb}
\end{equation}
where $q^2$ is the squared momentum of the leptons in the final state. This result has a deviation from the SM prediction by $2.5\sigma$. Similar measurements are available for $R(K^{*})$ by LHCb and Belle collaborations. While the LHCb Collaboration has reported \cite{Aaij:2017vbb}
\begin{equation}
R(K^*) = 
\begin{cases}
0.66^{+0.11}_{- 0.07} \pm 0.03,\ \ \text{$q^2 \in [0.045,1.1]$ {\it {\rm GeV}$^2$}},\\
0.69^{+0.11}_{- 0.07} \pm 0.05,\ \ \text{$q^2 \in [1.1,6]$ {\it {\rm GeV}$^2$}},
\label{eq:rkstlhcb}
\end{cases}
\end{equation}
Belle presented their first measurement \cite{Abdesselam:2019wac} of $R(K^{*})$ in $B^0$ and $B^+$ decays in April 2019 which reports
\begin{equation}
R(K^*) = 
\begin{cases}
0.52^{+0.36}_{- 0.26} \pm 0.05,\ \ \text{$q^2 \in [0.045,1.1]$ {\it {\rm GeV}$^2$}},\\
0.96^{+0.45}_{- 0.29} \pm 0.11,\ \ \text{$q^2 \in [1.1,6]$ {\it {\rm GeV}$^2$}}.
\label{eq:rkstbelle}
\end{cases}
\end{equation}
Although the measurements from Belle are compatible with the SM expectations, they have comparatively large uncertainties. Thus, considering the more precise results from LHCb, the anomaly in $R(K^*)$ stands at $\sim 2.4\sigma$.  

\begin{figure}[t!!]
\centering
\includegraphics[scale=0.45]{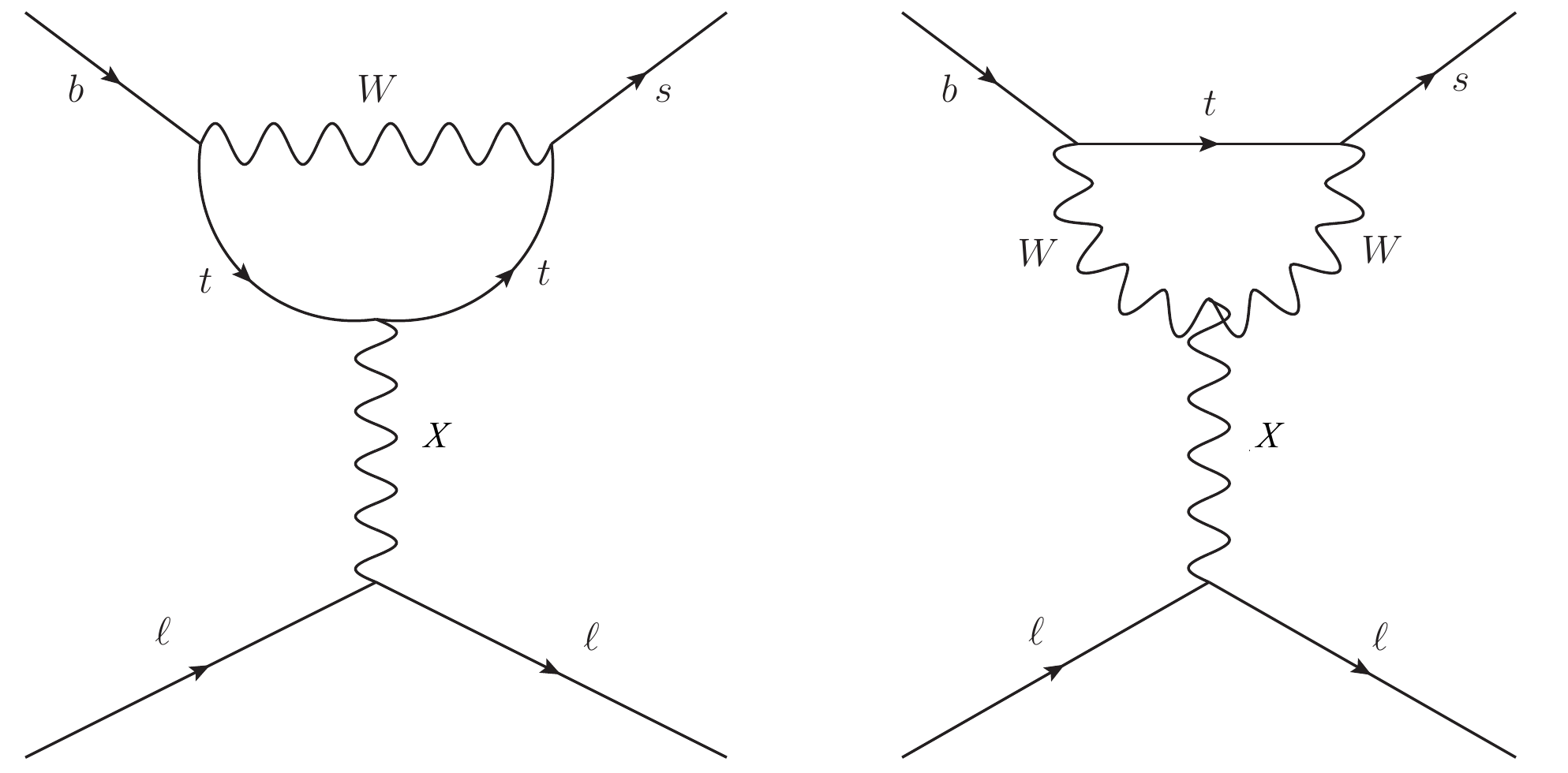}
\caption{Dominant diagrams contributing to $b \to s \ell^+ \ell^-$ decay.}
\label{Fig_btos}
\end{figure}

The effective Hamiltonian for the $b\to s$ transitions is given by \cite{Altmannshofer:2008dz}:
\begin{equation}
\mathcal{H}_{\text{eff}} = - \frac{4 G_F}{\sqrt{2}}V_{tb} V_{ts}^{\ast} \left(\sum_{i=1...6} C_i \mathcal{O}_i + 
\sum_{i=7,8,9,10,S,P} ( C_i \mathcal{O}_i + C_i^{\prime} \mathcal{O}_i^{\prime})\right) + h.c.
\label{Heff}
\end{equation}
where $O_i$ and $O_i^{\prime}$'s are the dimension six effective operators and $C_i$'s are the corresponding Wilson coefficients (WC). Although the semi-leptonic operators $\mathcal{O}_{9}^{(\prime)} \propto (\bar{s} \gamma_\mu P_{L(R)} b)(\bar{\mu} \gamma^\mu \mu)$ and $\mathcal{O}_{10}^{(\prime)} \propto (\bar{s} \gamma_\mu P_{L(R)} b)(\bar{\mu} \gamma^\mu \gamma_5 \mu)$ are relevant for the decay $b\to s \ell^+\ell^-$, the analysis with the very recent data suggests that $O_9$ is the only one operator scenario that can simultaneously explain all the data in $b\to s\ell\ell$ decays \cite{Bhattacharya:2019dot, Biswas:2020uaq}. However, there are a few two or three operator scenarios which can best explain the data at the moment, and that includes the combination $O_9$ and $O_{10}$ \cite{Bhattacharya:2019dot, Biswas:2020uaq}. In our model, the leading contributions to the Wilson coefficients will come from the diagrams shown in Fig.~\ref{Fig_btos}, and we will have contributions only in $C_9$ due to the vectorial coupling of the $X$ to the leptons. There will be contributions in both $b\to s\mu^+\mu^-$ and $b\to s e^+ e^-$ decays. As can be seen from eq.~\eqref{eq:zpint}, due to the absence of axial-vector coupling of $X$ to the leptons, we do not have contributions to $C_{10}$. Therefore, at the leading order, the new Wilson coefficient (WC) is given by  
\begin{equation}
C_9^{\ell,\rm NP} = \bigg(\frac{M_W^2 n_{\ell} g_{X} \epsilon}{ C_W S_W}\bigg)\bigg(1-\frac{4}{3}S_W^2\bigg)\bigg( \frac{1}{q^2 - M_{X}^2 + i \Gamma_{X} M_{X}}\bigg) \times C(x_t)
\label{eq:C9NP}
\end{equation}
with 
\begin{equation}
C(x_t) = \frac{x_t}{8}\left[\frac{6-x_t}{1-x_t} + \frac{3 x_t +2}{(1-x_t)^2} ln(x_t)\right] , \ \ \ \ x_t = \frac{m_t^2}{M_W^2}. 
\label{Cxt} 
\end{equation}
Here, $m_t$ and $M_W$ are the top quark and $W$-boson masses, respectively. The sine of the Weinberg angle is defined as $S_W = \text{sin } \theta_W$ and $C_W = \sqrt{1-S_W^2}$. Also, $n_{\ell} \equiv (n_1,n_2)$ depending upon the lepton flavour it contributes to.

As mentioned earlier, to build a UV complete theory, gauge anomalies should cancel, for which we need to introduce new heavy fermions in our theory in addition to the $X$ boson. It is important to note that there can be an additional contribution to $C_9$ due to the anomalous coupling of the longitudinal mode of $X$ boson with SM gauge bosons \cite{Dror:2017ehi, Dror:2017nsg}.  Depending on the masses of the heavy fermions, the contributions can be significant. However, such contributions from the Wess-Zumino terms will only occur if the new fermions have vectorial coupling with the SM gauge bosons and chiral coupling with the $U(1)_X$ gauge boson. As we will see later, in our construction of the toy models, we have added only three right-handed neutrinos which do not couple to the SM gauge bosons. We do not have any other exotic fermions in our models.  Hence, we will not have any such contributions as mentioned above from the longitudinal mode of $X$ boson.

Note that we are working in a model with the mass of $X$ in the GeV or sub-GeV range, in particular, we are focusing in the region $M_X > 2 m_{\mu}$. On the other hand for $B\to K^{(*)}\ell\ell$ decays, the allowed values of $q^2$ lie in the range $4m_{\ell}^2 < q^2 < (M_B-M_{K^{(*)}})^2$. In such a situation, one cannot Taylor expand the $X$ propagator in powers of $q^2/M_{X}^2$. Therefore, the new WC, as shown in eq.~\eqref{eq:C9NP} will have explicit $q^2$ dependence and in general, could be complex. Note that for the $X$-boson, we have introduced the Breit Wigner (BW) propagator. In this form of the propagator, we will get a finite analytic expression for the amplitude at the resonance region. This is because, around the mass of $X$, the zeroth-order propagator vanishes and the higher-order effects are leading, which is given by the imaginary part proportional to the $X$ decay width. The imaginary part will receive contributions from every particle into which $X$ can decay. In general, without a priori knowledge of all the decay channels of $X$, it is hard to predict its total decay width. However, we have considered a leptophilic $X$, and its primary decay channels are the dilepton final states, like $\ell^+\ell^-$ and $\nu\bar{\nu}$ with $\ell = e, \mu$. Hence, one needs to estimate the decay width $\Gamma_{X} \approx \Gamma(X \to \ell\ell) + \Gamma(X \to \nu\bar{\nu})$.  
 
In this model, there are free parameters which need to be constrained using the existing data. In particular, the constraints from low energy experiments, like neutrino trident production (NTP) bound, rare kaon decay $K^+ \to \nu_{\mu}\mu^+ X(\to \nu\bar{\nu})$, BaBar $4\mu$ channel search etc. along with cosmological observations of Big Bang nucleosynthesis (BBN) are important. As can be seen from \cite{Ilten:2018crw, Jho:2019cxq,Bauer:2018onh}, the current data allow a gauge coupling $n_{\ell} g_X \sim 0.0018$ for $M_{X} \sim 0.5$ GeV and it could be $\gsim 0.003$ for $M_{X} \gsim 1.0$ GeV\footnote{Note that the experimental bounds exist on the combined quantity $(n_\ell \times g_X)$ and therefore a proper rescaling with the lepton charge is required in order to correctly infer the bound on $g_X$.}. Depending on the values of the lepton charges the upper limits on $g_X$ would scale accordingly. For example, for $n_\ell = 2$, $g_X$ as large as 0.001 is allowed for $M_X = 0.5$ GeV, and it will be $\approx 0.0015$ for $M_X= 1$ GeV\footnote{Note that depending on the mass $M_X$, the bound obtained on the coupling $g_X$ in the refs. \cite{Ilten:2018crw, Jho:2019cxq,Bauer:2018onh} and from BaBar $4\mu$ channel search \cite{TheBABAR:2016rlg} will be little more relaxed in our case. The obtained bound depends on the assumption that the $Z'$ couples with all the charged leptons and neutrinos with the same strength, while in our case coupling strengths are not the same.}. On the other hand, the kinetic mixing parameter $\epsilon$ is constrained from neutrino-electron scattering experiments like CHARM-II, GEMMA and TEXONO, for details see \cite{Lindner:2018kjo}. Mixing strength $\gsim 10^{-3}$ is ruled out for gauge bosons of mass around the electroweak (EW) scale. For keV scale bosons, the bound is even tighter $\mathcal{O}(10^{-6})$. LEP II has put a lower bound on the ratio of new gauge boson mass to the new gauge coupling to be $M_{X}/g_{X} \ge 7 $ TeV \cite{Carena:2004xs}. However, since we are interested in the low mass of the gauge boson, bounds from hadron colliders like ATLAS and CMS will not be very relevant. Similarly, LEP bound is also not applicable in such low mass regime\footnote{As an example, one could see at the ref. \cite{Deppisch:2019ldi} for a detail of the direct search bounds on such a light gauge boson.}. With all these inputs, the $X$ decay width as mentioned above, will be of order ${\cal O}(10^{-9}$-$10^{-7})$ GeV for $g_X \in  (10^{-4},10^{-3})$, which is much smaller than $M_{X}$. In the limit $\frac{\Gamma_{X}}{M_{X}}\to 0$ (narrow-width approximation (NWA)), the BW becomes a delta distribution:  $\delta(q^2-M_{X}^2)$.  

\begin{figure}[t]
\begin{center}
	\subfigure[]
	{\includegraphics[scale=0.35]{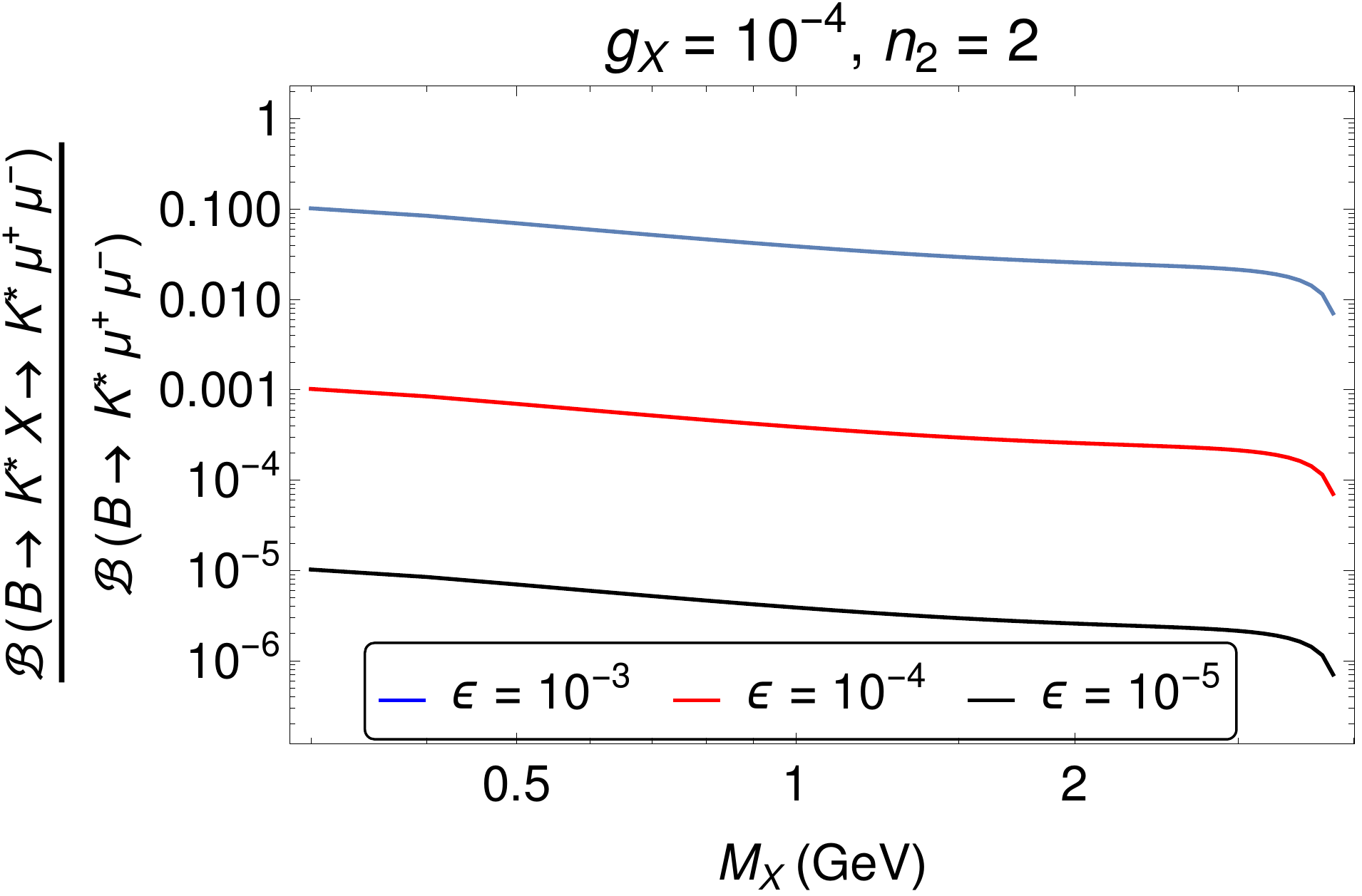}~~
	\label{fig:vareps}}
	\subfigure[]
	{\includegraphics[scale=0.35]{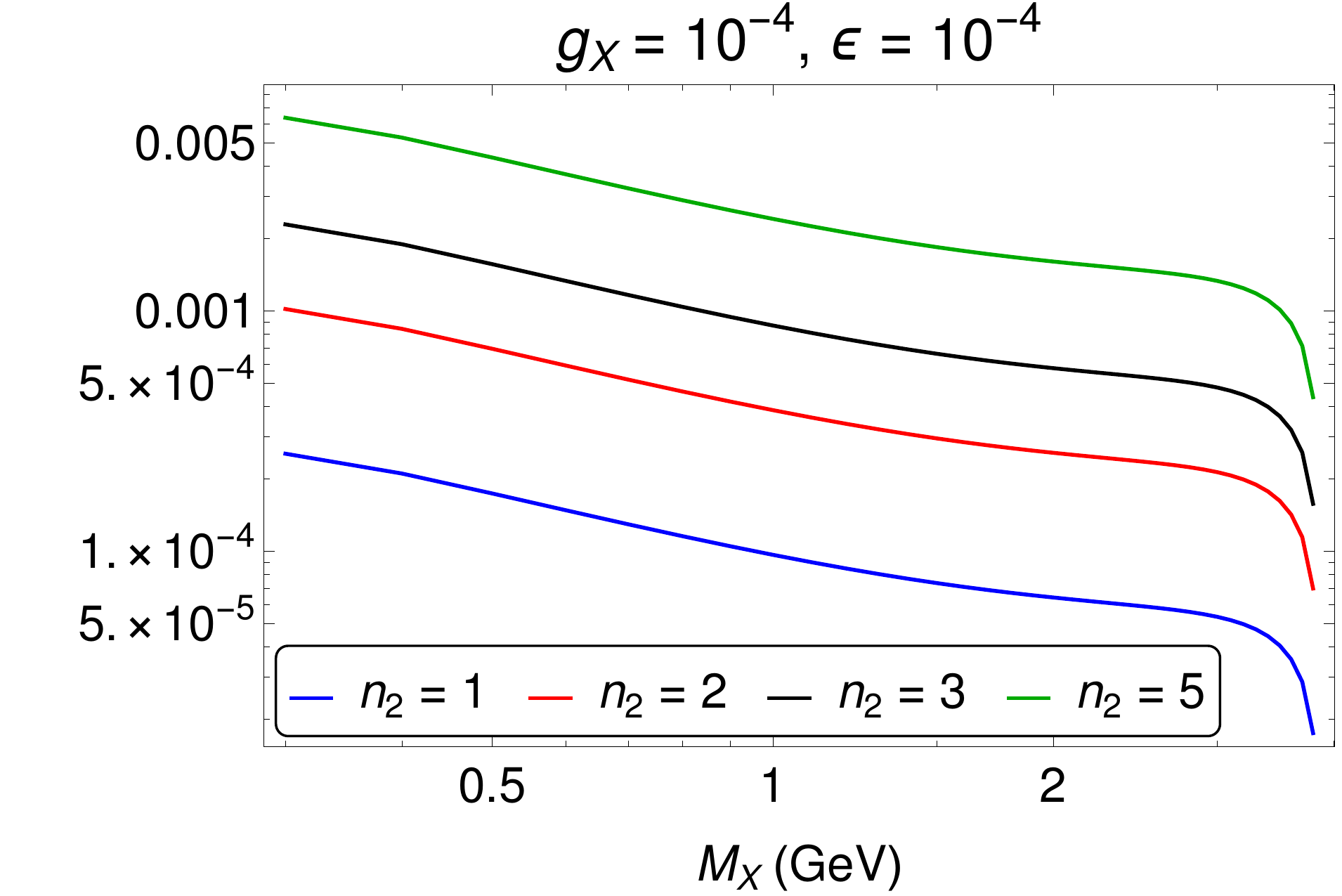}
	\label{fig:varchr}}
	\subfigure[]
	{\includegraphics[scale=0.35]{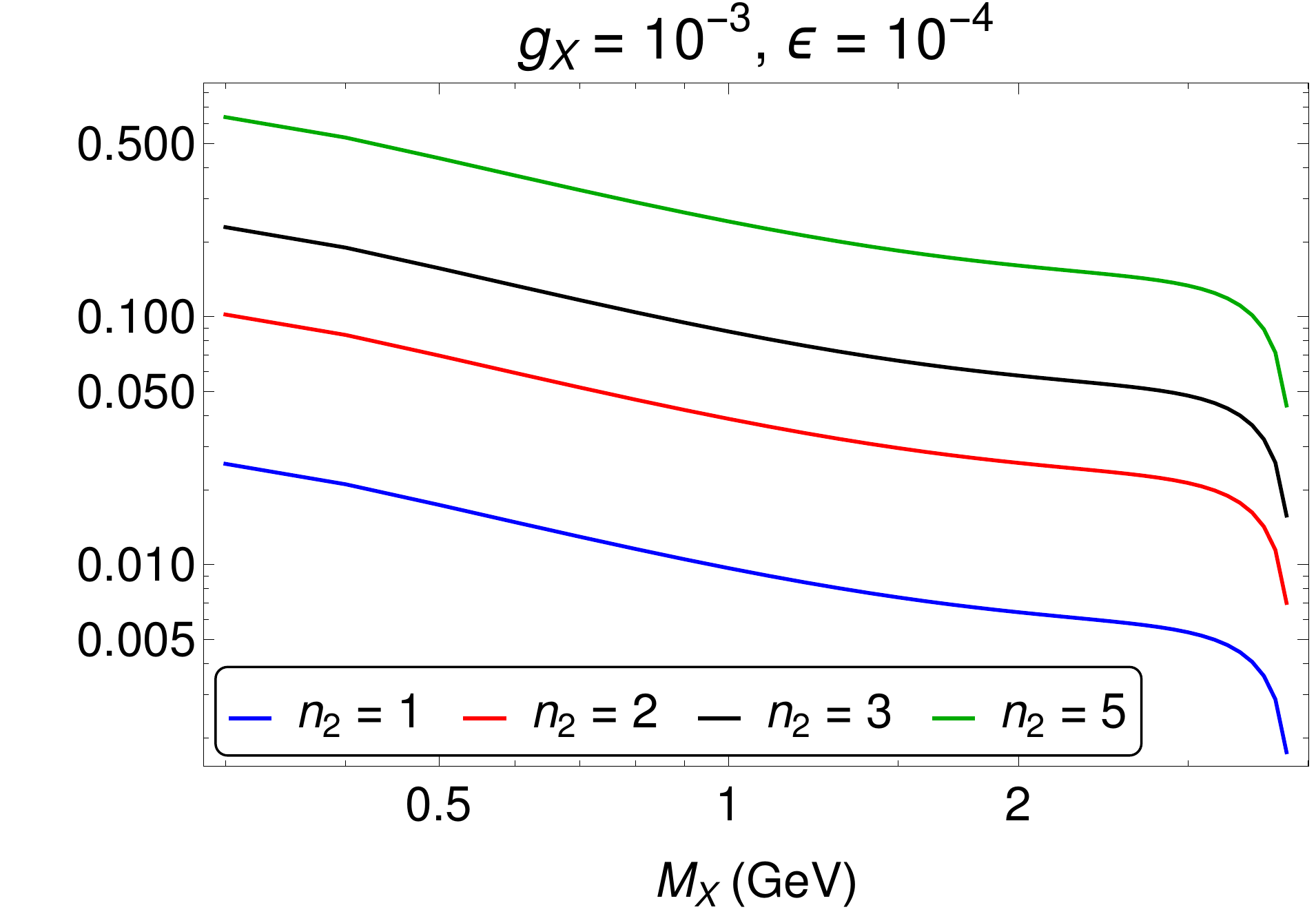}
	\label{fig:varchr2}}
\caption{Variation of $\mathcal{B}(B^0 \to  K^{*0}\chi (\mu^+ \mu^-))/\mathcal{B}(B^0 \to K^{*0}\mu^+ \mu^-)$ as a function of the mass $M_X$ for different values of (a) the mixing parameter `$\epsilon$' and (b) the $U(1)_X$ charge $n_i$. From Figs.~(b) and (c), we can check the dependence of the above ratio on the coupling $g_X$.}
\label{fig:lhcbmodel}
\end{center}
\end{figure}

LHCb has done a dedicated search for light hidden-sector bosons by measuring the branching fraction $\mathcal{B}(B^0 \to K^{*0} \chi(\mu^+ \mu^-)) = \mathcal{B}(B^0 \to K^{*0} \chi)\times \mathcal{B}(\chi\to\mu^+ \mu^-)$. Here, $\chi$ is the light boson in the hidden sector similar to $X$ in our case. Depending on the lifetime $\tau(\chi)$, LHCb has put bounds on the above mentioned branching fraction for a given mass range of $\chi$ \cite{Aaij:2015tna}. One can refer to Fig.7 of the supplemental material of reference \cite{Aaij:2015tna} in which the ratio $\mathcal{B}(B^0 \to  K^{*0}\chi (\mu^+ \mu^-))/\mathcal{B}(B^0 \to K^{*0}\mu^+ \mu^-)$ has been plotted as a function of $m(\chi)$ (with $214 \le m(\chi) \le 4350$ MeV) for different values of $\tau(\chi)$ including $\tau(\chi) = 0$. Here, the branching fraction $\mathcal{B}(B^0 \to K^{*0}\mu^+ \mu^-)$ is defined for $1.1 < m^2(\mu^+ \mu^-) < 6.0$ GeV$^2$. Note that if we choose the lifetime $\tau(\chi) = 1000$ ps, which corresponds to a very small decay width of $X$, the ratio as mentioned above could be of order one. However, the bounds on the same ratio will be $\le {\cal O}(10^{-2})$ (at 95\% Confidence Level (CL)) for $\tau(\chi) = 10$ ps, which is even the case in the limit $\tau(\chi) \to 0$. The width $\Gamma_{X} \approx 10^{-9}$ GeV  corresponds to a lifetime $\approx 10^{-4}$ ps which is close to zero. 

In our model, we have estimated $\mathcal{B}(B^0 \to  K^{*0} X (\mu^+ \mu^-))/\mathcal{B}(B^0 \to K^{*0}\mu^+ \mu^-)$ within the accessible ranges of $M_{X}$. The normalisation $\mathcal{B}(B^0 \to K^{*0}\mu^+ \mu^-)$ has been measured by LHCb for ${1.1 < q^2 < 6.0}$ GeV$^2$ \cite{Aaij:2013iag}, which is given by $(1.6 \pm 0.3)\times 10^{-7}$. The dependences of this ratio on different model parameters like $\epsilon$, the charge $n_2$ and the coupling $g_X$ are shown in Fig.~\ref{fig:lhcbmodel}. A close inspection of Figs.~\ref{fig:varchr} and \ref{fig:vareps} suggests that if we choose $\epsilon \lsim 10^{-4}$, for values of $n_2$ as large as 5, the constraints from LHCb will be satisfied within the accessible ranges of $M_{X}$. However, even though for $g_X = 10^{-3}$, $n_2 \sim 5$ is allowed by the LHCb constraints as shown in Fig.~\ref{fig:varchr2}, it will not be able to satisfy other low energy experimental bounds mentioned previously. To satisfy the low energy bounds for $n_2 = 5$, we need $g_X \lsim 0.5 \times 10^{-4}$. In the rest of our analysis, we will consider $\epsilon \approx 10^{-4}$. Note that even for a relatively large gauge coupling ($\sim 10^{-3}$) we can still be able to satisfy the upper bound provided by LHCb. 

\subsection{Anomalous Magnetic Moments}\label{sec:muong2x}
\begin{figure}[t]
	\begin{center}
		\includegraphics[scale=0.6]{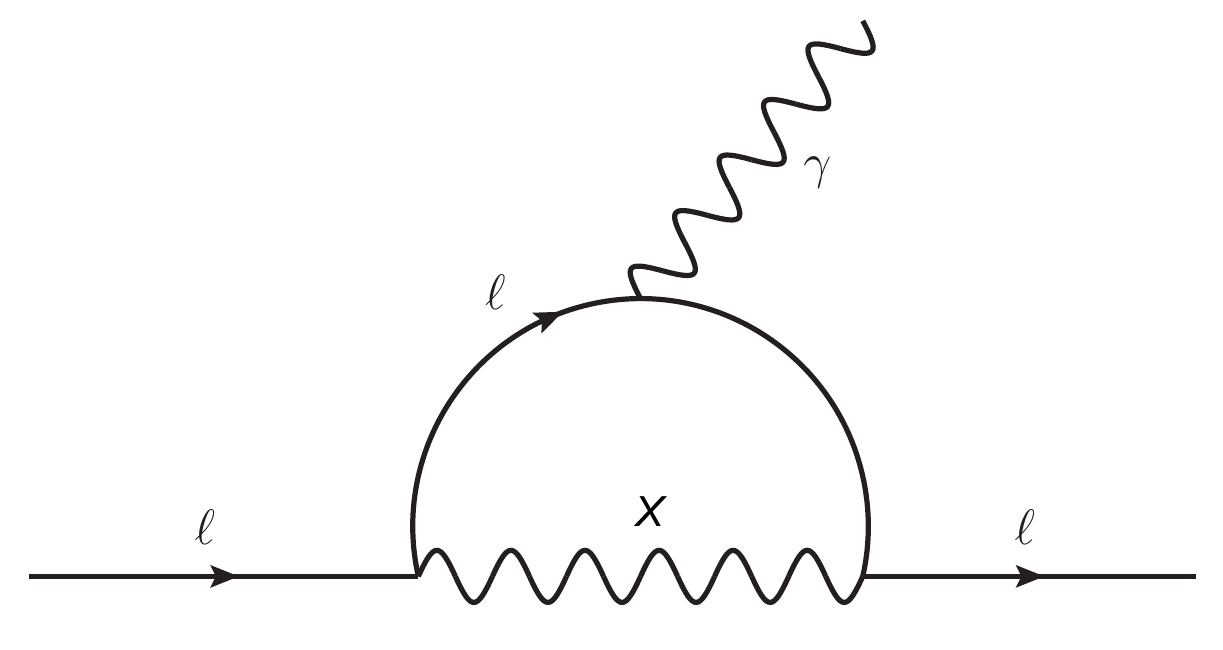}
		\caption{Diagram contributing to the anomalous magnetic moment of charged leptons.}
		\label{fig:g2}
	\end{center}
\end{figure}

Another important observable which could be useful to put tight constraints on the model parameters is the anomalous magnetic moments of muon or electron. As one can see from  eq.~\eqref{eq:zpint}, since we do not have lepton-flavour violating couplings of $X$, the gauge boson mediated diagram will not contribute to decays like $\tau \to \mu \gamma, \mu \to e \gamma$ etc. 

\begin{figure}[t]
	\begin{center}
		\includegraphics[scale=0.4]{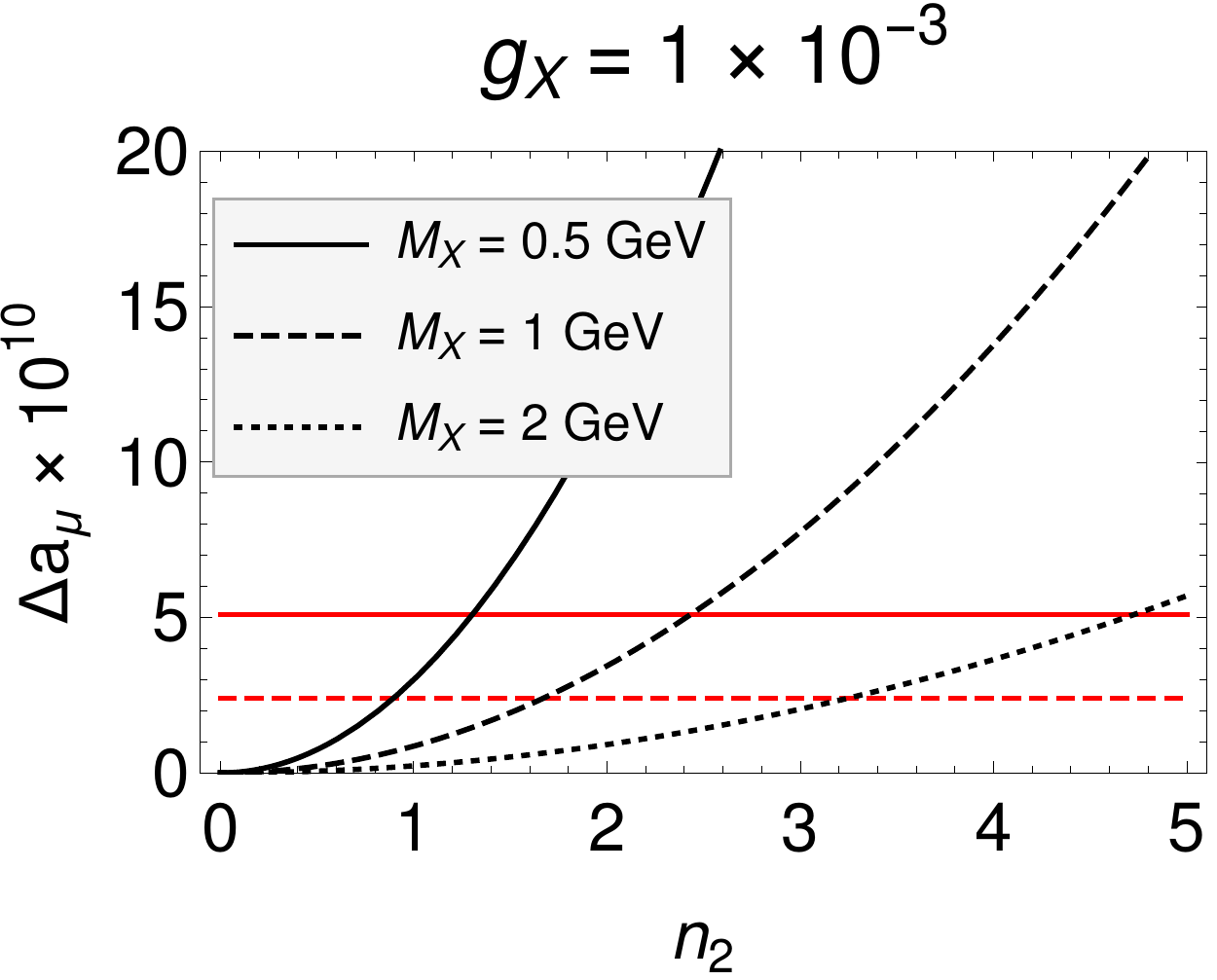}~~
		\includegraphics[scale=0.4]{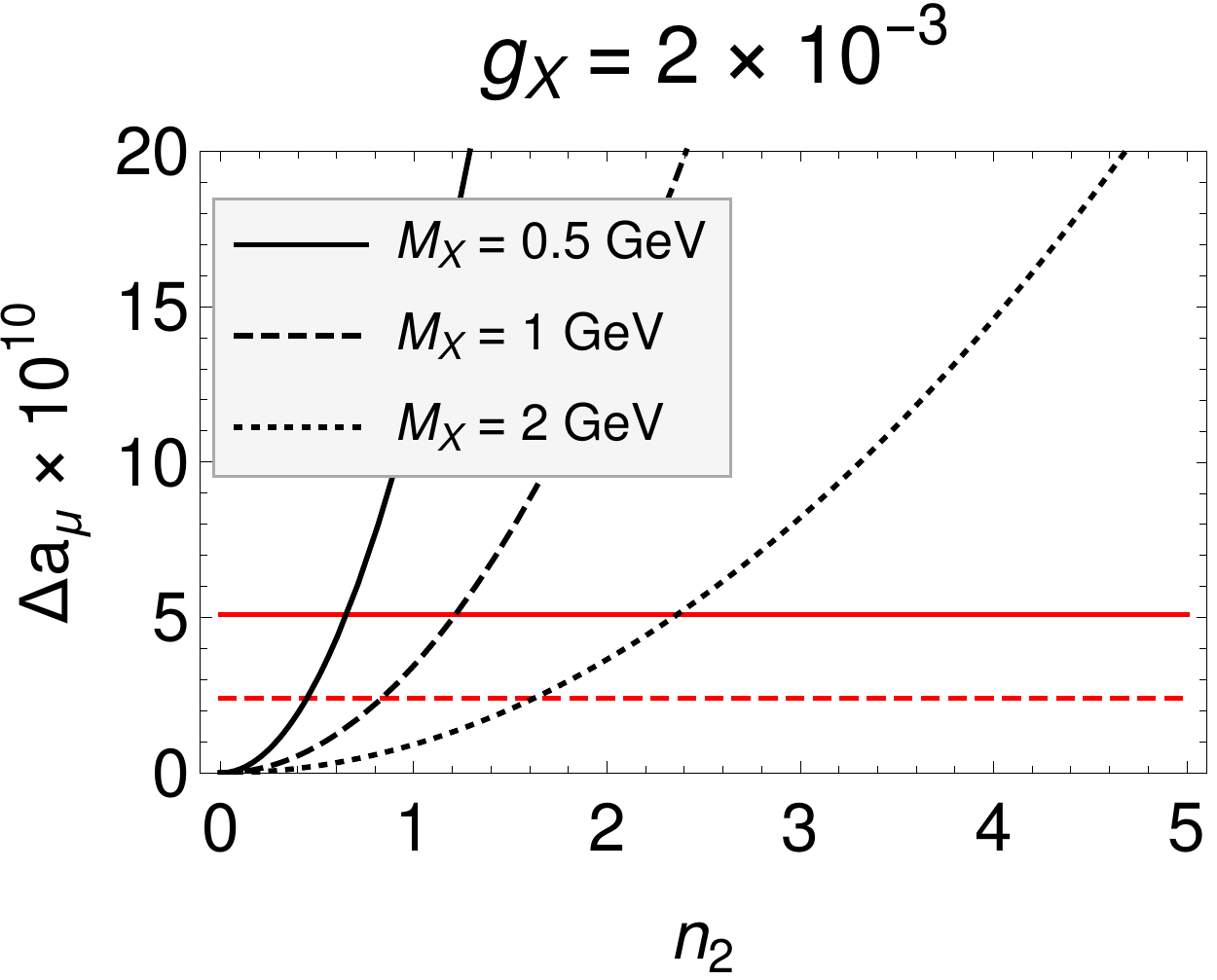}~~
		\includegraphics[scale=0.4]{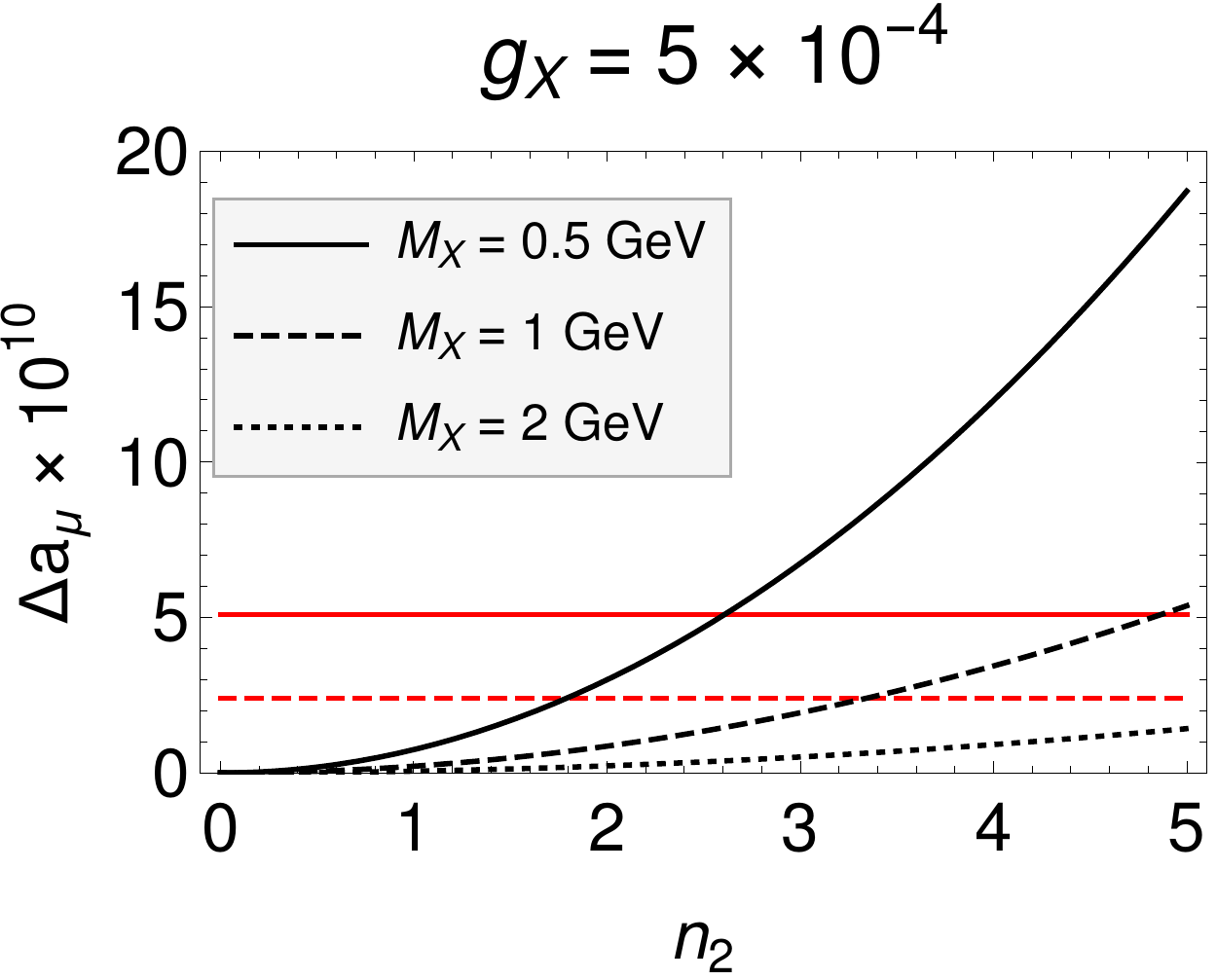} \\
		\caption{The dependencies of the muon anomalous magnetic moment ($\Delta a_{\mu}$) on the different new physics model parameters. The red-dashed and solid lines represent the 3$\sigma$ lower limit of the measured values of $\Delta a_{\mu}$ given in eqs.\eqref{eq:delamu_old} and \eqref{eq:delamu_new} respectively.}
		\label{fig:muong2zp}
	\end{center}
\end{figure}

The effective vertex of photon with any charged particle is given by: 
\begin{equation}
{\bar u}(p^{\prime}) e \Gamma_\mu u(p) = {\bar u}(p^{\prime})\left[e \gamma_{\mu} F_1(q^2) + \frac{i e \sigma_{\mu\nu} q^{\nu}}{2 m_f} F_2(q^2)+ ...\right] u(p).
\end{equation}
The factor $g_{\mu} \equiv 2 (F_1(0) + F_2(0))$, and the anomalous magnetic moment is given as $a_{\mu} \equiv F_2(0) \ne 0$ (since $F_1(0) = 1$ at all order). In our model, the diagram that will contribute to muon and electron anomalous magnetic moments is given in Fig.~\ref{fig:g2}. In our model, the contribution to $\Delta a_{\ell}$ is given by 
\begin{equation}
\Delta a_\ell^{(X)} = \frac{n_\ell^2 g_{X}^2}{8\pi^2} \frac{m_\ell^2}{M_{X}^2} \int_0^1 dx \frac{2x^2 (1-x)}{(1-x) + x^2   r_\ell (x) }
\label{g2_Zprime}
\end{equation}
where $r_\ell (x) = \left(\frac{m_\ell^2}{M_X^2}\right)$, $\ell \equiv e, \mu$ and $n_\ell (= n_i)$ denotes the $U(1)_X$ charge of the lepton. Our analytical expression can be compared with the one obtained in \cite{Lindner:2016bgg}. Note that the contributions in $\Delta a_{\ell}$ for both $\ell = \mu$ and $e$ are positive; however, in the case of electron magnetic moment, the expectation is negative. Also, as compared to the requirement, the contribution in electron anomalous magnetic moment is negligibly small. The dependences of $\Delta a_{\mu}$ on various model parameters are shown in Fig. \ref{fig:muong2zp}. We can easily explain the excess in $\Delta a_{\mu}$ for values of $g_{X}$ of order ${\cal O}(10^{-3})$, and the data prefers a value of $M_{X} \lsim 1$ GeV. In such situation, the value of $n_2$ need not be $>>1$. However, if we choose $g_{X} \approx 10^{-4}$ then in order to explain the excess in muon $(g-2)$, we need relatively larger values of $n_2 (>> 1)$.  

We have already pointed out in the introduction that the current measurement of the fine structure constant poses a negative $\sim 2.4 \sigma$ deviation in anomalous magnetic moment of electron from its theory prediction \cite{Parker_2018}:
\begin{equation}
\Delta a_{e} = - 8.8 (3.6)\times 10^{-13}.
\label{eq:delae}
\end{equation}
In electron anomalous magnetic moment, the contribution from the diagram in Fig.~\ref{fig:g2} will be positive and is given by 
\begin{equation}
\Delta a_{e}^{(X)} = 2.11 \times 10^{-15}
\label{eq:ae_X}
\end{equation} 
for $M_X = 1$ GeV, $g_X = 0.001$ and $n_1 = 1$. Hence, we can not explain the current trend of data in $\Delta a_{e}$ with only an additional $U(1)_X$ gauge boson. 

\subsection{Combined parameter spaces}

In this subsection, we discuss the constraints obtained on the model parameters from a simultaneous analysis of the observables in $b\to s\ell\ell$ decays and $\Delta a_{\mu}$. As we can see from eqs. \eqref{eq:rkstlhcb} and \eqref{eq:rkstbelle}, data are available in two different $q^2$ regions, one for  $q^2 \in [0.045,1.1]$ {\it {\rm GeV}$^2$} (low-$q^2$) and the other for $q^2 \in [1.1,6]$ {\it {\rm GeV}$^2$} (high-$q^2$). In our analysis, we have considered the inputs from $R(K^{(*)})$, $\mathcal{B}(B\to K^{(*)} \mu^+\mu^-)$ (in both the $q^2$ regions) and $\Delta a_{\mu}$. For $R(K^{(*)})$, we have not considered the Belle data since it has significant errors, and we have considered the LHCb data on it at their 2$\sigma$ CL. Similarly, $\Delta a_{\mu}$ has been considered in its 3$\sigma$ CL interval.

\begin{figure}[t]
	\centering
	{{\includegraphics[width=0.32\textwidth]{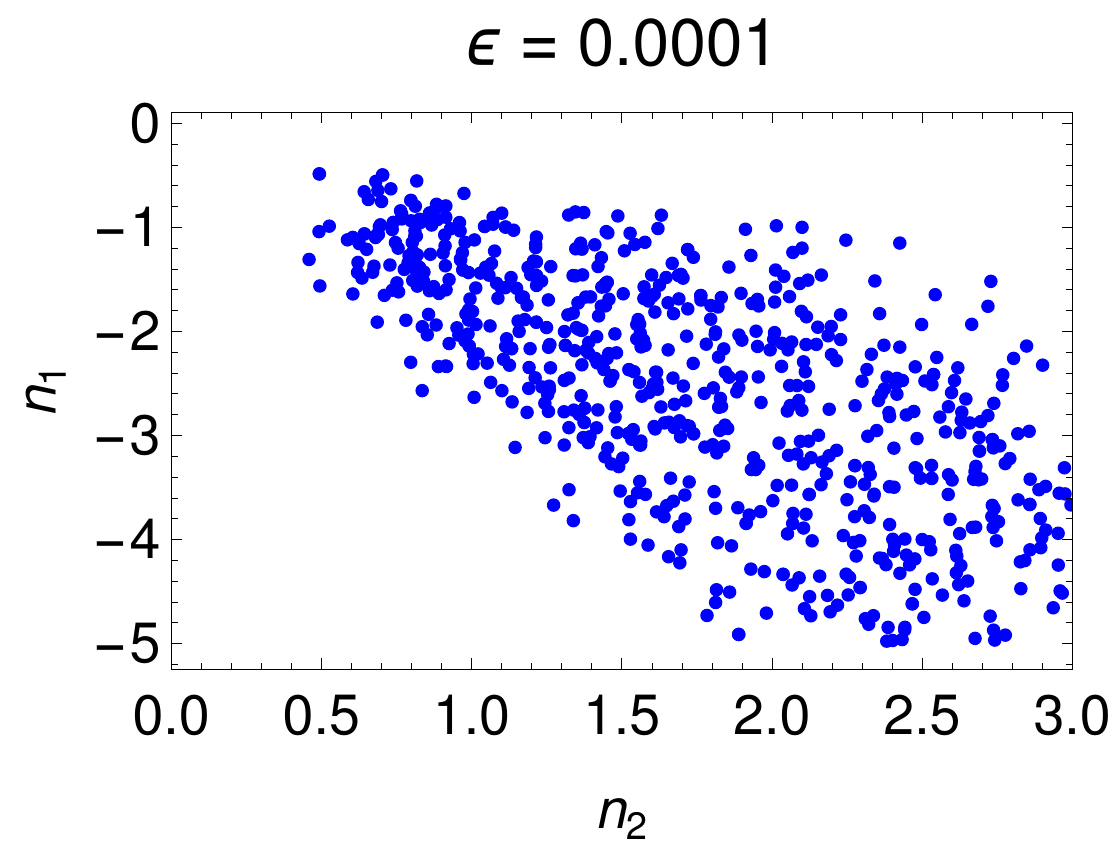}}~
		{\includegraphics[width=0.32\textwidth]{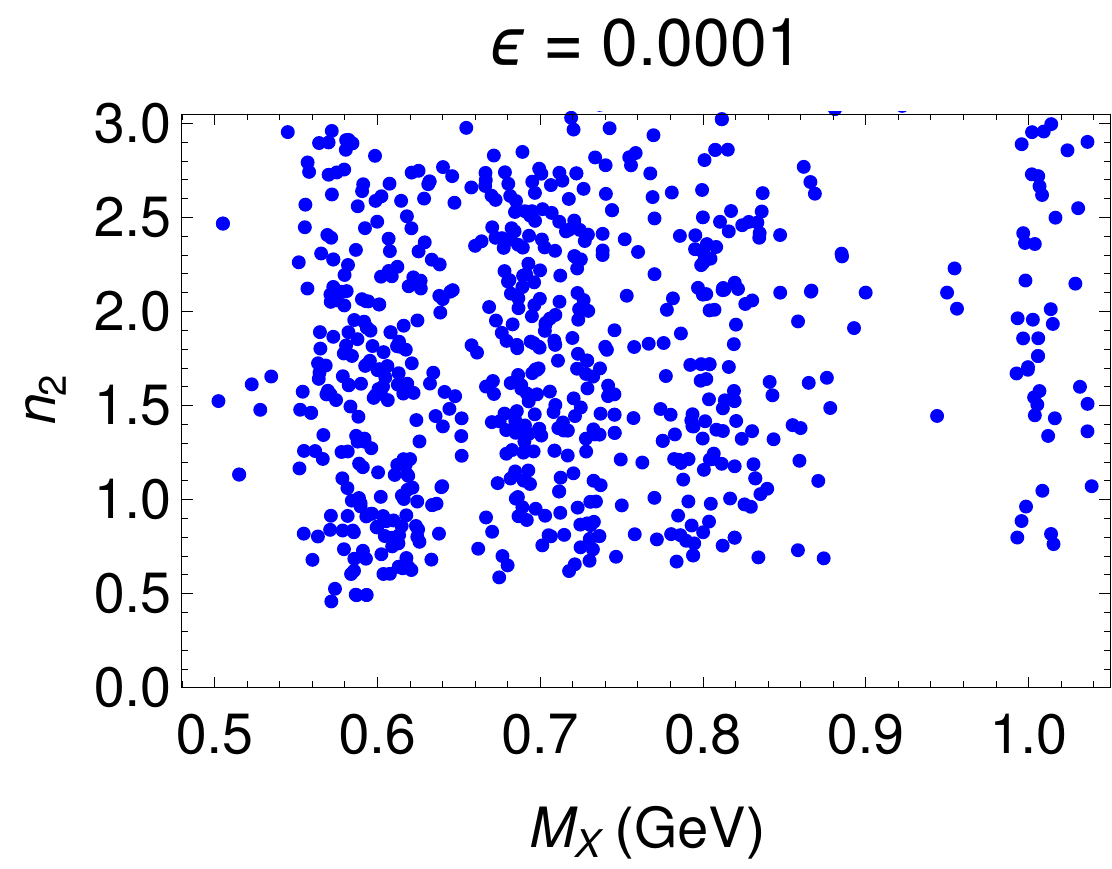}}~
		{\includegraphics[width=0.32\textwidth]{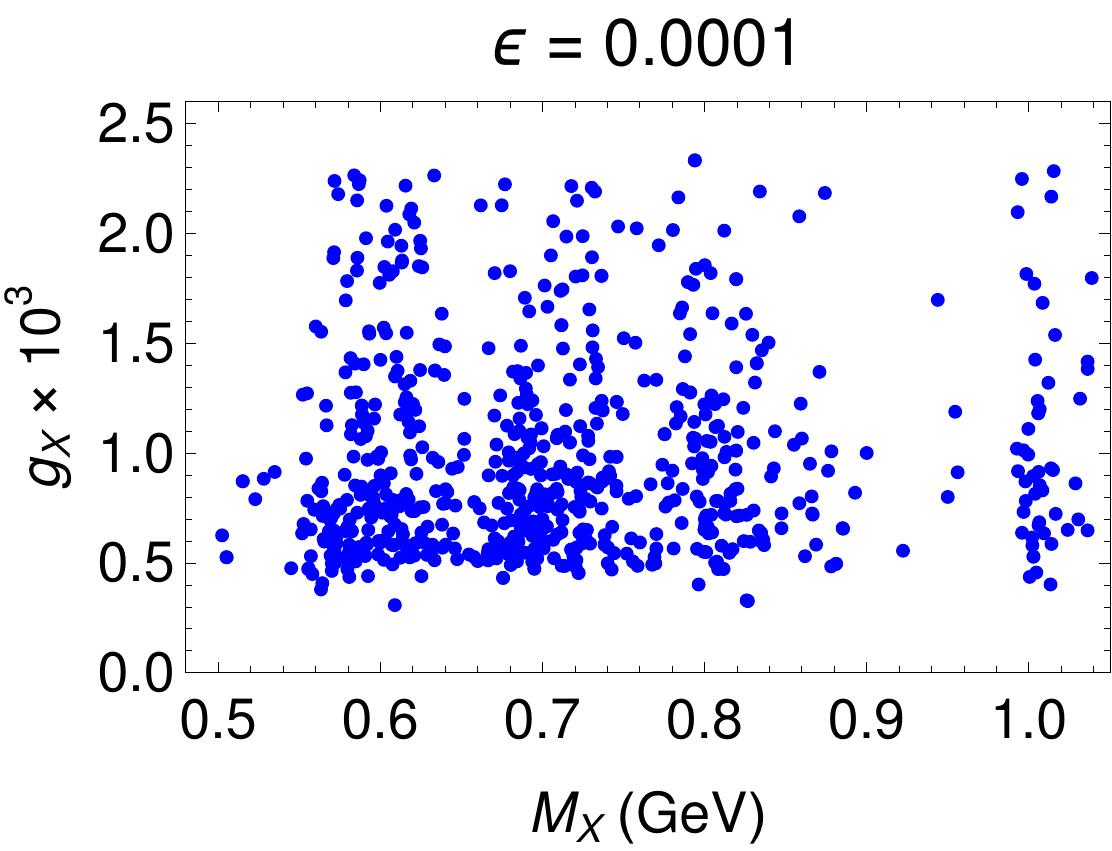}\label{fig:conshighq2}}}\\
	\caption{The allowed parameter spaces for $n_1$, $n_2$, $g_X$ and $M_{X}$ which passes the constraints from $R(K^{(*)})$, the anomalous magnetic moment of the muon and $\mathcal{B}(B\to K^{(*)}\mu\mu)$. Note that the observables in $b\to s\ell\ell$ decays are defined in high ( $1.1 < q^2 < 6.0 $ (GeV$^2$)) as well as in the low-$q^2$ ($0.045 < q^2 < 1.1$ (GeV$^2$)) regions.}
	\label{fig:allscan}
\end{figure}

As mentioned earlier, the mixing parameter $\epsilon$ plays a crucial role in constraining the other relevant new parameters. We have noted that when $\epsilon \approx 10^{-4}$ the allowed regions of the other parameters are more relaxed than the one obtained for $\epsilon \approx 5\times 10^{-4}$. We scan the parameters over the following intervals: $0.5 \le M_X \le 1.5$ (in GeV), $-5 \le n_1 \le 5$,  $-5 \le n_2 \le 5$, $0.1 \le g_{X}(\times 10^{3}) \le 3$. Here, we would like to mention that the low mass regions $0.22 < M_{X} < 0.5$ (GeV) are also allowed by the data as discussed above; in the next section, we will show it in a specific scenario. The allowed parameter spaces for $\epsilon= 1\times 10^{-4}$ are shown in Fig.~\ref{fig:allscan}. Note that $n_2$ and $n_1$ have a nice correlation, higher positive values of $n_2$ prefers higher negative values of $n_1$. Also, within our chosen parameter values, only negative values of $n_1$ are allowed. For a fixed value of $n_2$, a wide range of values of $n_1$ is allowed. However, as expected, the scenario $n_1 = n_2$ is excluded. Here, we have shown only the positive values of $n_2$, which are allowed by the data. The allowed values of $n_2$ are symmetrically distributed about the origin along the $n_2$-axis. In addition, we see that for $\epsilon \approx 1\times 10^{-4}$, within the given range of $M_{X}$, the allowed values of the coupling $g_{X}$ lies in between $ 0.5 \times 10^{-3}$ and  $2 \times 10^{-3}$. To be conservative, we have not considered values of $g_{X}$ larger than $2\times 10^{-3}$ since other low energy observables constrain higher values, for details see \cite{Ilten:2018crw, Jho:2019cxq,Bauer:2018onh,TheBABAR:2016rlg}.  

\section{The extension of $U(1)_X$ with additional degrees of freedom}
\label{sec4}

\begin{figure}[t]
\centering
\includegraphics[scale=0.4]{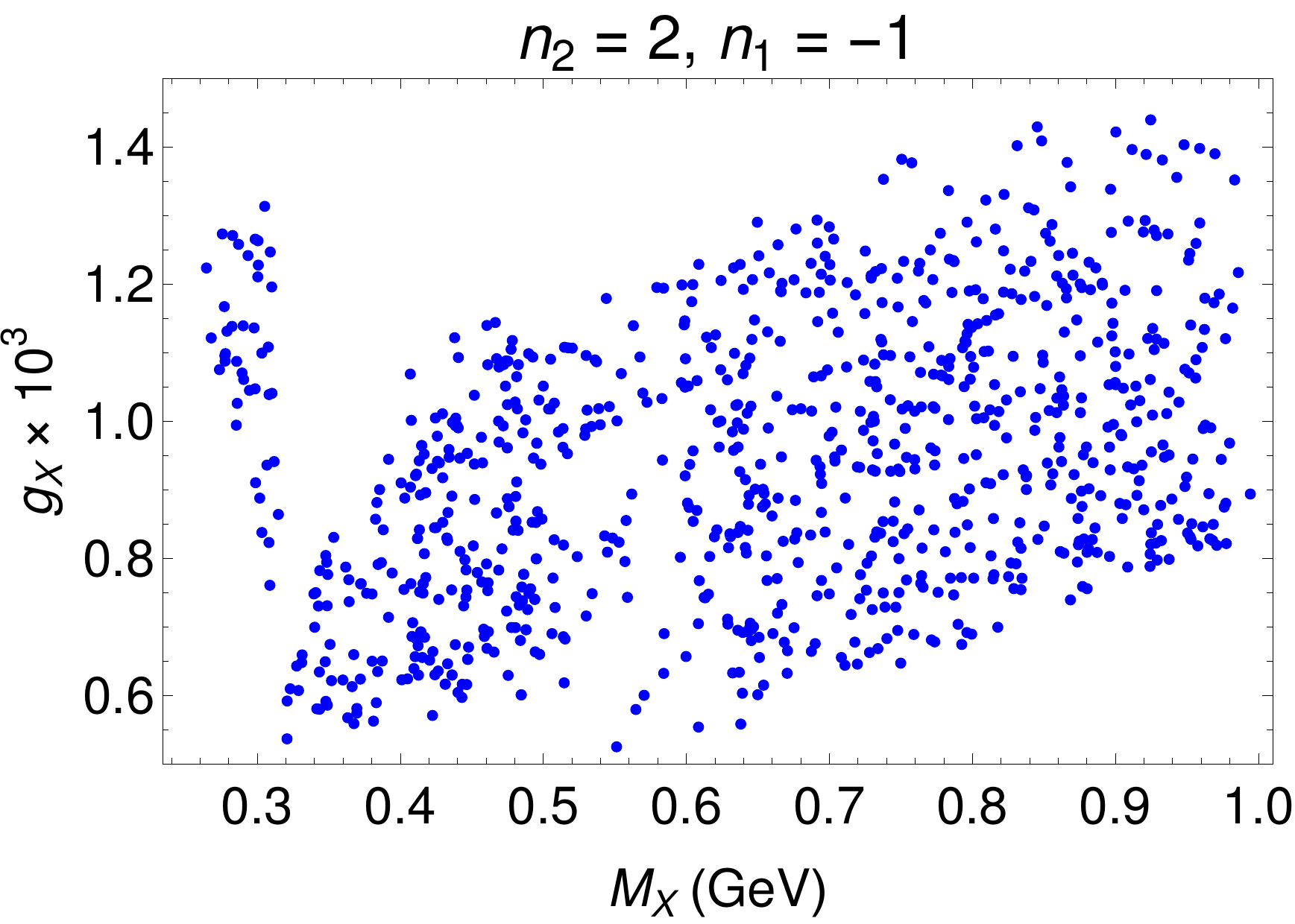}
\caption{Correlation between $M_X$ and $g_X$ constrained from all low energy flavour data and muon anomalous magnetic moment for $n_2$ and $n_1$ fixed at 2 and -1 respectively.}
\label{fig:Mxgx}
\end{figure}

A certain combination of $n_1,n_2$ would lead to a particular extension of the SM with chiral fermions. While such an extension is not unique, we stick to minimal possible extensions in order to address the problems discussed earlier. Therefore we can now proceed towards making a specific choice for these parameters in order to complete our model in a way that the extension is minimal. We have already been able to constrain $n_1$ and $n_2$ from low energy data while $n_3$ remains unconstrained. One can easily see that the minimal way to cancel the anomalies would be to add three chiral singlet fermions with $U(1)_X$ charges $n'_1 = -n_1$, $n'_2 = -n_2$ and $n'_3 = -n_3$. This will make the sum of the charges as well as sum of the cubes of the charges equal to zero. The three fermions can be considered to be 3 right-handed neutrinos (RHNs). As a benchmark scenario we choose $(n_1,n_2,n_3) = (-1,2,-1)$ which is in good agreement with the flavour data as shown in Fig.~\ref{fig:allscan}. For $n_2=2$,$n_1=-1$, the correlation between $M_X$ and $g_X$ is shown in Fig.~\ref{fig:Mxgx}, here we have shown the region $0.25 \lsim M_X \lsim 1.0$ (GeV). For these values of $[n_2, n_1]$ and for $g_X = 0.001$, within the allowed ranges of $q^2$ and $M_X$ the numerical values of the WCs $\Delta C_9^{\mu}$ and $\Delta C_9^{e}$ will lie in between $[-0.827, -1.83]$ and $[0.413, 0.91]$, respectively. These values of the WCs are consistent with the result (within 2$\sigma$ CI) obtained from a global fit to all the available data in $b \to s \ell\ell$ decays considering NP effects in both the muon and electron final states \cite{Hurth:2020rzx}.

Since the light gauge boson also couples to SM neutrinos, our model will have contributions to both exclusive and inclusive rare B-meson FCNC decay to invisible final states. The present upper limits on such modes are \cite{pdg2018}
\begin{equation}
\begin{aligned}
\mathcal{B}(B^+ \to K^+ \nu \bar{\nu}) & < 1.6 \times 10^{-5}, \\
\mathcal{B}(B^0 \to K^{0*} \nu \bar{\nu}) & < 1.8 \times 10^{-5}.
\end{aligned}
\end{equation}
In Table.~\ref{tab:BtoKnunu}, we have specified the SM and NP contributions to the branching fractions of these rare decay modes considering only central values of form factors and other decay parameters \cite{Buras:2014fpa, Calcuttawala:2017usw}. Our choice of the light gauge boson mass and mixing modifies the SM prediction of the branching fraction of the exclusive decay channels by $\sim (1-3)\%$ only while the inclusive $B \to X_s \nu \bar{\nu}$ branching is enhanced by upto $\sim 20\%$. At present, the predicted branching fractions are well within the current experimental limit. Note that no experimental bounds are available on $\mathcal{B}(B \to X_s \nu \bar{\nu})$.  

\begin{table}[!h]
\footnotesize
	\centering
	\renewcommand{\arraystretch}{1.5}
	\begin{tabular}{|c|c|c|c|}
		\hline
		 & $\mathcal{B}(B^+ \to K^+ \nu \bar{\nu}) \times 10^6$ &  $\mathcal{B}(B^0 \to K^{0*} \nu \bar{\nu}) \times 10^6$ & $\mathcal{B}(B \to X_s \nu \bar{\nu}) \times 10^6$\\
		\hline
		\bf SM & \bf 3.90 & \bf 9.12 & \bf 28.11  \\
		$M_X = 0.3$ GeV & 4.02 & 9.39 &  34.07 \\
		$M_X = 0.6$ GeV & 3.94 & 9.15  & 28.51 \\
		$M_X = 0.9$ GeV & 3.95 & 9.49  & 35.23\\
		\hline
	\end{tabular}
	\caption{Predictions for the SM and NP branching fractions for the rare B-meson decay to a pair of neutrinos. The NP branching fractions are mentioned for three different light gauge boson masses considering $g_X = 10^{-3}$, $\epsilon = 10^{-4}$ and $U(1)_X$ charges as mentioned above. }
	\label{tab:BtoKnunu}
\end{table}

One can make the fermion content richer by adding more chiral fermions with appropriate charges that satisfy the anomaly cancellation requirements. However, we would like to have a plausible explanation for the neutrino masses, and at the same time, we want to keep our model minimal. Therefore, we extend our model with only three RHNs. All these fermions couple directly to SM leptons via SM Higgs (due to equal and opposite $U(1)_X$ charges of right and left handed leptons while SM Higgs remains chargeless under it), and therefore we cannot consider one of them to be our DM candidate. One can, of course, add a Dirac fermion on top of this which will not contribute to any anomaly and assign this to be the DM. But such a scenario will be ad-hoc and less motivating since the DM does not arise naturally from the anomaly cancellation requirements. Also, its mass remains a free parameter without being connected to the scale of $U(1)_X$ symmetry breaking. Thus we need to look beyond this minimal solution by extending the particle content further\footnote{One can consider one of the RHNs to have very tiny Yukawa couplings with leptons and become a candidate for sterile neutrino DM. We do not consider this possibility here, for details of such scenarios please refer to the review article \cite{Adhikari:2016bei}.}. On the other hand, imposing a discrete $\mathcal{Z}_2$ symmetry on the new chiral fermions (at least in one of them) will help us in forbidding their direct coupling to SM fermions and SM Higgs. In non-minimal or UV complete version of such minimal scenarios, it is possible to realise such $\mathcal{Z}_2$ symmetry as a remnant after spontaneous symmetry breaking of $U(1)_X$ \cite{Borah:2012qr, Adhikari:2015woo, Nanda:2017bmi, Barman:2019aku, Biswas:2019ygr, Nanda:2019nqy}. Our minimal setup here will enable us to have a DM candidate without adding new fermions apart from the RHNs. Under such a scenario, there are two possibilities with the different origin of light neutrino masses but with almost the same DM phenomenology, which we discuss in the following section.

\section{Toy Models}
\label{sec5}
In the following subsections, we discuss the toy models which have been built considering the $U(1)_X$ charge assignments of the SM leptons and new chiral fermions as described in the previous section i.e. $(n_1, n_2, n_3) = (-1,2,-1)$ and $(n'_1, n'_2, n'_3) = (1, -2, 1)$. We consider the additional fermions (namely, $N_1, N_2$ and $N_3$) to be right-handed, hence, their $U(1)_X$ charges will be the sign-flipped version of $(n'_1, n'_2, n'_3)$ i.e. $(-1, 2, -1)$. Based on how we are imposing the $\mathcal{Z}_2$ symmetry on the new chiral fermions, one can come up with different  models, and here we will discuss two such toy models.

\subsection{Toy Model I}

\begin{table}[t]
	\begin{center}
		\renewcommand{\arraystretch}{0.8}
		\begin{tabular}{|c|c|c|c|}
			\hline
			Particles & $SU(3)_c \times SU(2)_L \times U(1)_Y$ & $U(1)_X $  & $\mathcal{Z}_2$ \\
			\hline
			$Q_L=\begin{pmatrix}u_{L}\\
			d_{L}\end{pmatrix}$ & $(3, 2, \frac{1}{6})$ & 0  & +\\
			$u_R$ & $(3, 1, \frac{2}{3})$ & 0 & + \\
			$d_R$ & $(3, 1, -\frac{1}{3})$ & 0  & +\\
			$L_1 = \begin{pmatrix}\nu_{e}\\
			e\end{pmatrix}_L$ & $(1, 2, -\frac{1}{2})$ & $-1$ & + \\
			$L_2 = \begin{pmatrix}\nu_{\mu}\\
			\mu \end{pmatrix}_L$ & $(1, 2, -\frac{1}{2})$ & $2$ & + \\
			$L_3 = \begin{pmatrix}\nu_{\tau}\\
			\tau \end{pmatrix}_L$ & $(1, 2, -\frac{1}{2})$ & $-1$  & +\\
			$e_R$ & $(1, 1, -1)$ & $-1$ & + \\
			$\mu_R$ & $(1, 1, -1)$ & $2$ & + \\
			$\tau_R$ & $(1, 1, -1)$ & $-1$ & + \\
			$H_1$ & $(1,2,\frac{1}{2})$ & 0 & +\\
			\hline
			$N_{1R}$ & $(1, 1, 0)$ & $-1$ & - \\
			$N_{2R}$ & $(1, 1, 0)$ & $2$ & - \\
			$N_{3R}$ & $(1, 1, 0)$ & $-1$ & - \\
			$H_2$ & $(1,2,\frac{1}{2})$ & 0 & -\\
			$\varphi_1$ & $(1,1,0)$ & 2 & + \\
			$\varphi_2$ & $(1,1,0)$ & 4 & + \\
			\hline
		\end{tabular}
	\end{center}
	\caption{Particle content for Toy model I.}
	\label{table2}
\end{table}

\subsubsection{Particle Content}
In this scenario, we consider that all generations of the RHNs, $N_i$ ($i=1,2,3$), to be odd under a discrete $\mathcal{Z}_2$ symmetry while all the SM particles are even. Thus to write a Yukawa term for the RHNs with the SM leptons, we would require an additional Higgs doublet ($H_2$) which is also odd under this discrete symmetry. The unbroken $\mathcal{Z}_2$ symmetry prevents $H_2$ from acquiring a non-zero vacuum expectation value (vev), and it remains inert. However, it plays a crucial role in neutrino mass generation by the radiative seesaw mechanism \cite{Ma:2006km}, which has been described later. To give mass to the chiral fermions, we require at least two singlet neutral scalars with non-zero $U(1)_X$ charges which we choose to be 2 and 4, respectively. The lightest singlet neutral fermion can be a suitable DM candidate since the $\mathcal{Z}_2$ symmetry protects its decay into other lighter particles. However, since $N_2$ is unique from the other singlet fermions in terms of its $U(1)_X$ charge, so we consider this to be our DM candidate and ensure that it is the lightest among all the $\mathcal{Z}_2$ odd fermions. In Table~\ref{table2}, we have shown the entire particle content alongside with their respective charges with respect to different symmetries of the model.

\subsubsection{Lagrangian and Scalar Mass Spectrum}
In the set up given above, the total Lagrangian can be written as

\begin{equation}
\begin{aligned}
\mathcal{L}_{\rm Tot} &= \mathcal{L}_{\rm SM} + \mathcal{L}_S - \mathcal{L}_Y + \frac{i}{2} \sum \limits_{i=1}^3 \bar{N_i}\slashed{\partial}N_i + i g_{X} \sum \limits_{i=1}^{3} \bigg[ n_i \hspace{0.1cm} ( \bar{\ell}^L_i\gamma^\mu \ell^L_i + \bar{e}^R_{i} \gamma^\mu e^R_i ) + n'_i \hspace{0.1cm} \bar{N}_{iR} \gamma^\mu N_{iR}\bigg] X_{\mu} \\& - \frac{1}{4} X_{\mu\nu}X^{\mu\nu} + \frac{\epsilon}{4}B_{\mu \nu} X^{ \mu \nu}
\end{aligned}
\label{Ltot1}
\end{equation}
where $n_i, n'_i$ are the $U(1)_X$ charges of the SM lepton generations and RHN generations, respectively.
The relevant Yukawa interactions are given by :
\begin{equation}
-\mathcal{L}_Y \supset    \sum \limits_{i,j} Y_{ij}^\ell \bar{L}_i H_1 e_{jR} + \sum \limits_{i,j} Y_{ij} \bar{L}_i \tilde{H_2} N_j + Y_{kk} \bar{L}_k \tilde{H_2} N_k + \sum \limits_{i,j} Y^{\varphi}_{ij} \bar{N_i^c} N_j \varphi_1 + Y^{\varphi}_{kk} \bar{N_k^c} N_k \varphi^{\dagger}_2
\label{eq:LYuk1}
\end{equation}
where $i,j,k$ are the generation indices with $i,j = (1,3)$, while $k=2$ and $\tilde{H_2} = i \sigma_2 H_2^*$. The scalar Lagrangian $\mathcal{L}_S$ can be written as:
\begin{equation}
\mathcal{L}_S = (D_\mu H_1)^\dag(D^\mu H_1) +(D_\mu H_2)^\dag(D^\mu H_2) +(D_\mu \varphi_1)^\dag(D^\mu \varphi_1) +(D_\mu \varphi_2)^\dag(D^\mu \varphi_2) - V(H_1,H_2, \varphi_1, \varphi_2),
\label{eq:massbosons}
\end{equation}
where the covariant derivative is given by
\begin{equation}
D_\mu= \bigg(\partial_\mu + i g \frac{\tau^a}{2}W_\mu^a + i g^\prime Y B_\mu - i G_X X_\mu \bigg).
\label{eq:CovDer}
\end{equation}
Here, $G_X = (g_X X + g^\prime \epsilon Y)$ and $(Y,X)$ are the hypercharges related to $U(1)_Y$ and $U(1)_X$ gauge groups respectively. In the above mentioned scenario, $X=4$ and $2$ for $\varphi_2$ and $\varphi_1$, respectively, while $X = 0$ for $H_1$ and $H_2$. As defined earlier, $\epsilon$ is the kinetic mixing parameter.  The scalar potential $V(H_1,H_2, \varphi_1, \varphi_2)$ is defined as
\begin{equation}
\begin{aligned}
V(H_1,H_2, \varphi_1, \varphi_2) &=  \mu_1^2|H_1|^2 +\mu_2^2|H_2|^2 +\frac{\lambda_1}{2}|H_1|^4+\frac{\lambda_2}{2}|H_2|^4+\lambda_3|H_1|^2|H_2|^2  +\lambda_4|H_1^\dag H_2|^2  \\ & + \bigg\{\frac{\lambda_5}{2}(H_1^\dag H_2)^2 + \text{h.c.}\bigg\}  + \mu_3^2|\varphi_1|^2 + \mu_4^2|\varphi_2|^2 + \frac{\lambda_6}{2}|\varphi_1|^4 +\frac{\lambda_7}{2}|\varphi_2|^4  \\& + \lambda_8 (\varphi_1^\dag \varphi_1)(\varphi_2^\dag \varphi_2)  + \lambda_9 |H_2|^2 |\varphi_1|^2 + \lambda_{10} |H_2|^2 |\varphi_2|^2 + \lambda_{\varphi_1} |H_1|^2 |\varphi_1|^2  \\& +\lambda_{\varphi_2} |H_1|^2 |\varphi_2|^2 + \bigg\{\delta \hspace{0.1cm} \varphi_1 \varphi_1 \varphi_2^\dag  + \text{h.c.}\bigg\},
\end{aligned}
\label{scalarpot1}
\end{equation}
with the doublet and singlet scalars after the electroweak symmetry breaking (EWSB) defined as 
\begin{equation}
H_1=\begin{pmatrix} w^\pm \\  \frac{ v +h^{'} + i z}{\sqrt 2} \end{pmatrix} , H_2=\begin{pmatrix} H^\pm\\  \frac{H^0+iA^0}{\sqrt 2} \end{pmatrix}, 
\varphi_1 = \bigg(\frac{v_1 + s_1^{'} + i A_1^{'}}{\sqrt{2}}\bigg), \varphi_2 = \bigg(\frac{v_2 + s_2^{'} + i A_2^{'}}{\sqrt{2}}\bigg).
\end{equation}
After spontaneous symmetry breaking, all the scalars apart from $H_2$ acquires a vev and is responsible for giving mass to other particles. In order to spontaneously break the electroweak symmetry as well as $U(1)_X$, we must have $\mu_1^2 < 0$, $\mu_3^2 < 0$ and $\mu_4^2 < 0$. Also since the inert doublet does not acquire a vev, $\mu_2^2 > 0$. Here, the term proportional to $\delta$ in the scalar potential \eqref{scalarpot1} will play an important role in determining the mass of pseudo-scalars like $A_2'$. The potential minimization conditions are given by  

\begin{eqnarray}
\mu_1^2 &=& -\frac{1}{2}\bigg(\lambda_1 v^2 + \lambda_{\varphi_1}v_1^2 + \lambda_{\varphi_2}v_2^2\bigg), \nonumber \\
\mu_3^2 &=& -\frac{1}{2}\bigg(\sqrt{2}v_2 \delta + \lambda_6 v_1^2 + \lambda_{8}v_2^2 + \lambda_{\varphi_1}v^2\bigg), \\
\mu_4^2 &=& -\frac{1}{2}\bigg(\lambda_7 v_2^2 + \lambda_{8}v_1^2 + \lambda_{\varphi_2}v^2\bigg) +\bigg(\frac{\sqrt{2} v_1^2 \delta}{2 v_2}\bigg). \nonumber
\label{MinCond1}
\end{eqnarray}

The gauge boson mass term can be obtained from the kinetic terms in eq.~\eqref{eq:massbosons} which is given by 
\begin{equation}
\mathcal{L}_{mass} = M_{W}^2 W^+_{\mu} W^{-\mu} + \frac{1}{2} M^2_{Z^0} Z^0_{\mu} Z^{0\mu} - \Delta^2 Z^0_{\mu} X^{\mu} + \frac{1}{2} M^2_{X} X_{\mu} X^{\mu}, 
\label{eq:Lgaumass}
\end{equation}
with 
\begin{equation}
\begin{aligned}
&M_{W}^2 = \frac{1}{4}g^2v^2, \\ &
M_\gamma^2 = 0, \\&
M_{Z^0}^2 = \frac{1}{4}(g^2 + {g^\prime}^2)v^2, \\ &
\Delta^2 = \frac{1}{4} v^2 g' \epsilon \sqrt{g^2 + {g^\prime}^2} , \\& 
M_{X}^2 = \frac{1}{4} {g^\prime}^2 v^2 \epsilon^2   + 4 g_{X}^2 (v_1^2 + 4 v_2^2).
\end{aligned}
\label{eq:GBMass1}
\end{equation}
Note that $H_2$ does not acquire a vev; hence, it does not play any role in the mass generation of the gauge bosons or fermions.  We obtain the masses of $W$-boson, $Z$-boson and photon as in case of SM. The $X$ boson mass has been obtained as a combination of the vevs of the singlet scalars and the vev of $H_1$; the contributions from $H_1$ is suppressed by the factor $\epsilon^2$.

In eq.~\eqref{eq:Lgaumass}, to obtain the masses of the neutral gauge bosons, we need to carry out the standard electroweak rotation as given below  
\begin{eqnarray}
W_\mu^3 &=& S_W A_\mu + C_W Z_\mu^0 \\
B_\mu &=& C_W A_\mu - S_W Z_\mu^0,
\label{Weinberg_Rot}
\end{eqnarray}
where $A_{\mu}$ is the photon field. Note that after the symmetry breaking there will be a remaining mixing between $ Z_\mu^0$ and $X_\mu$, which can be written as:
\begin{equation}
M_{GB}^2  = \begin{pmatrix}
 M_{Z^0}^2 & -\Delta^2  \\
  -\Delta^2 & M_X^2
\end{pmatrix}.
\label{eq:GBMassMat1}
\end{equation}
Here, we have neglected the mixing between the photon and the new gauge boson.
The masses of the physical heavy gauge bosons ($Z,Z^\prime$) can be obtained after diagonalising the above matrix by a rotation, and the masses are given by
\begin{eqnarray}
M_Z^2 &=& \frac{1}{2}\bigg( M_{Z^0}^2 + M_X^2 + \sqrt{(M_{Z^0}^2 - M_X^2)^2 + 4 \Delta^4}\bigg), \\
M_{Z^\prime}^2 &=& \frac{1}{2}\bigg( M_{Z^0}^2 + M_X^2 - \sqrt{(M_{Z^0}^2 - M_X^2)^2 + 4 \Delta^4}\bigg). 
\end{eqnarray}
In the limit that $M_X << M_{Z^0}$ and mixing parameter $\epsilon << 1$, we obtain the masses as
\begin{eqnarray}
M_Z^2 &\simeq &  M_{Z^0}^2, \\
M_{Z^\prime}^2 &=& M_X^2 - \bigg(\frac{\Delta^4}{M_{Z^0}^2}\bigg) \approx M_{X}^2,\ \ \  \text{since $\Delta^4 \propto \epsilon^2$ (in our case, $\epsilon^2 \approx 10^{-8}$) }
\end{eqnarray}
and the mixing angle is given by :
\begin{equation}
\text{tan } 2\zeta = \left(\frac{2 \Delta^2}{M_{Z^0}^2 - M_X^2}\right).
\end{equation}

On the other hand, the mass mixing matrix for the CP even and $\mathcal{Z}_2$ even neutral scalars $(h^{'}, s_1^{'}, s_2^{'})$ is given by :
\begin{equation}
M_{hs}^2 = \begin{pmatrix}
\lambda_1 v^2 & \lambda_{\varphi_1}v v_1 & \lambda_{\varphi_2}v v_2 \\
\lambda_{\varphi_1}v v_1 & \lambda_6 v_1^2 + \frac{v_2 \delta}{\sqrt{2}} & v_1 (\sqrt{2} \delta+ \lambda_8 v_2) \\
\lambda_{\varphi_2}v v_2 & v_1 (\sqrt{2} \delta + \lambda_8 v_2) & -\bigg(\frac{\sqrt{2}v_1^2 \delta - 2 \lambda_7 v_2^3}{2 v_2^3}\bigg)
\end{pmatrix}
\label{eq:CPevenMassMat1}
\end{equation}
The physical scalars $(h,s_1,s_2)$ are obtained after diagonalising the above mass mixing matrix and they are related to the unphysical ones by an orthogonal transformation. We consider a general real orthogonal $3\times 3$ rotation matrix $\mathcal{O}$ with three mixing angles (no phase) for diagonalising the above mentioned mass mixing matrix as
\begin{equation}
\begin{pmatrix} h\\ s_1 \\s_2 \end{pmatrix} = \mathcal{O}^T \begin{pmatrix}
h^{'} \\ s_1^{'} \\ s_2^{'} \end{pmatrix}
= \begin{pmatrix}
c_{\alpha_{12}} c_{\alpha_{13}} & s_{\alpha_{12}} c_{\alpha_{13}} & s_{\alpha_{13}} \\ 
-s_{\alpha_{12}} c_{\alpha_{23}} - c_{\alpha_{12}} s_{\alpha_{23}} s_{\alpha_{13}} & c_{\alpha_{12}} c_{\alpha_{23}} - s_{\alpha_{12}}s_{\alpha_{23}} s_{\alpha_{13}} & s_{\alpha_{23}} c_{\alpha_{13}} \\ 
s_{\alpha_{12}} s_{\alpha_{23}} - c_{\alpha_{12}} c_{\alpha_{23}} s_{\alpha_{13}} & -c_{\alpha_{12}} s_{\alpha_{23}} - s_{\alpha_{12}}c_{\alpha_{23}}s_{\alpha_{13}} & c_{\alpha_{23}}c_{\alpha_{13}}
\end{pmatrix}^T
\begin{pmatrix}
h^{'} \\ s_1^{'} \\ s_2^{'} \end{pmatrix}
\end{equation}
where $c_{\alpha_{ij}} \equiv \text{cos }\alpha_{ij}$ and $s_{\alpha_{ij}} \equiv \text{sin } \alpha_{ij}$. In order to make the notation simpler, we redefine the angles as $\alpha_{12} \equiv \alpha_1$, $\alpha_{13} \equiv \alpha_2$ and $\alpha_{23} \equiv \alpha_3$. Another important variable is the ratio between the vevs $v$ and $v_1$ which we have defined as $\tan\beta = \frac{v_1}{v}$. In general, to keep the analysis simple, we can assume that the mixing of $s_2$ with $s_1$ and $h$ are negligibly small. In such situation, we need to focus only on the mixing between $s_1$ and $h$, i.e $s_{\alpha_1}$ or $c_{\alpha_1}$. There are studies on the singlet scalar extension of the SM, and bounds are available on the respective model parameters like $\tan\beta$ and $s_{\alpha_1}$; for example, see \cite{Basso:2010jm, Robens:2015gla, Bojarski:2015kra}. These studies took into account the bounds from various experimental measurements like the precision observables S, T and U parameters, $W$-mass, LEP and LHC bounds. Alongside, they have considered various theory inputs, like perturbative unitary constraints on scalar self-interactions, vacuum stability, etc. All these studies suggest that for $300 \le M_{s_1} \le 800$ (in GeV), one can safely assume $|\sin{\alpha_1}| \le 0.3$ and $\tan\beta > 1$. Note that in our model, we have two singlet scalars, and as discussed above we have more free parameters. In general, we can expect that the bounds as mentioned above will be little more relaxed in case of our model parameters. However, to be on the safe side, we have used these bounds in our analysis. This will help us to constrain a few of the other model parameters. In our analysis, we have considered $|\text{sin } \alpha_1| \lsim 0.2$ and $\tan\beta = 2.0$  and $M_{s_1} = 500$ GeV \cite{Robens:2015gla, Bojarski:2015kra} which is even more conservative. The corresponding values of $M_{s_2}$ can be obtained after the evaluation of $v_2$ from eq.~\eqref{eq:GBMass1}. 

The mass mixing matrix for the CP odd and $\mathcal{Z}_2$ even neutral scalars $(A_1^{'}, A_2^{'})$ is given by
\begin{equation}
M_{AA}^2 = \begin{pmatrix}
-2\sqrt{2}v_2 \delta & \sqrt{2} v_1 \delta  \\
\sqrt{2} v_1 \delta & -\bigg(\frac{v_1^2 \delta}{\sqrt{2} v_2}\bigg)
\end{pmatrix}.
\label{eq:CPoddMassMat1}
\end{equation}
After diagonalizing this matrix with an orthogonal transformation we will obtain one massless goldstone ($A_1$) corresponding to the gauge boson of $U(1)_X$ and another massive physical CP odd scalar ($A_2$) of mass $\big(-\frac{2\sqrt{2}v_1 \delta}{s_\gamma^2}\big)^{1/2}$. Here, $s_\gamma \equiv \text{sin } \gamma$, where $\gamma$ is the mixing angle between the physical and unphysical CP odd scalars. It is evident from this expression that the dimensionful coupling $\delta$ has to be negative. Also here, for simplicity, we can limit our discussion to the value $s_{\gamma} << 1$.

The masses of the neutral and charged inert scalars are given by : 
\begin{eqnarray}
M_{H^0}^2 &=& \frac{1}{2}\bigg(2 \mu_2^2 + \lambda_L v^2 + \lambda_{9} v_1^2 + \lambda_{10} v_2^2\bigg)\nonumber \\
M_{A^0}^2 &=& \frac{1}{2}\bigg(2 \mu_2^2 + \lambda_A v^2 + \lambda_{9} v_1^2 + \lambda_{10} v_2^2 \bigg)\\
M_{H^\pm}^2 &=& \mu_2^2 + \frac{1}{2}\bigg( \lambda_3 v^2 + \lambda_9 v_1^2 + \lambda_{10} v_2^2 \bigg) \nonumber
\label{inertMasses}
\end{eqnarray}
where $\lambda_L = (\lambda_3 +\lambda_4 + \lambda_5)$ and $\lambda_A = (\lambda_3 +\lambda_4 - \lambda_5)$. To summarise, in Appendix \ref{sec:Const1}, we have presented various couplings in terms of the relevant physical masses, vevs and the mixing angles. These are the most general relations from which one can obtain the approximate relations for small mixing angle. The coupling strength of the interaction between $H_1$ and $H_2$ is defined by $\lambda_L = \lambda_3 + \lambda_4 + \lambda_5$. In an inert two Higgs doublet model (2HDM) where $H^0$ is considered as a suitable DM candidate, the bound on this type of coupling is given by $\lambda_L \lsim 6\times 10^{-3}$ \cite{Belanger:2015kga}. We have not explored this possibility. In our study, the doublet $H_2$ is relevant for the neutrino mass generation and the required coupling is $\lambda_5$ which we have treated as free parameter.


\begin{figure}[t]
	\centering
	\includegraphics[scale=0.38]{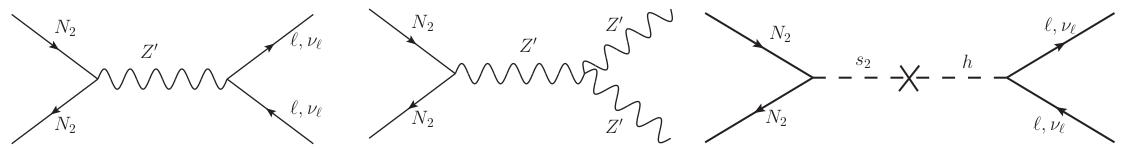}\\
	\includegraphics[scale=0.35]{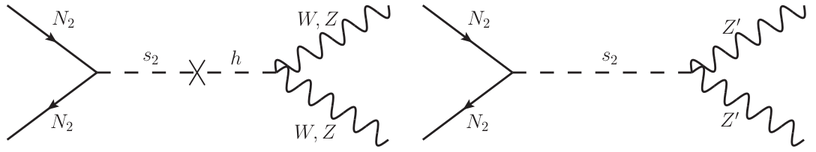}\\
	\includegraphics[scale=0.3]{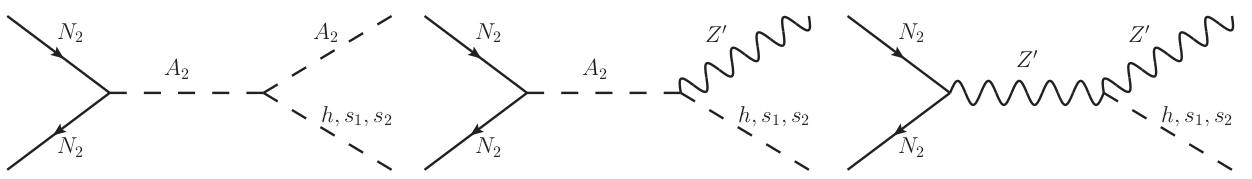}	
	\caption{The diagrams giving leading contributions to the relic abundance.}
	\label{fig:dmann}
\end{figure}

\begin{figure}[htbp]
\begin{center}
	\subfigure[]
	{\includegraphics[scale=0.39]{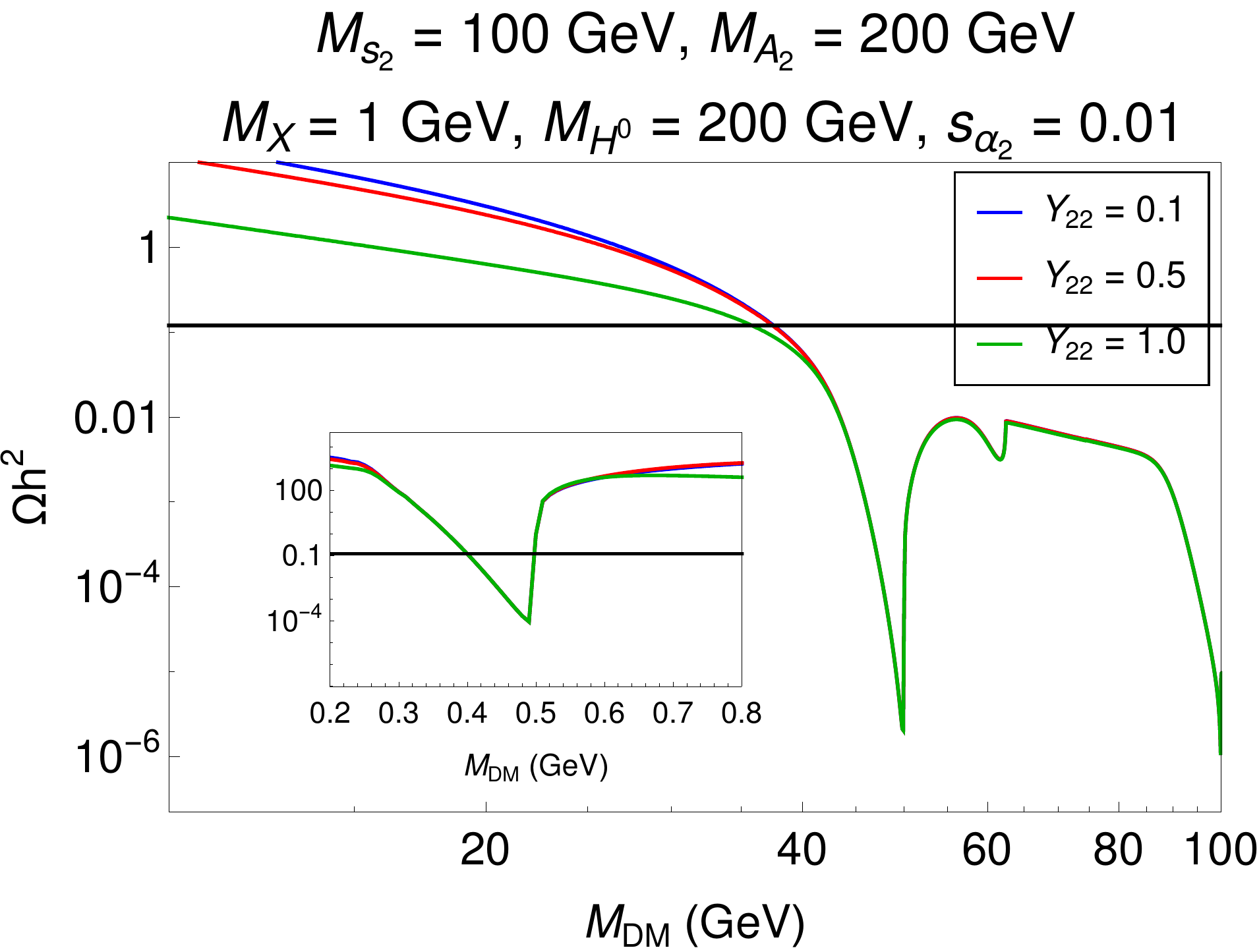}
		\label{fig:relicY22}}~~
		\subfigure[]
		{\includegraphics[scale=0.39]{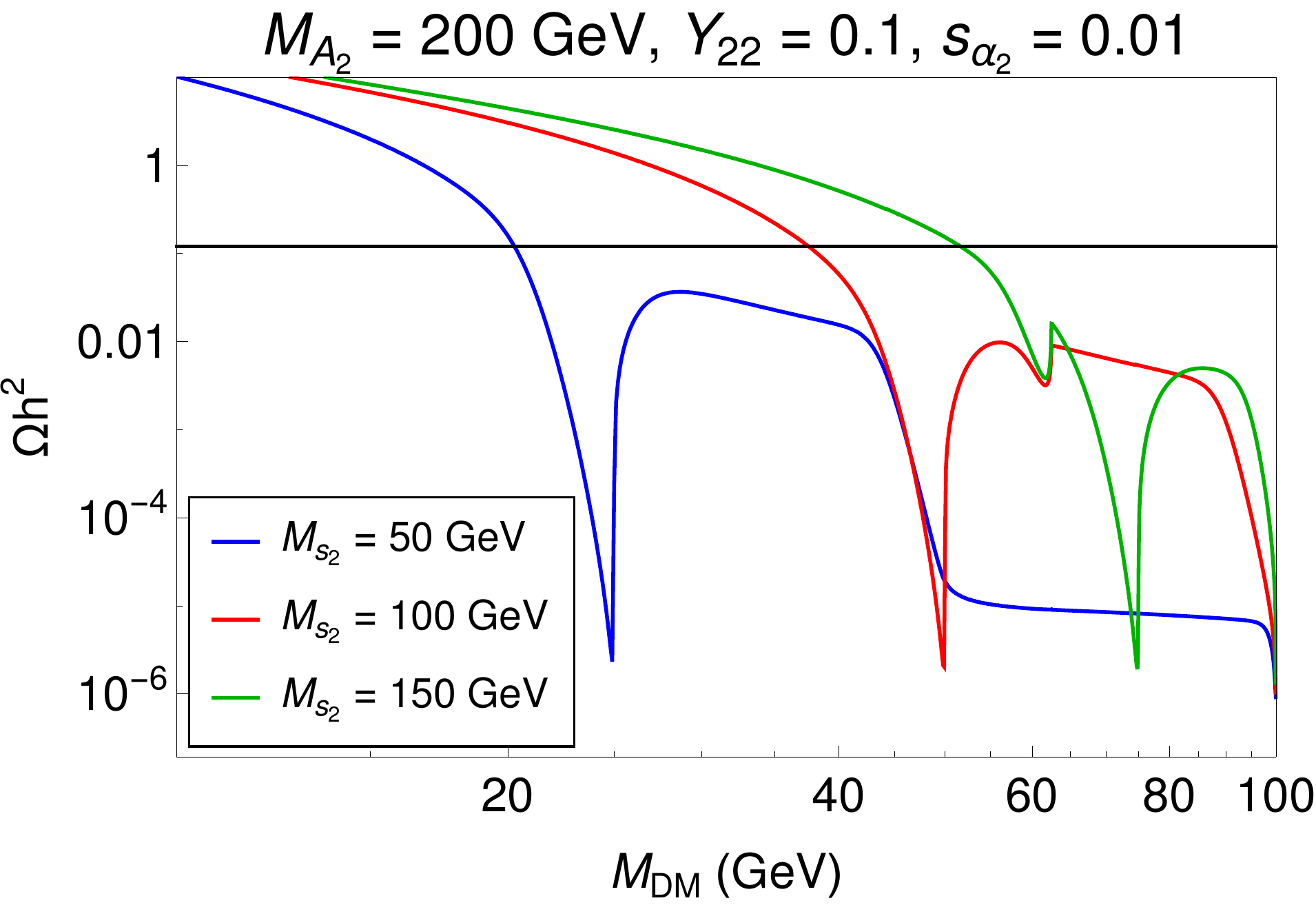}
			\label{fig:relicMs2}}~~ \\
			\subfigure[]
			{\includegraphics[scale=0.39]{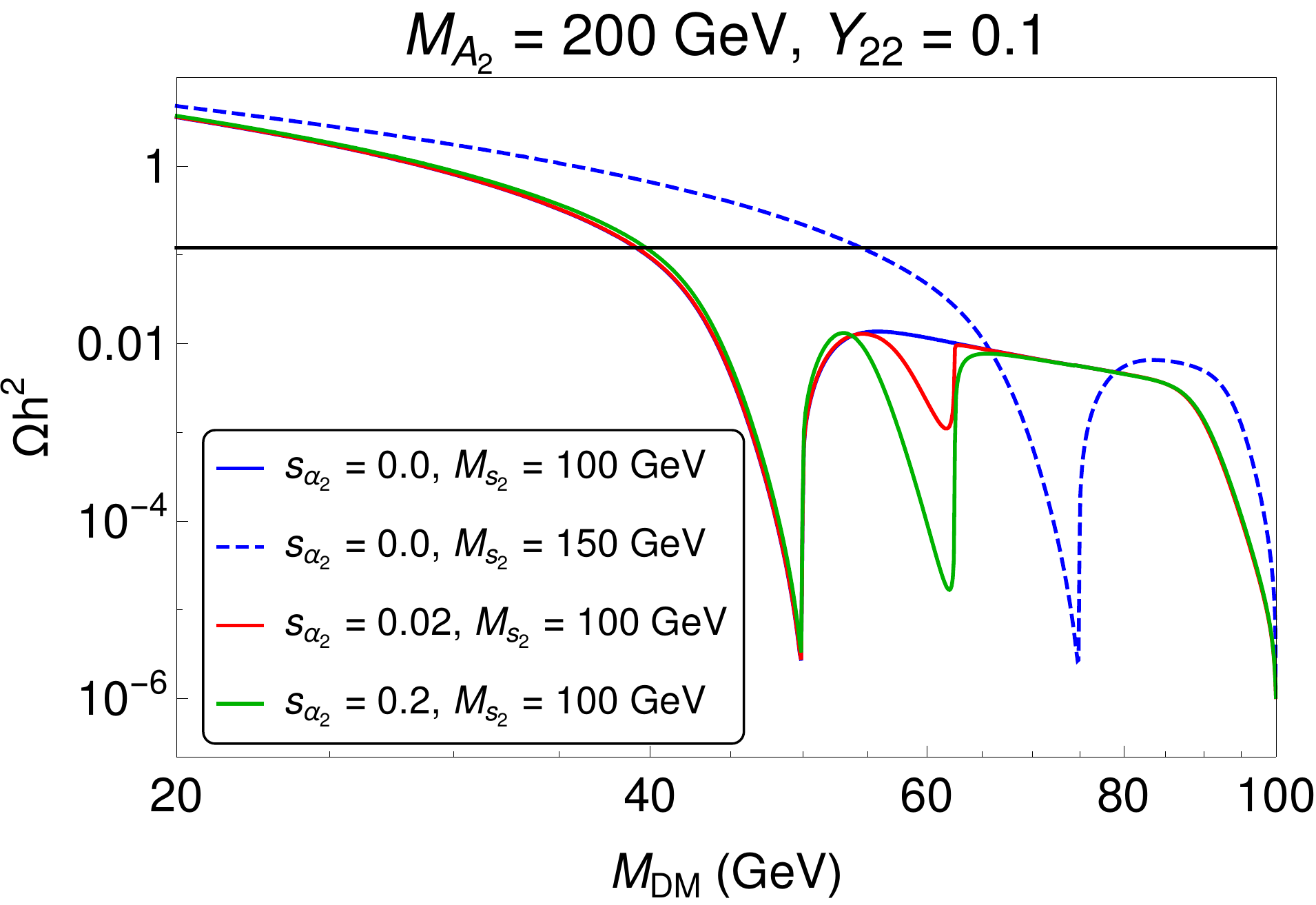}
				\label{fig:relics13}}~~
	\subfigure[]
	{\includegraphics[scale=0.41]{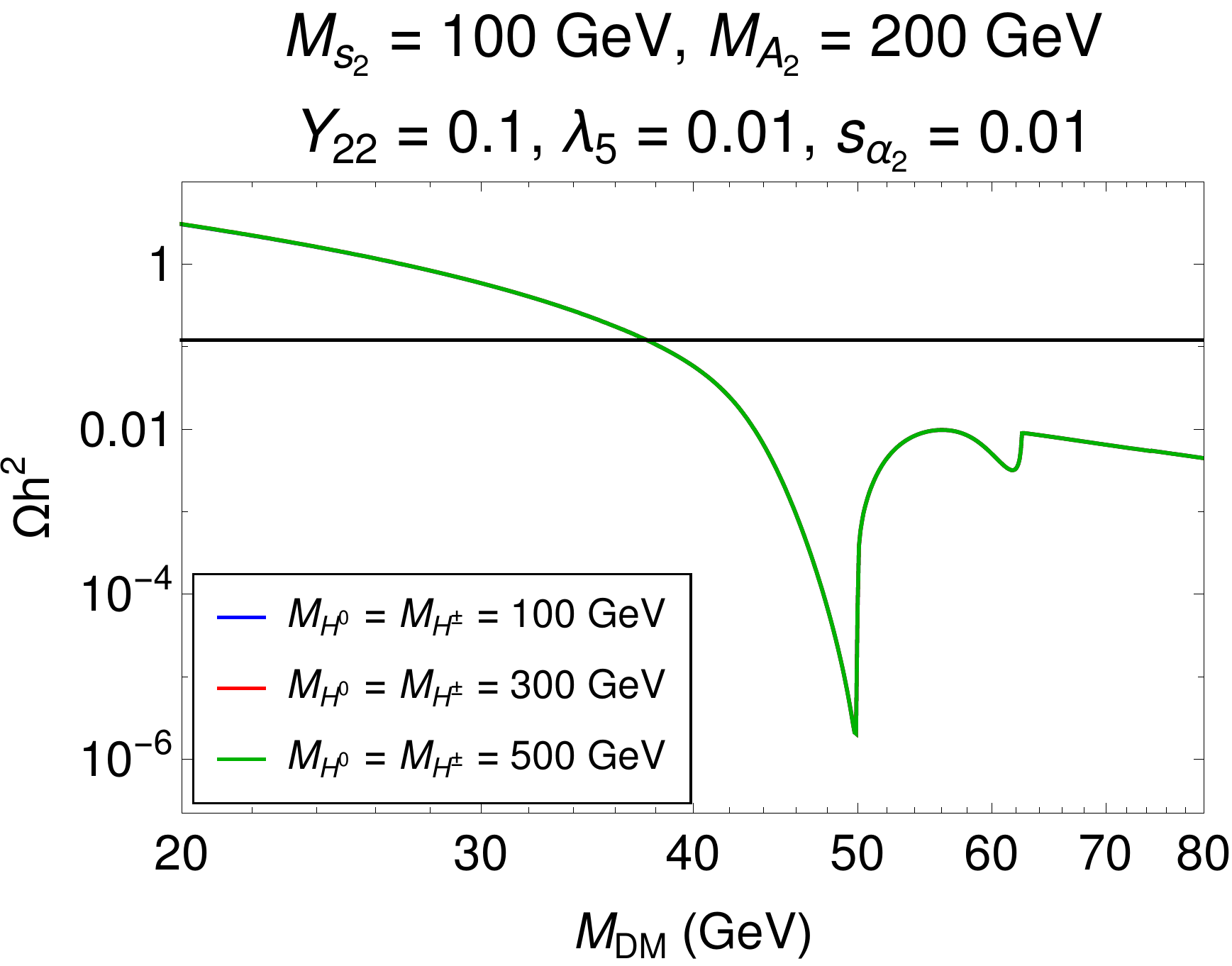}
		\label{fig:relicMH0}}~~ 
	\caption{Relic density vs dark matter mass (in GeV) for different values of Yukawa coupling $Y_{22}$ (\ref{fig:relicY22}), $M_{s_2}$ (\ref{fig:relicMs2}), mixing angle $s_{\alpha_2}$ (\ref{fig:relics13}) and masses of inert scalars (\ref{fig:relicMH0}). The solid black line in each figure denotes the Planck observed relic abundance of DM. In Fig.~\ref{fig:relicY22} we have also shown the variation of the relic for sub-GeV masses of the DM in the inset. }
	\label{fig:relicVsDM}
	\end{center}
\end{figure}

\begin{figure}[t]
\begin{center}
	\subfigure[]
	{\includegraphics[scale=0.45]{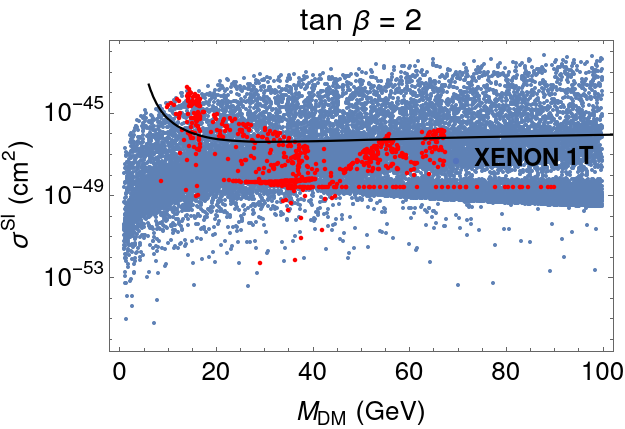}
		\label{fig:DDscan}}~~
	\subfigure[]
	{\includegraphics[scale=0.4]{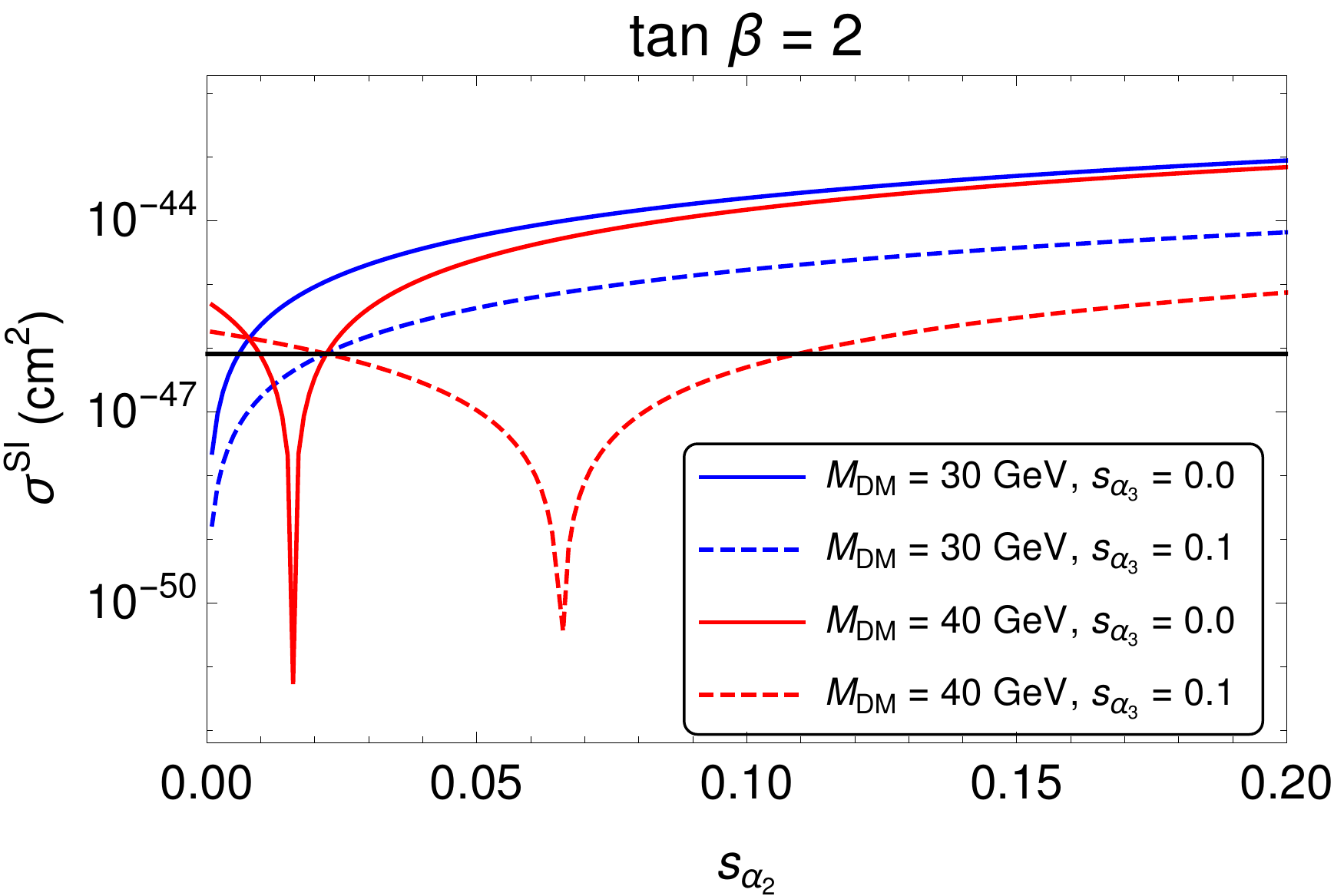}
		\label{fig:DDs1323}}~~
	\caption{(a) The allowed regions of the DM mass satisfying the bounds on relic density and the spin-independent direct detection cross section ($\sigma^{SI}$) of DM  from  XENON 1T, please see the text for other details. The allowed DM masses near $M_{Z'}/2$ are not shown in this plot. (b) Dependencies of $\sigma^{SI}$ on $s_{\alpha_2}$ and $s_{\alpha_3}$ within the allowed ranges of $M_{DM}$.}
	\label{fig:DDrelic_toy1}
	\end{center}
\end{figure}

\subsubsection{DM Phenomenology}
\label{sec:toy1DM}
We adopt the thermal DM paradigm where DM gets produced in the early Universe thermally followed by its freeze-out from the thermal bath which decides its present day abundance. The relic abundance of DM can be computed by solving the appropriate Boltzmann equation and the model parameters can be constrained by comparing the calculated relic with observed abundance which, in terms of density 
parameter $\Omega$ and $h = \text{Hubble Parameter}/(100 \;\text{km} ~\text{s}^{-1} 
\text{Mpc}^{-1})$, is conventionally reported as \cite{Aghanim:2018eyx}:
$\Omega h^2 = 0.120\pm 0.001$
at 68\% C.L. We solve the Boltzmann equation numerically using \texttt{micrOMEGAs} \cite{Belanger:2013oya} where the model information has been supplied to \texttt{micrOMEGAs} using \texttt{FeynRules}
\cite{Alloul:2013bka}. 

In our model, we choose $N_2$ as the DM candidate which is supposed to be the lightest RHN. Note that its $U(1)_X$ charge is different from the other two RHNs. The dominant contributions to the relic abundance of DM will come from the annihilation diagrams shown in Fig.~\ref{fig:dmann}. There are a few other diagrams which are shown in Fig.~\ref{fig:subann} in Appendix~\ref{sec:SubDM} whose contributions in the DM relic abundance will be sub-leading\footnote{See \cite{Borah:2018smz, Mahanta:2019gfe} for scenarios where such contributions can be important. As we can see, the mediators of the DM interactions are the following: $X \equiv Z'$, $s_2$, $H_0$, $H^{\pm}$ and $A_2$.  The scalar $s_2$ does not interact directly with the SM fermions, and it decays into them via mixing with the SM Higgs ($h$). In general, $s_2$ can mix with $s_1$ as well. However, those diagrams will be highly (doubly) suppressed because $s_1$ will decay to SM fermions or gauge bosons via its mixing with $h$. Also, when the mass of the neutral inert scalars or the other RHNs are close to the mass of $N_2$, there would be several other co-annihilation channels that may contribute to the relic abundance as shown in Fig.~\ref{fig:subcoann}. However, we have checked that those contributions are negligible compared to the one given by annihilation diagrams in Fig.~\ref{fig:dmann}. Note that in the low DM mass range of our interest, the efficient coannihilation processes will require the scalars from inert Higgs doublet to be also in the low mass regime which is in tight constraints with LEP data.}

In this model, apart from $M_X (\equiv M_{Z^{'}})$ and $g_X (\equiv g_{Z'})$ the other parameters that are relevant for DM phenomenology are $M_{N_2}$, $M_{H^{0}}$, $M_{H^{\pm}}$, $M_{A^{0}}$, $Y_{22}$, and $Y_{22}^{\varphi} \approx \frac{\sqrt{2} M_{N_2}}{v_2}$, respectively. The co-annihilation diagrams in Fig.~\ref{fig:subcoann} are sensitive to $Y_{ij}$ (with $i,j=1$ or 3). Therefore, the relic density is almost insensitive to these parameters since the contributions from these diagrams are suppressed. Considering the bounds from the low energy data in the rest of our analysis, we have fixed the mass $M_{Z^{'}}$ at 1 GeV; also, we have set $g_{Z^{'}} \approx 10^{-3}$. As mentioned earlier, with a particular choice of $\tan\beta$ one can fix the value of $M_{s_1}$ since $M_h$ is known. Once this is done the allowed values $M_{s_2}$ can be fixed from eq.~\eqref{eq:GBMass1}. In this regard, the perturbativity of the scalar couplings will also play an important role. Since we have chosen $\tan\beta \approx 2$ and $M_{s_1} \approx 500$ GeV, the corresponding values of $M_{s_2}$ and $M_{A_2}$ will be limited to $\lsim 200$ GeV. Accordingly, the mass of DM will be restricted because $s$-channel annihilations are the dominant annihilation process for the DM.

In Fig.~\ref{fig:relicVsDM}, we have shown the variation of the relic abundance with $M_{DM} = M_{N_2}$ for different choices of the other model parameters as mentioned above. The sensitivities of the relic abundance to $Y_{22}$, $M_{s_2}$, $M_{H^0} (M_{H^{\pm}})$ and $s_{\alpha_2}$ are shown in Figs.~\ref{fig:relicY22}, \ref{fig:relicMs2}, \ref{fig:relicMH0} and \ref{fig:relics13}, respectively. Note that with the increasing values of $M_{DM}$, the relic density decreases to a minimum value at the resonances, and it starts increasing again as the DM mass moves away from the respective resonances. In this model, we have a couple of such resonances; the first one is at $M_{DM} \sim M_{Z^\prime}/2$ (Fig.~\ref{fig:relicY22}) which is the annihilation via the gauge boson $Z^\prime$. Notice that for values of $M_{DM}$ close to this resonance (on both sides) the relic density satisfy its measured value. The other resonance peaks are at $M_{DM} \sim M_{s_2}/2$ and $\sim M_h/2$, respectively. Fig.~\ref{fig:relicMs2} shows the pattern of the changes in variations of relic density with $M_{DM}$ for different values of $M_{s_2}$. Note that for $M_{s_2} < M_h$ the bound on relic density is satisfied for DM masses close to, but less than $M_{s_2}/2$. In such scenario, the relic density is under-abundant at or near $M_{DM}\sim M_h/2$. On the contrary, when $M_{s_2} > M_h$ the relic density is satisfied at a DM mass close to $M_{h}/2$, provided we assume that there is a mixing between $h$ and $s_2$, i.e  $s_{\alpha_2} \ne 0$. 

As discussed earlier, we have restricted our analysis to the small values of $s_{\alpha_2} (\approx 0.01)$. Later we will see that such a restriction will be useful to evade stringent bounds on the direct detection of Higgs portal DM from XENON-1T experiment \cite{XENON1T}. However, we have shown the dependences of the relic density on $s_{\alpha_2}$ in Fig.~\ref{fig:relics13}. As expected, in the no-mixing scenario, the resonance peak due to the Higgs mass vanishes. In such situation, when $M_{s_2} > M_h$, the relic density will be satisfied for values of DM mass close to $M_{s_2}/2$ instead at $M_{h}/2$, which is the case for non-zero $s_{\alpha_2}$. Note that there will be another resonance peak at $M_{DM} \sim M_{A_{2}}/2$, at or around which the relic density will be much lower than the existing bound. One can also see from the Figs.~\ref{fig:relicY22} and \ref{fig:relicMH0} that the co-annihilations with inert scalars do not play much role since the relic abundance do not change with change in the value of coupling $Y_{22}$ and the masses $M_{H_0}$ or $M_{H^{\pm}}$. This is precisely due to the fact that the scalar masses are much larger than DM mass making the coannihilation processes inefficient.

In Fig.~\ref{fig:DDscan}, we have shown the allowed ranges of the DM mass obtained from a scan with the constraints from relic density projected against the upper limit on direct detection cross-section $\sigma^{SI}$ (spin-independent) of DM from XENON 1T experiment \cite{XENON1T}. To generate this plot we consider $\tan\beta = 2$, and the values of the other relevant parameters are the following: $s_{\alpha_1} \sim 0.01, 0 < s_{\alpha_3} < 0.01$, $0 < s_{\alpha_2} < 0.01$, $ 20 \leq M_{s_2} \leq 200$ GeV and $ M_{s_1} \sim 500$ GeV. In this model, the spin-dependent direct detection cross-section is highly suppressed; hence we have not considered it for a numerical study. Note that the allowed values of the DM mass lies in between 5 and 90 GeV. It is evident that there will be an allowed region near $M_{Z'}/2$ which is not shown in this plot. The dependencies of $\sigma^{SI}$ on $s_{\alpha_2}$ and $s_{\alpha_3}$ are shown in Fig.~\ref{fig:DDs1323}. As expected, the large values of $s_{\alpha_3}$ allows the relatively larger values of $s_{\alpha_2}$, however, we will stick to the low values of both like $s_{\alpha_2} \approx s_{\alpha_3} = 0.01$.        

\subsubsection{Additional contribution to $b \to c \ell \bar{\nu_{\ell}}$}

There are a couple of important observables associated with $b\to c\ell{\bar\nu_{\ell}}$ decays. Among them, $R(D^{(*)})= \frac{\mathcal{B}({\bar B}\to D^{(*)}\tau{\bar\nu_{\tau}})}{\mathcal{B}({\bar B}\to D^{(\ast)}\ell{\bar\nu_{\ell}})}$ (with $\ell = e \text{ or } \mu$) are useful for the test of lepton universality. Significant deviations from their respective SM predictions will be a clear signal for the lepton universality violating (LUV) new physics. For the last couple of years special attention has been given to these modes, both theoretically and experimentally. For an update of SM predictions and the relevant measurements the reader can look at \cite{hflav,Gambino:2019sif,Bordone:2019vic,Jaiswal:2020wer}. A certain degree of discrepancy has been found between the predictions and their respective measurements. Also, in both $R(D)$ and $R(D^*)$, the measured values are higher than the respective predictions. The experimental world averages of these observables are \cite{hflav}
\begin{equation}
\begin{aligned}
& R(D) = 0.340\pm 0.027 \pm 0.013, \\ &
R(D^*) = 0.295 \pm 0.011 \pm 0.008,
\end{aligned}
\end{equation}
while the SM expectations read \cite{Jaiswal:2017rve,hflav,Jaiswal:2020wer}
\begin{equation}
\begin{aligned}
& R(D)_{SM} = 0.299 \pm 0.003, \\ &
R(D^*)_{SM} = 0.252 \pm 0.006.
\end{aligned}
\end{equation}
At the moment the measurements of $R(D)$ and $R(D^*)$ exceeds the respective SM predictions by 1.4$\sigma$ and 3$\sigma$, respectively. It could be little more if we consider the experimental correlation between $R(D)$ and $R(D^*)$ which is $-0.38$. The experimental averages as mentioned above include the most recent results from Belle \cite{Abdesselam:2019dgh}
 \begin{equation}
  \begin{aligned}
  & R(D) = 0.307\pm 0.037 \pm 0.016, \\ &
  R(D^*) = 0.283 \pm 0.018 \pm 0.014;
  \end{aligned}
 \end{equation}
here, $R(D)$ is consistent with the respective SM prediction, however, $R(D^*)$ is 1.5$\sigma$ away from its SM prediction \cite{Jaiswal:2020wer}. In principle, we don't need a large new physics contribution to explain the current excesses. 

In a model-independent effective theory approach, the Hamiltonian describing the $b\to c\ell \bar\nu_{\ell}$ transitions with all possible four-fermion operators in the lowest dimension is given by 
\be
{\cal H}_{eff} = \frac{4 G_F}{\sqrt{2}} V_{cb} \Big[( \delta_{\ell\tau} + \mathcal{C}_{V_1}^{\ell}) {\cal O}_{V_1}^{\ell} + 
\mathcal{C}_{V_2}^{\ell} {\cal O}_{V_2}^{\ell} + \mathcal{C}_{S_1}^{\ell} {\cal O}_{S_1}^{\ell} + \mathcal{C}_{S_2}^{\ell} {\cal O}_{S_2}^{\ell}
+ \mathcal{C}_{T}^{\ell} {\cal O}_{T}^{\ell}\Big],
\label{eq:heffbtoc}
\ee
where the operator bases are defined as 
\begin{align}
{\cal O}_{V_1}^{\ell} &= ({\bar c}_L \gamma^\mu b_L)({\bar \tau}_L \gamma_\mu \nu_{\ell L}),~~~ 
{\cal O}_{V_2}^{\ell} = ({\bar c}_R \gamma^\mu b_R)({\bar \tau}_L \gamma_\mu \nu_{\ell L}),~~~ 
{\cal O}_{S_1}^{\ell} = ({\bar c}_L  b_R)({\bar \tau}_R \nu_{\ell L}) \nn, \\
& \qquad~~~~ {\cal O}_{S_2}^{\ell} = ({\bar c}_R b_L)({\bar \tau}_R \nu_{\ell L}),~~~ 
{\cal O}_{T}^{\ell} = ({\bar c}_R \sigma^{\mu\nu} b_L)({\bar \tau}_R \sigma_{\mu\nu} \nu_{\ell L}),
\label{eq:oprbtoc}
\end{align}
and the corresponding Wilson coefficients (WC) are given by $\mathcal{C}_W^\ell$ ( $W =V_1,V_2,S_1,S_2,T$ ). In the above-mentioned basis neutrinos are assumed to be left handed. The other theory details and the results of the model-independent new physics analysis on these modes can be seen in \cite{Bhattacharya:2015ida,Bhattacharya:2016zcw,
Bhattacharya:2018kig,Jaiswal:2020wer} and the references therein. 
 
\begin{figure}[t]
\begin{center}
	\subfigure[]
	{\includegraphics[scale=0.7]{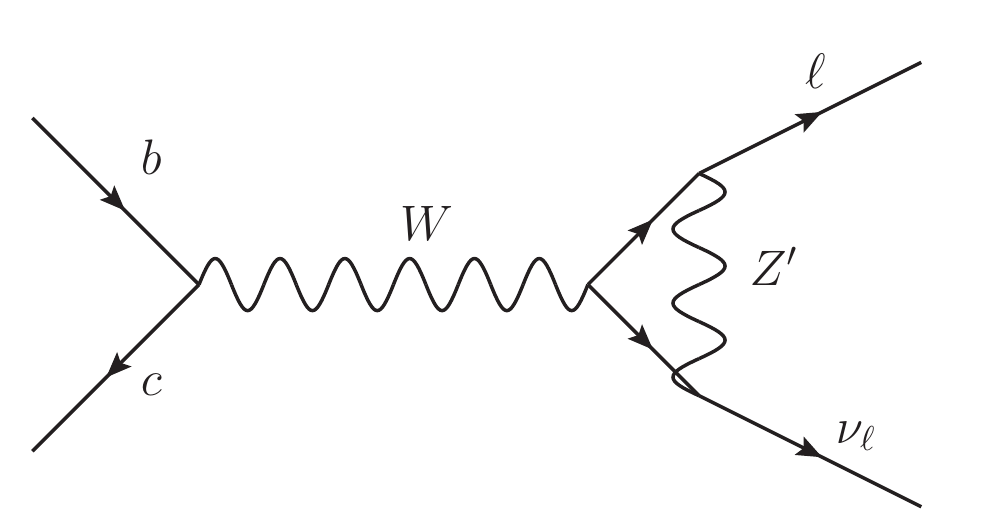}
		\label{fig:diabtoc1}}
	\subfigure[]
	{\includegraphics[scale=0.7]{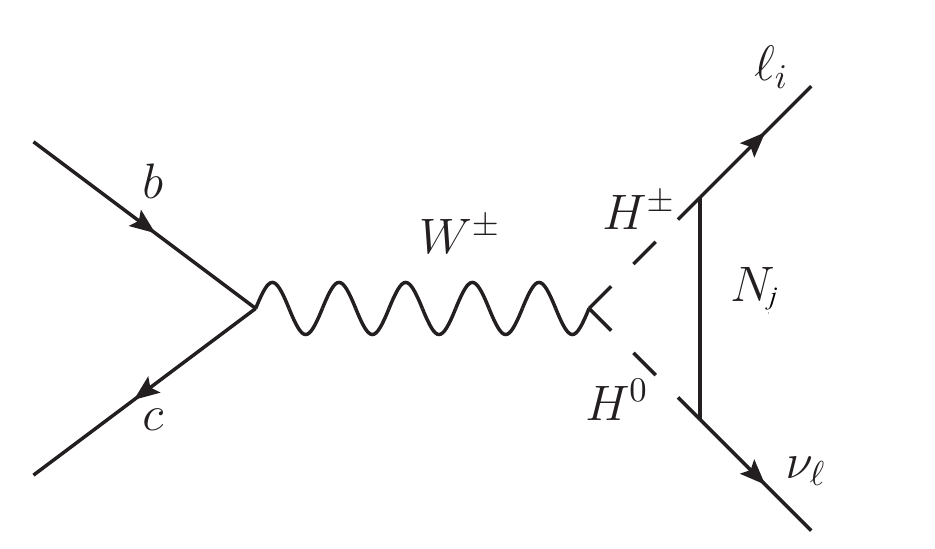}
		\label{fig:diabtoc2}}
	\caption{Diagrams contributing to flavour changing charge current process $b \to c \ell {\bar\nu_\ell}$.}
	\label{fig:diabtoc}
\end{center}
\end{figure}

We have noticed that due to the coupling of the $Z^{'}$ to the lepton families, there would be a vertex correction diagram contribution to the channel $b \to c \tau \bar{\nu_\tau}$ as shown in Fig.~\ref{fig:diabtoc1}. However, that contribution is not sufficient to explain the anomalies in $R(D^{(*)})$. With the addition of the inert scalar doublet and RHNs, we will have additional diagrams as shown in Fig.~\ref{fig:diabtoc2}. The lepton vertex is modified due to the loop corrections coming from the scalars $H^{\pm}, H^{0}$ and $N_i$, $(i=1,2,3)$. Hence one can obtain a bound on the Yukawa couplings $Y_{ij}$ of eq.~\eqref{Ltot1} and masses of $H^{\pm}, H^{0}$ and RHN from the semi-leptonic $b\to c$ decays.

The diagram given in Fig.~\ref{fig:diabtoc2} will contribute to $\mathcal{C}_{V_1}$ of eq.~\eqref{eq:heffbtoc} which is the WC of the four-fermion operator $\mathcal{O}_{V_1}$ in eq.~\eqref{eq:oprbtoc}. The following is the corresponding mathematical expression :
\begin{equation}
\mathcal{C}_{V_1}^{i}= \frac{Y_{ij}^2}{32 \pi^2} \int_0^1 dx \int_0^{1-x} \text{ln } \Delta_{WHH}^j,
\label{CV1}
\end{equation}
and 
\begin{equation}
\Delta_{WHH}^j = x M_{N_j}^2 + (1-x -z)M_{H}^2 + z M_{H^\pm}^2.
\label{WHH}
\end{equation}
Here $(i,j)$ denote the generation index of the lepton and RHN respectively and $M_H$ denotes the mass of $H^0$ or $A^0$ running in the loop. Hence depending on the generation of RHN running in the loop, we would have different contributions to $\mathcal{C}_{V_{1}}$ corresponding to each lepton flavour.

\begin{figure}[t]
\begin{center}
	\subfigure[]
	{\centering \includegraphics[scale=0.38]{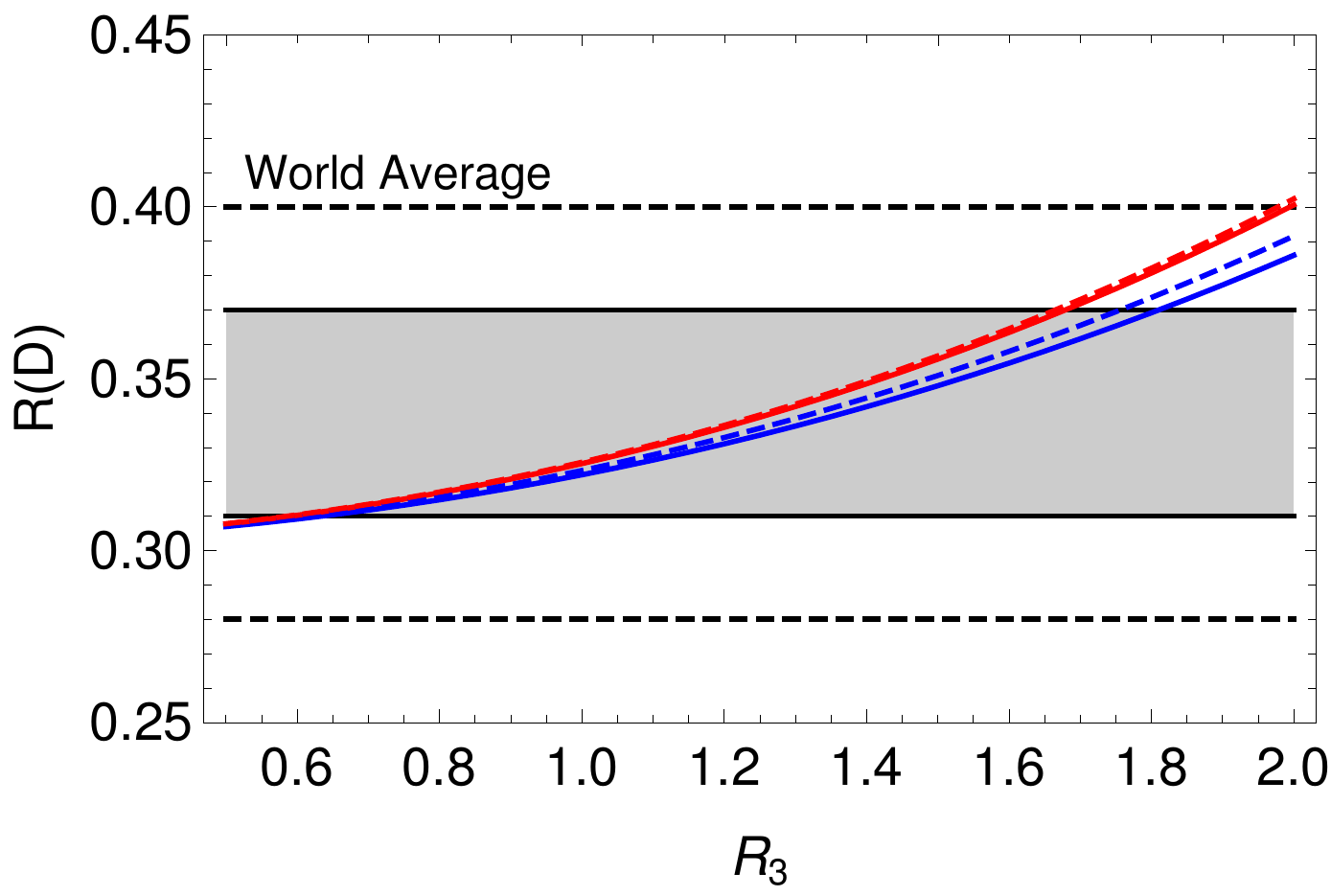}
		\label{fig:RDplt}}
	\subfigure[]
	{\centering \includegraphics[scale=0.44]{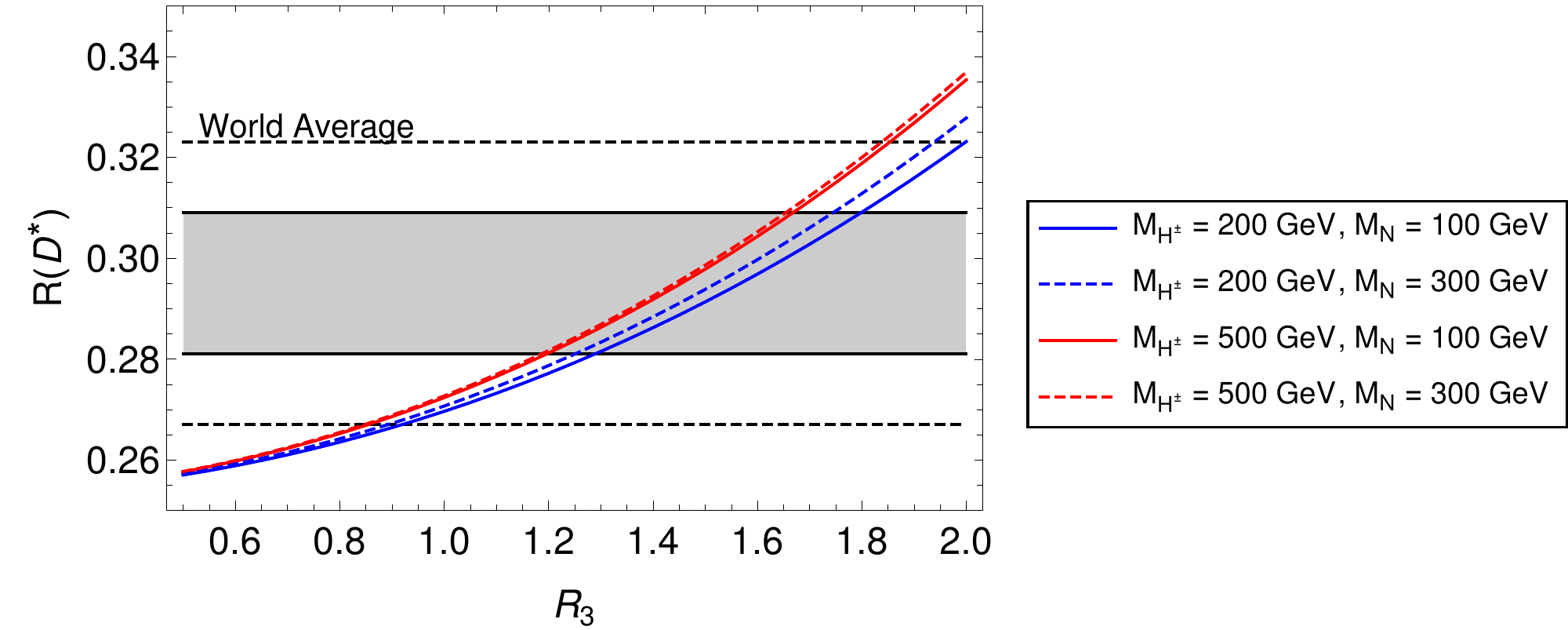}
		\label{fig:RDstplt}}
	\caption{Variation of $R(D)$ and $R(D^*)$ with coupling $R_3$ for two different values of $M_{H^\pm}$ and $M_N$ as shown by the legends. The grey bands are the respective $1\sigma$ ranges of the world averages of these observables \cite{hflav}. The dashed lines are the similar $2\sigma$ ranges of the data. We have plotted these for $M_{H^0} = M_{H^\pm}$ and $R_1 = 0.2$.}
	\label{fig:RDRDstplt}
\end{center}
\end{figure}

As shown earlier, $N_2$ is our DM candidate, hence, the corresponding diagram will contribute only to ${\bar B} \to D^{(*)} \mu\bar\nu_{\mu}$ decays and will be proportional to $Y_{22}^2$. There will be other diagrams with $N_1$ and $N_3$ which can contribute simultaneously to both ${\bar B} \to D^{(*)} \tau\bar\nu_{\tau}$ and ${\bar B} \to D^{(*)} e\bar\nu_{e}$ decays. Since the loop factor as mentioned above is sensitive to $M_{N_i}$,  if we consider the masses of $N_1$ and $N_3$ equal, for simplicity, then the total NP WC for the tau mode would be proportional to $R_3^2 \equiv (Y_{31}^2 + Y_{33}^2)$ while that for the electron mode would be proportional to $R_1^2 \equiv (Y_{11}^2 + Y_{13}^2)$. If we assume that $Y_{13} << Y_{11}$ or $Y_{31} << Y_{33}$ then we can write down the following approximate relations: $R_3^2\approx Y_{33}^2$ and $R_1^2\approx Y_{11}^2$. Since the semi-leptonic branching fractions of the $B$ meson into the light lepton channels are precisely measured \cite{Belle2016_BtoD, Belle2019_BtoDst} and SM consistent, global fits of the NP Wilson coefficients to $R(D), R(D^*)$ data \cite{Jaiswal:2020wer} is done by considering NP only in the $\tau$ decay modes. However, since we have contributions to all semi-leptonic decay channels, we consider NP in both the numerator and denominator of $R(D), R(D^*)$ while also ensuring that the contribution to the light lepton modes does not overshoot the experimental limits on their branching fractions. In Fig.~\ref{fig:RDRDstplt} we have shown the variation of $R(D)$ and $R(D^*)$ with the coupling $R_3$ for two different values of $M_{H^\pm} = M_{H^0}$ and $M_N$ as shown by the blue and red legends. Since the measured value of $R(D)$ has a large error, a value of $R_3 \gsim 0.6$ can easily explain the observed data in its 1$\sigma$ interval. On the other hand, to explain $R(D^*)$ within its 1$\sigma$ range we need a value of $R_3 \gsim 1.2$; however, to explain the data at its 2$\sigma$ range we need $R_3 \gsim 1.0$. Note that in these plots, we have not included the errors in the respective SM predictions.

\begin{figure}[t]
	\centering
	\subfigure[]
	{\includegraphics[scale=0.38]{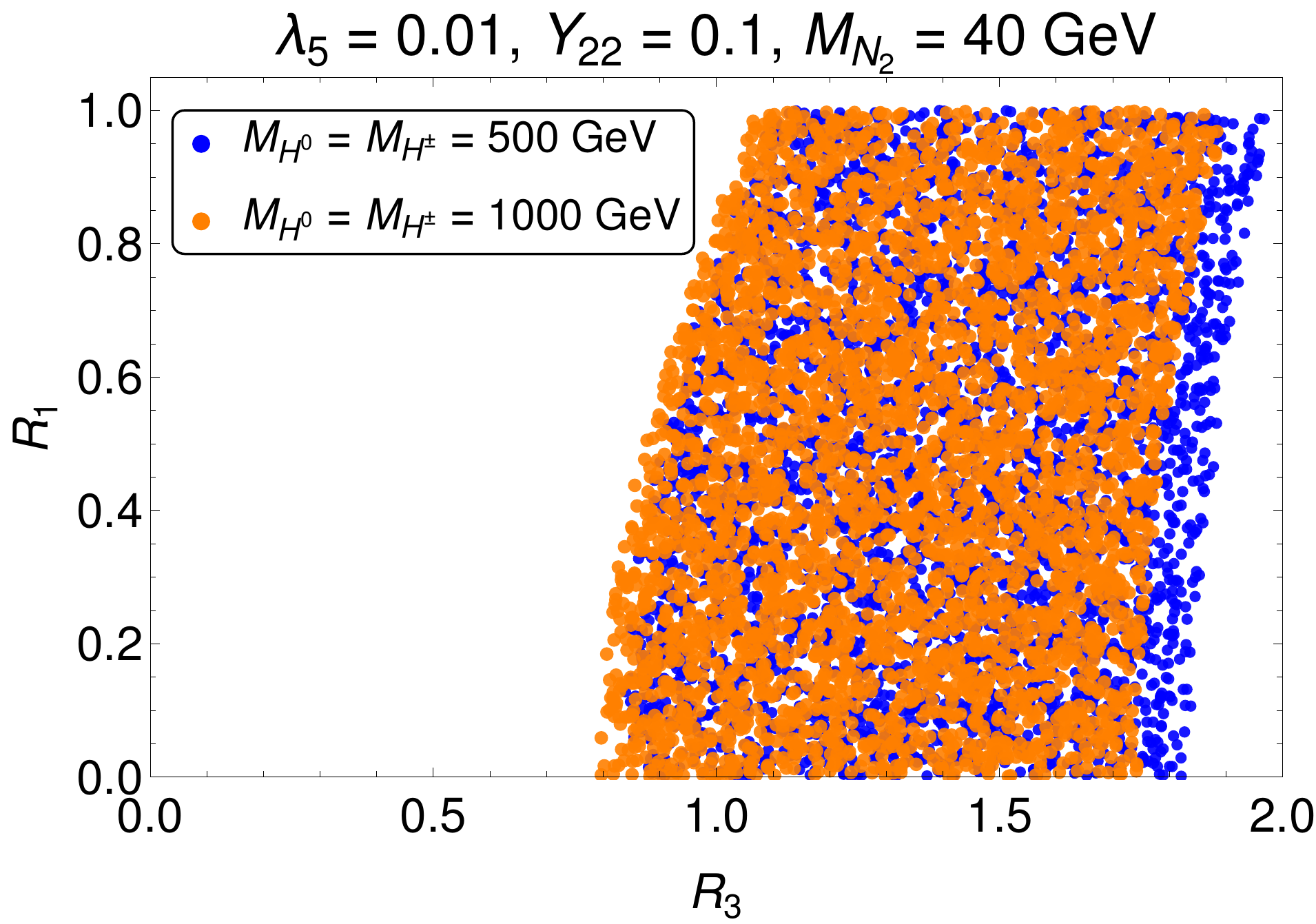}
		\label{fig:R1R3scan}}
	\subfigure[]
	{\includegraphics[scale=0.38]{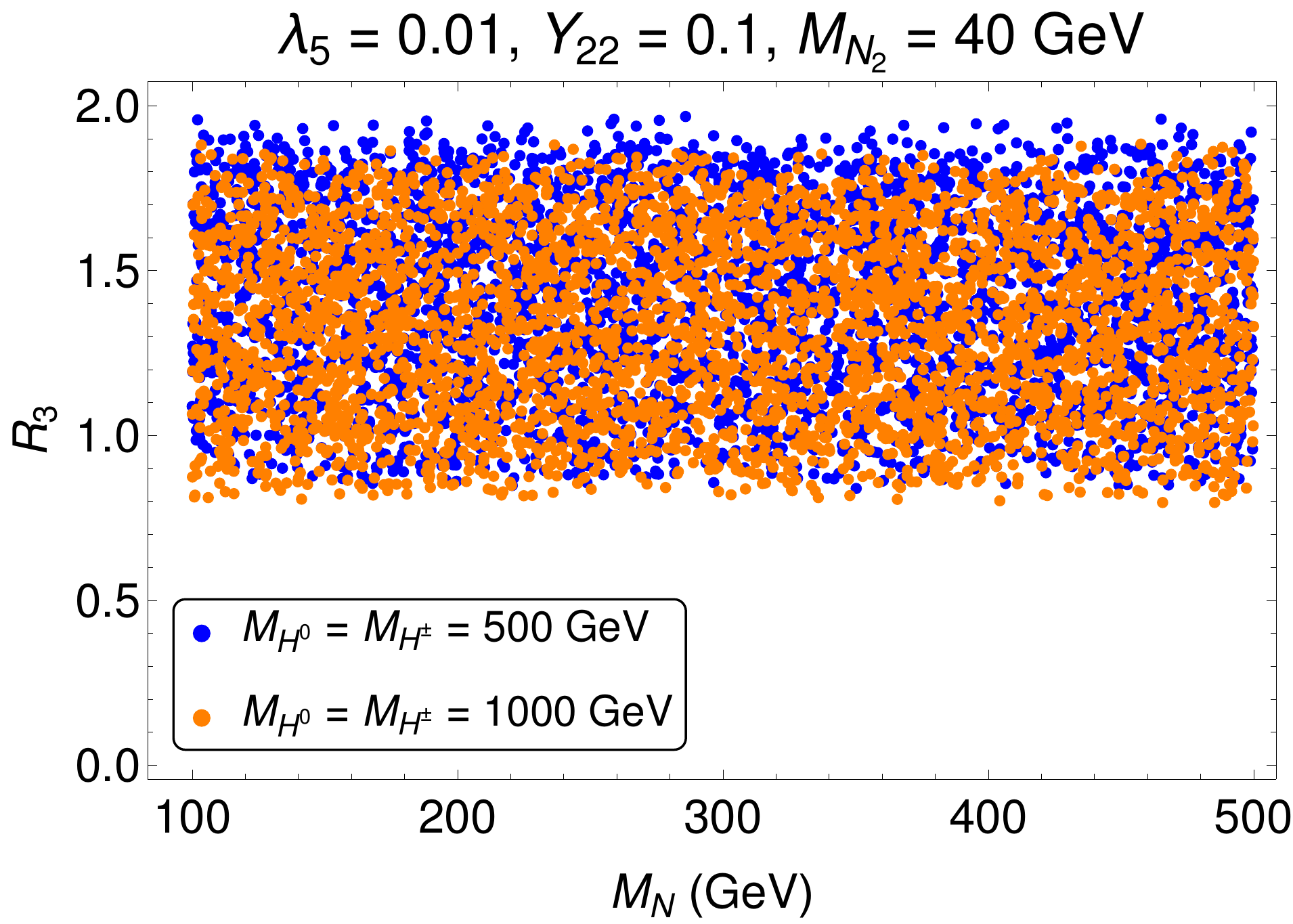}
		\label{fig:R3MNscan}}
	\caption{In the left plot, the region bounded by the points is the parameter space of $R_1$ and $R_3$ that satisfies $R(D),R(D^*)$ and $\mathcal{B}({\bar B} \to D^{(*)} \ell {\bar\nu_\ell})$ constraints in their $2\sigma$ CL and $\mathcal{B}(B_c \to \tau \nu) < 30\%$ for $M_{H^\pm} = 500$ GeV (blue) and $M_{H^\pm} = 1000$ GeV (orange). The other relevant parameters have been fixed as shown in the plot label. The correlation between $R_3$ and $M_N$ is shown in the right plot.}
	\label{fig:RDRDstscan}
\end{figure}

We perform a parameter space scan of $R_1, R_3$ and $M_N (\equiv M_{N_1} = M_{N_3})$ by fixing $Y_{22} = 0.1$ and the masses of the inert scalars as shown in Fig.~\ref{fig:RDRDstscan}. The blue and the orange points are the allowed regions for $M_{H^\pm} = 500$ GeV and $1000$ GeV respectively, when the RHN masses are varied between $100-500$ GeV. All these allowed points satisfy the experimental constraints on $R(D), R(D^*)$ and the branching fraction of ${\bar B}\to D^{(*)} \ell \bar\nu_\ell$ at their respective $2\sigma$ confidence interval (CI). These parameter spaces also satisfy the bound $\mathcal{B} (B_c \to \tau \nu) < 30\%$ and the corresponding expression in terms of $\mathcal{C}_{V_{1}}$ can be seen in \cite{Alonso:2016oyd}. From Fig.~\ref{fig:R1R3scan} it is evident that the data prefers $R_3 > R_1$. Also, as expected, athough it is allowed, we don't necessarily need large values for $R_1$ and the data allows a solution like $R_1 \approx 0$ while $R_3 \gsim 1$. In our model, $\mathcal{C}_{V_1}$ is positive; hence, the new contribution will interfere constructively with the SM and increase the relevant branching fractions from their SM predictions. Notice that  there are minimal dependencies of $R_{D^{(*)}}$ on $M_{H^{\pm}}$ and $M_{H^0}$. However, as can be seen from Fig.~\ref{fig:R3MNscan}, it is almost independent of $M_{N}$.

Similar type of diagrams as given in Figure \ref{fig:btoc} with the replacement $c \rightarrow u$ will contribute to $b\to u \tau\bar{\nu_{\tau}}$ processes like $B^{\pm}\to \tau^{\pm}\nu_{\tau}$, ${\bar B^0} \to \pi^{+}\tau^-\bar\nu_{\tau}$ decays. We have checked that the required values of the parameters for an explanation of the data in $b \to c \tau \bar{\nu_\tau}$ decays can accommodate the current observation $\mathcal{B}(B^{\pm}\to \tau^{\pm}\nu_{\tau}) = (1.09 \pm 0.24) \times 10^{-4}$ \cite{pdg2018}. Similarly, our model will contribute to $s\to u \tau \bar\nu_{\tau}$, $c\to s \tau \bar\nu_{\tau}$ decays which will lead to semileptonic and purely leptonic decays of $K$ and $D/D_s$-mesons, respectively.

\begin{figure}[t]
\centering
\includegraphics[scale=0.65]{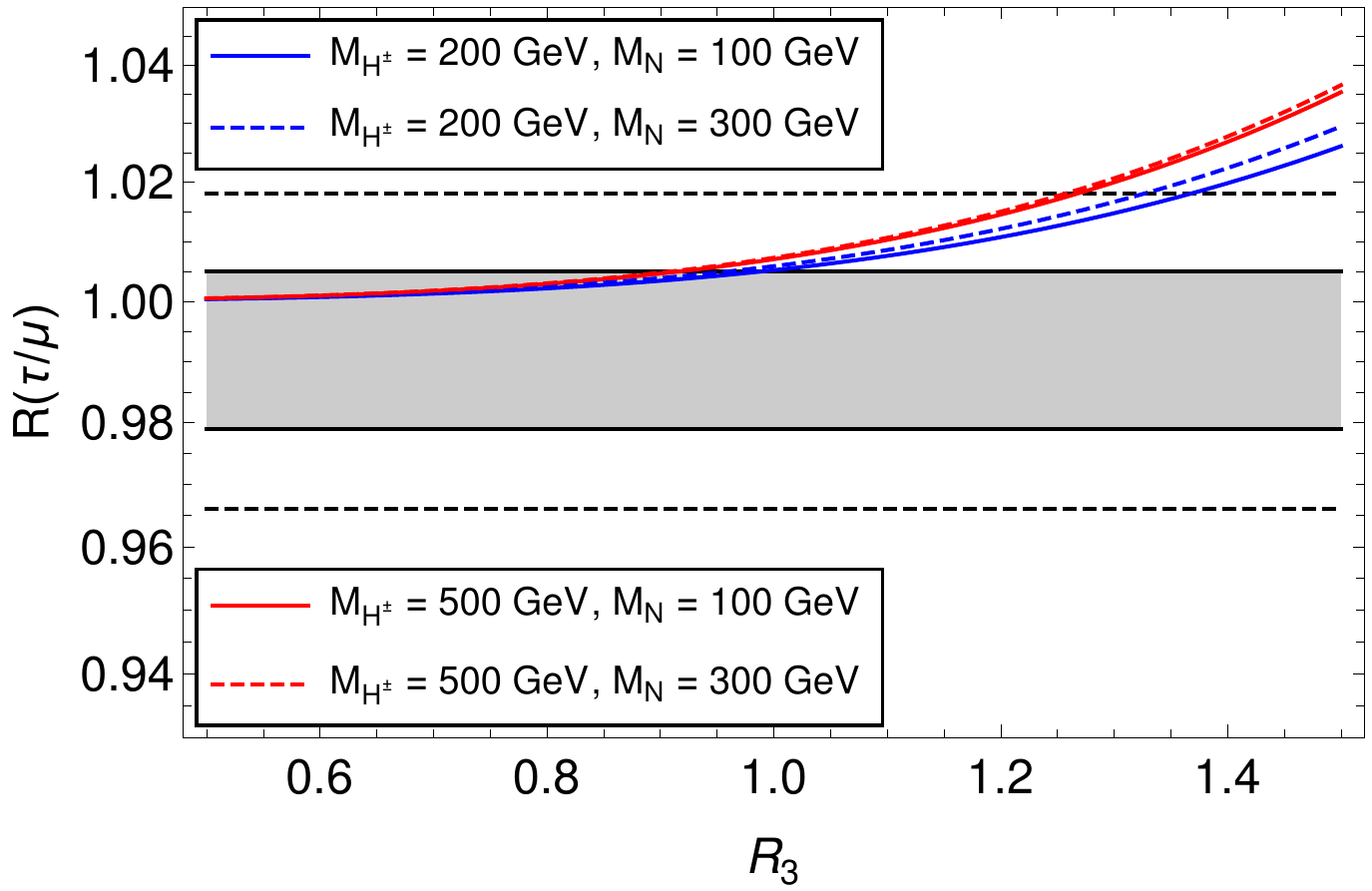}
\caption{Dependence of lepton flavour universality test ratio $R(\tau/\mu)$ on parameter $R_3$ for different values of scalar and heavy neutrino masses as shown in the legends keeping $Y_{22}$ fixed at 0.1. }
\label{Rtaumu}
\end{figure}

The readers may note that in our model the contributions to the semileptonic decays mentioned above will modify the $W$-$\ell$-$\nu_\ell$ vertex. Therefore, in this respect, the ratio $R(\tau/\mu) = \frac{\mathcal{B}(W \to \tau \nu_\tau)}{\mathcal{B}(W \to \mu \nu_\mu)}$ will be a good probe for such kind of NP effects. The most recent measurement of this ratio by ATLAS Collaboration \cite{Aad:2020ayz}
\begin{equation}
R(\tau/\mu) = 0.992 \pm 0.013,
\end{equation} 
is by far most precise and is in well agreement with the SM expectation. Therefore, it is important to ensure that our modification of the W-vertex does not overshoot this result. In Fig.~\ref{Rtaumu}, we have shown the variation of this ratio with $R_3$ for some fixed values of the inert Higgs and RHN masses and $Y_{22} = 0.1$. If we consider the data withing its 1-$\sigma$ CI then $R_3 > 1$ is not allowed, however within the 2-$\sigma$ range of $R(\tau/\mu)$ the values like $R_3 \approx 1.4$ is allowed. Therefore, we still have some part of the parameter space shown in Fig.~\ref{fig:RDRDstscan} which is not excluded by this lepton flavour universality test; to conclude it further we have to wait for more precise data.  

\subsubsection{Anomalous Magnetic Moment and LFV}
\begin{figure}
	\centering
	\includegraphics[scale=0.6]{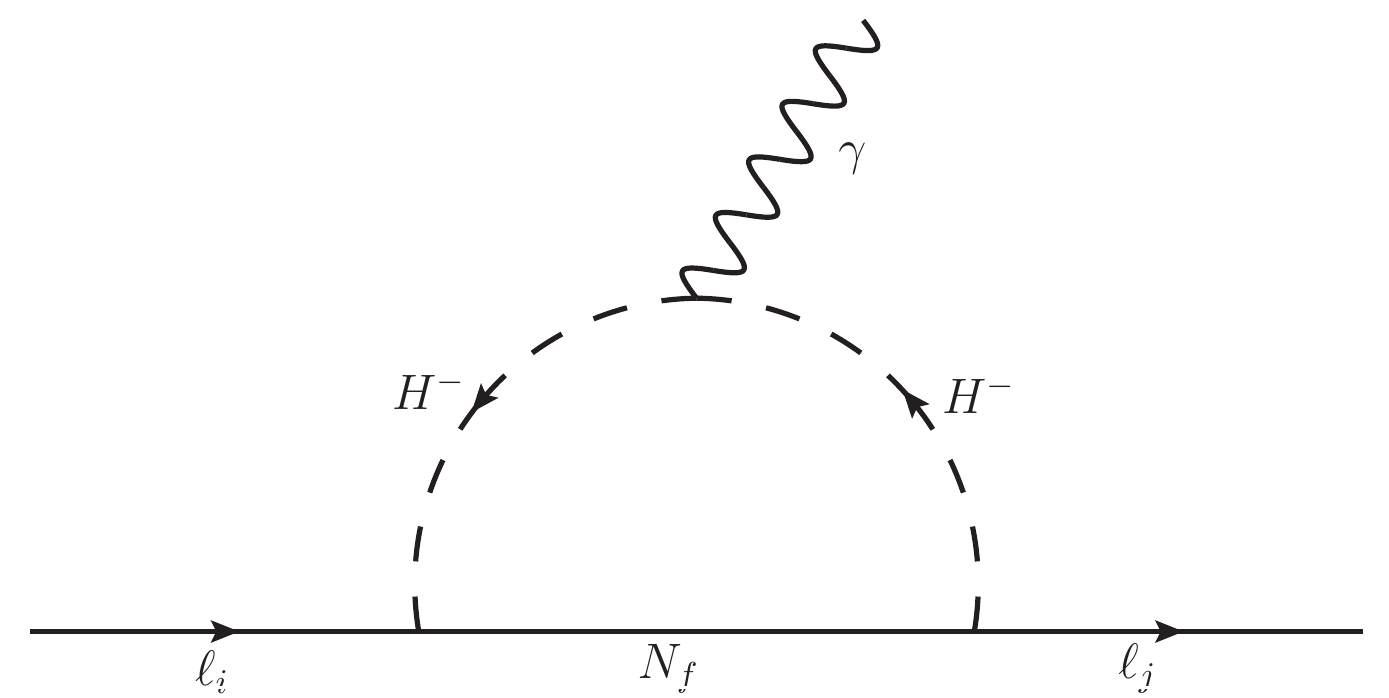}
	\caption{One loop charged Higgs contribution to lepton anomalous magnetic moment and $\tau \to e \gamma$.}
	\label{fig:LFV_chrgdsc}
\end{figure}

\paragraph{Magnetic moments :-}
In toy model-I, the additional contributions to anomalous magnetic moments will come from the type of diagram given in Fig.~\ref{fig:LFV_chrgdsc} (for $i=j$) with charged Higgs and RHNs in the loop. The contribution is given by
\begin{equation}
\Delta a_\ell^{(H^\pm, N_R)} = -\frac{|Y_{ij}|^2}{8\pi^2} \frac{m_\ell^2}{M_{H^\pm}^2} \int_0^1 dx \frac{2x^2 (1-x)}{x + (1-x) (1-\lambda_j^{-2}x) \lambda_j^2 r_\ell\text{\footnotesize{(H$^\pm$)}}}
\label{eq:g2Hpm}
\end{equation}
where $(i,j)$ denotes the generation of the lepton and RHN respectively, $\lambda_j = \frac{M_{N_j}}{m_\ell}$ and $r_\ell${\footnotesize{(H$^\pm$)}} $= (\frac{m_\ell}{M_{H^\pm}})^2$. In case of muon anomalous magnetic moment, for $M_{N_2} = 40$ GeV and $Y_{22}=0.1$, we will obtain the following from eq.~\eqref{eq:g2Hpm}  
\begin{equation}
\Delta a_{\mu}({H^{\pm}})=
\begin{cases}
   -1.10\times 10^{-11},~~~~~\text{For $M_{H^{\pm}} = 200$ GeV}, \\
  -1.86\times 10^{-12},~~~~\text{For $M_{H^{\pm}} = 500$ GeV}, \\
  -4.70\times 10^{-13}, ~~~~\text{For $M_{H^{\pm}} = 1000$ GeV},
\end{cases}
\label{value:muong2Hpm}
\end{equation}
which are suppressed compared to the gauge boson mediated diagram for it, which is shown in Fig.~\ref{fig:g2} (with $X$ in the loop), by two or three orders in magnitude. On the contrary, as shown in eq.~\eqref{eq:ae_X}, the contribution to electron anomalous moment from the same diagram is negligibly small. Therefore, by extending the symmetry of the SM by an abelian $U(1)_X$ gauge group without additional degrees of freedom, we cannot explain the observed discrepancy in the electron magnetic moment.  

\begin{figure}[t]
	\centering
	\includegraphics[scale=0.5]{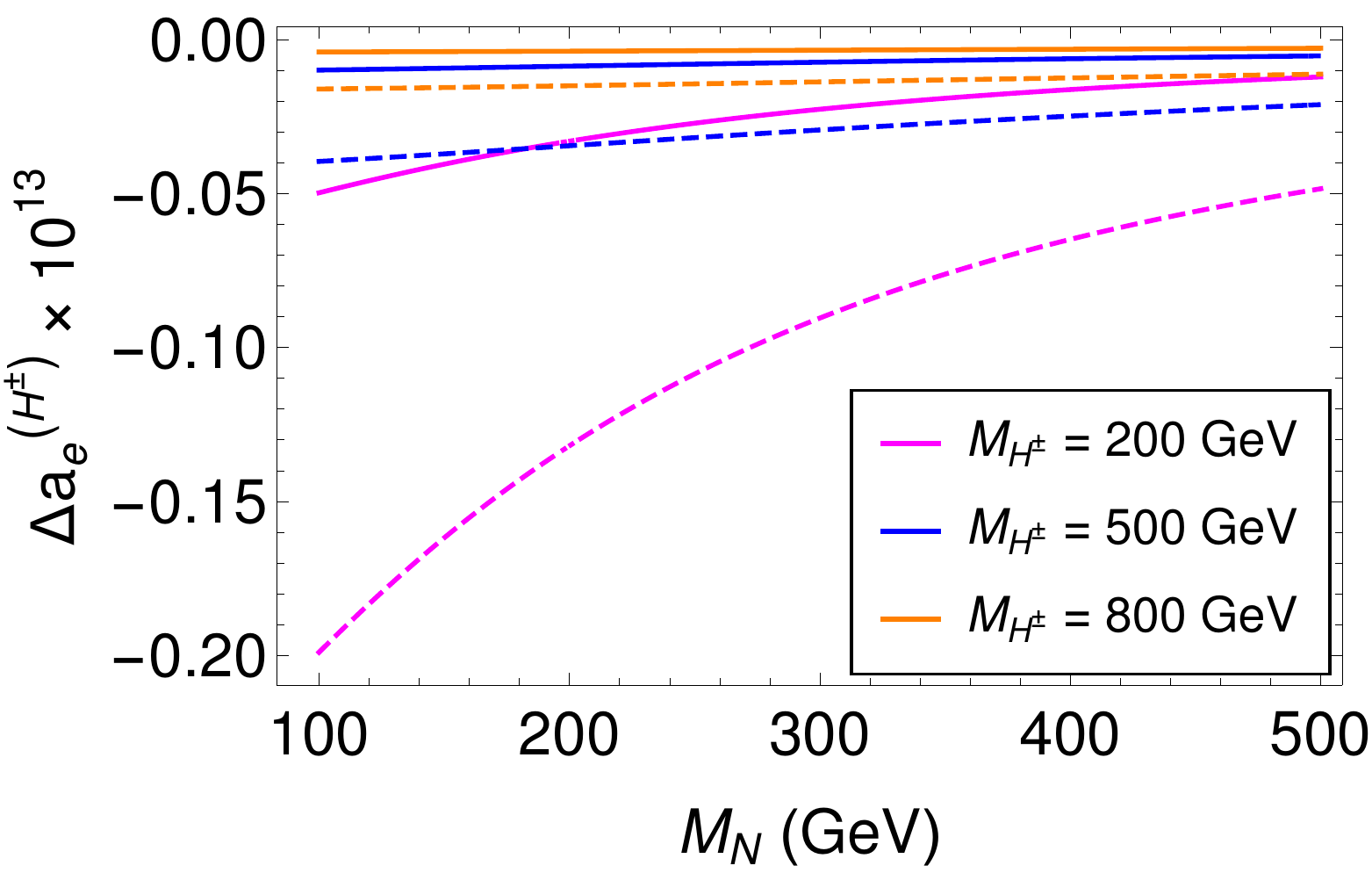}
	\caption{Plot shows the variation of the electron magnetic moment with the mass of RHNs for three different values of charged scalar mass $M_{H^\pm}$ as denoted by the legends. We have plotted each curve for two different values of $R_1$ : $R_1 = 0.5$ (solid) and $R_1 = 1.0$ (dashed).}
	\label{fig:electron_g2}
\end{figure}

Note that $\Delta a_{e}$ (eq.~\eqref{eq:delae}) has a significant error, and at 3-$\sigma$ CI it is consistent with zero. Therefore, it would be too early to prejudge the potential impact of new physics on this observable. Here, we will show that our model has the potential to predict negative values $\Delta a_{e}$ although it is difficult to explain the data in one or two-$\sigma$ CI. From eq.~\eqref{eq:g2Hpm} it is clear that $\Delta a_{e}$ is sensitive to our predefined variable $R_1 \equiv \sqrt{Y_{11}^2 + Y_{13}^2}$. In Fig.~\ref{fig:electron_g2}, we show the variation of $\Delta a_e$ with $M_{N}$ for two different values of $R_1$ and three values of $M_{H^\pm}$. The chosen values of $R_1$, $M_N$ and $M_{H^\pm}$ explain the data on $R(D),R(D^*)$ as previously shown in Fig.~\ref{fig:RDRDstscan}. From the plot we see that the contribution is negative and small and we can not reach the present experimental limit within its 1 or 2-$\sigma$ CI. To get a large negative contribution we need small values of $M_{H^{\pm}}$ and $M_N$, and a large value of $R_1$ ($R_1 > 1$). However, $R_1 >> 1$ is not allowed by $\mathcal{B}(B \to D^{(*)} \ell \nu)$ data. Therefore, within the allowed parameter space of $R_1$ and $M_N$, we have a contribution to the electron anomalous magnetic moment which is negative and of order $\mathcal{O}(10^{-14})$ and within the $3\sigma$ range of the experimental data.

\paragraph{Lepton flavour violation:-}
The same one loop diagram given in Fig.~\ref{fig:LFV_chrgdsc} will also contribute to the LFV process $\tau \to e \gamma$. Therefore, one must ensure that the contribution is within the current experimental limit $\mathcal{B}(\tau \to e \gamma) < 3.3 \times 10^{-8}$\cite{pdg2018}. However, in our model there will not be any contribution to $\tau \to \mu \gamma$ or $\mu \to e \gamma$. The expression for the partial decay width $\Gamma(\ell_i \to \ell_j \gamma)$ for the diagram in Fig.~\ref{fig:LFV_chrgdsc} is given by \cite{Lavoura:2003xp} :
\begin{equation}
\Gamma(\ell_i \to \ell_j \gamma) = \frac{\alpha}{4}\frac{|Y_{if}^* Y_{jf}|^2}{(16\pi^2)^2}\frac{m_i^5}{M_{H^\pm}^2} \mathcal{A}(r)^2
\end{equation}
where,
\begin{equation}
\mathcal{A}(r) = \frac{2r^2 -5r -1}{12(r-1)^3} - \frac{r^2 \text{log }r}{2(r-1)^4}
\end{equation}
and $r \equiv \left(\frac{M_{N_f}^2}{M_{H^\pm}^2}\right)$.

In the above expression for the decay width we have a combination $|Y_{11}Y_{31}|$ or $|Y_{33}Y_{13}|$ depending on whether $N_1$ or $N_3$ runs in the loop. So we can constrain the allowed values of these product couplings from the experimental upper limit on the branching fraction of $\tau \to e \gamma$. As we have seen earlier, if we assume that the off-diagonal Yukawas are much smaller in value than the diagonal ones, the data on  $B \to D^{(*)} \ell \nu_\ell$ allow $R_3 \equiv Y_{33} \sim 1$ and $R_1 \equiv Y_{11} \sim 1$ for $M_N$ in the range $(100-500)$ GeV or more. 
Therefore, in general, the magnitude of the product couplings as mentioned above could be small even if we assume $Y_{33} \sim Y_{11}\approx 1$. 

\begin{figure}[t]
	\centering
	\subfigure[]
	{\includegraphics[scale=0.45]{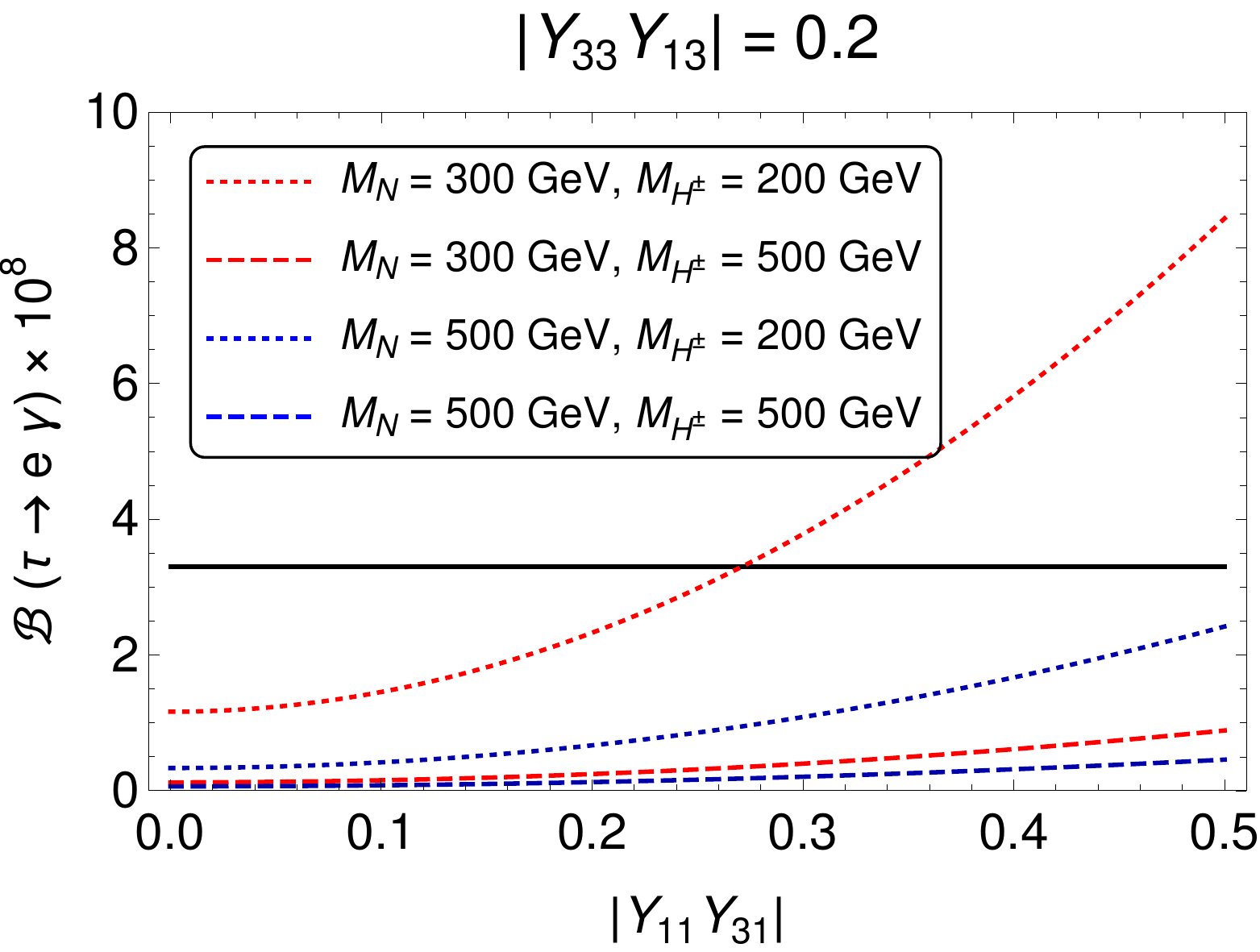}
		\label{fig:TautoEGam1}}
	\subfigure[]
	{\includegraphics[scale=0.45]{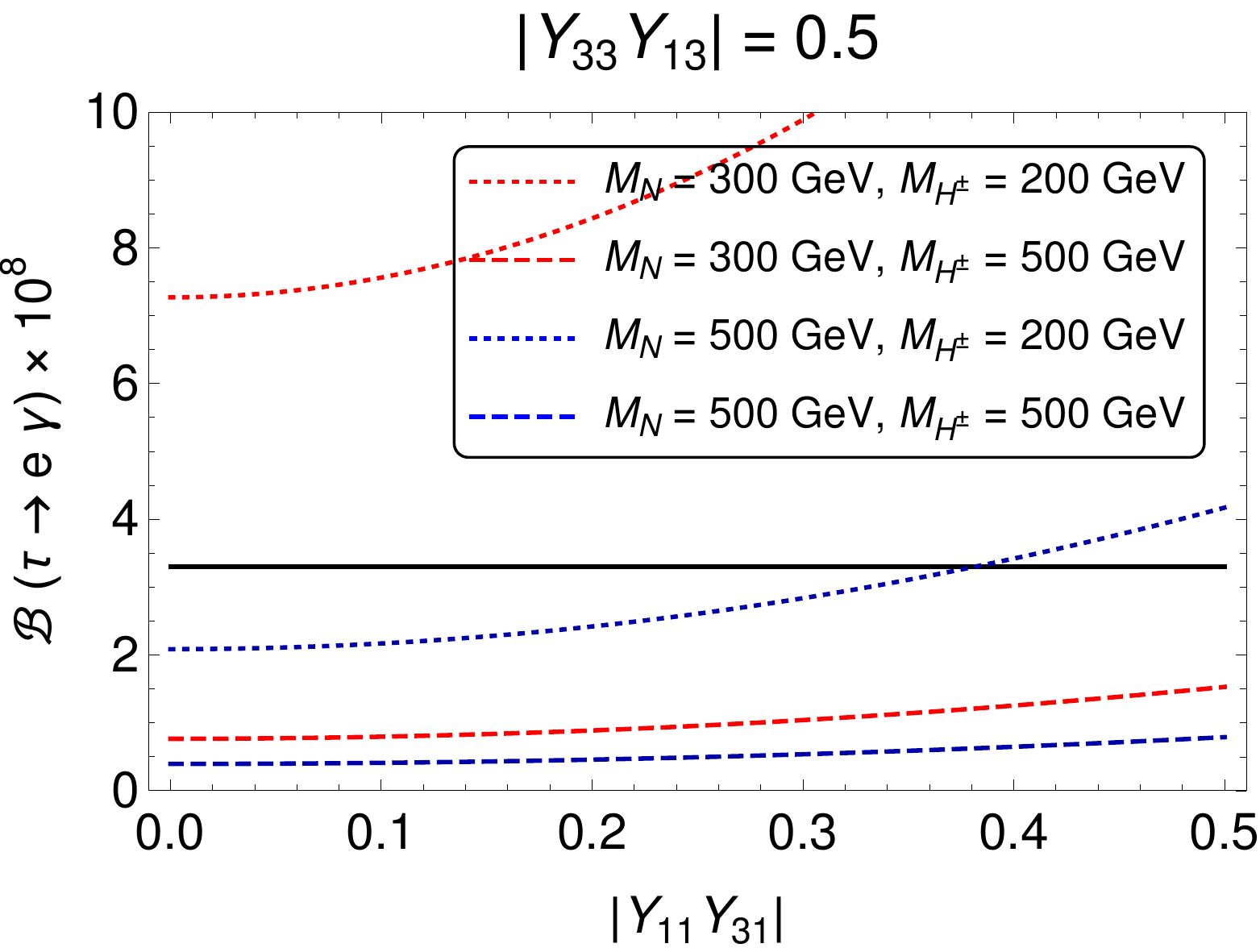}
		\label{fig:TautoEGam2}}
	\caption{Variation of $\mathcal{B}(\tau \to e \gamma$) with the coupling combination $|Y_{11}Y_{31}|$ for two different values of RHN mass $M_N \equiv M_{N_1} = M_{N_3}$ (red and blue). For a fixed $M_{N}$, we have also shown the variation with respect to the charged Higgs mass $M_{H^\pm}$  by dotted and dashed legends. The black solid line is the experimental upper limit on the branching fraction.}
	\label{fig:TautoEgam}
\end{figure}

In Figs.~\ref{fig:TautoEGam1} and \ref{fig:TautoEGam2}, we have shown the variation of $\mathcal{B}(\tau \to e \gamma)$ with the product coupling $|Y_{33}Y_{13}|$ for different values of $M_{H^{\pm}}$ and $M_N$. Also, these two figures are generated for two discrete values of $|Y_{11}Y_{31}|$, which will be helpful to understand the dependence of $\mathcal{B}(\tau \to e \gamma)$ on this coupling. Notice that for low values of the masses, both the product couplings are tightly constrained. Masses like $M_{H^{\pm}} \sim 200$ GeV and $M_N \sim 200$ GeV are allowed for values of the product couplings about $0.2$ or less which are perfectly consistent with all the other observations as mentioned earlier. For higher values of the masses more higher values of the the product couplings are allowed.


\subsubsection{Neutrino Mass Generation}

\begin{figure}[t] 
	\centering
	\includegraphics[scale=0.55]{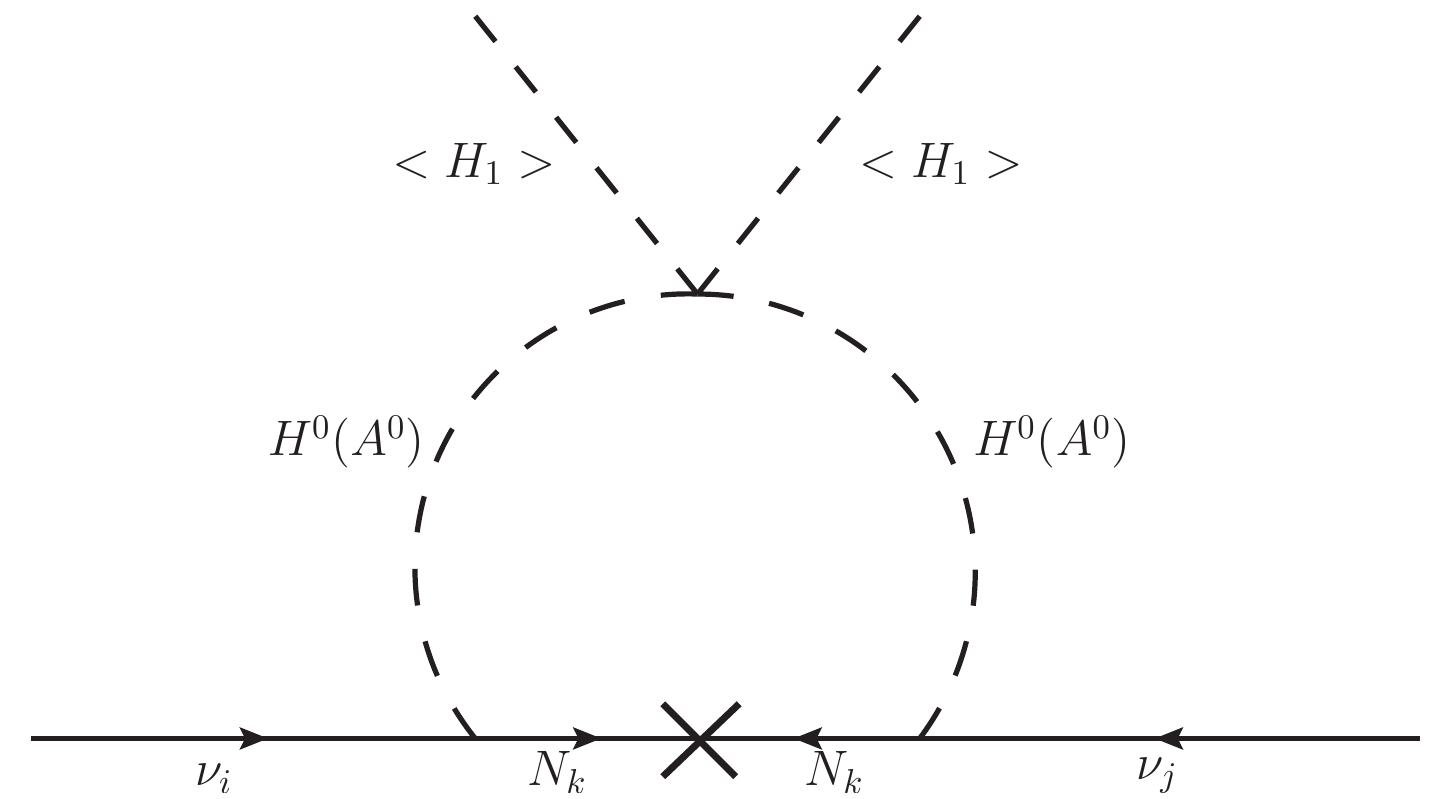}
	\caption{One loop neutrino mass generation mechanism.}
	\label{Scotogenic}
\end{figure}

The neutrino mass will be generated by radiative scotogenic mechanism in a way similar to the original proposal of \cite{Ma:2006km} as depicted in Fig.~\ref{Scotogenic} and will be mainly moderated by the mass splitting between $H^0$ and $A^0$. The one-loop contribution is given by :

\begin{equation}
(M_\nu)_{ij} = \sum \limits_{k} \frac{Y_{ik}Y_{jk}M_k}{32\pi^2} \bigg(\frac{M_{H^0}^2}{M_{H^0}^2-M_k^2} \text{ ln }\frac{M_{H^0}^2}{M_k^2} - \frac{M_{A^0}^2}{M_{A^0}^2-M_k^2} \text{ ln }\frac{M_{A^0}^2}{M_k^2}\bigg)
\label{numasstm1}
\end{equation}
where, $M_k$ is the mass of the RHN $N_k$ running in the loop.
The Majorana mass matrix, however, has the following texture :
\begin{equation}
M_R = \begin{pmatrix}
\frac{v_1}{\sqrt{2}}Y^\varphi_{11} & 0 &\frac{v_1}{\sqrt{2}}Y^\varphi_{13}\\
0 & \frac{v_2}{\sqrt{2}} Y^\varphi_{22} & 0 \\
\frac{v_1}{\sqrt{2}}Y^\varphi_{31} & 0 & \frac{v_1}{\sqrt{2}}Y^\varphi_{33}
\end{pmatrix}.
\label{mrtm1}
\end{equation}
From the expression of the inert scalar masses in eq.~\eqref{inertMasses}, one can immediately see that $(M_{H^0}^2 - M_{A^0}^2) = \lambda_5 \hspace{0.1cm}v^2$. Thus by tuning the parameter $\lambda_5$, one can obtain the correct light neutrino masses. However, it is important to ensure that the Yukawa couplings involved in the expression of light neutrino mass are consistent with the upper bound on the sum of the light neutrino masses, $\sum \limits_i m_i \leq 0.12$ eV \cite{Aghanim:2018eyx}, as well as oscillation data on the neutrino mass squared differences and mixing angles \cite{deSalas:2017kay,Esteban:2018azc}. Hence it is convenient to rewrite the Yukawa couplings in terms of the light neutrino parameters in order to automatically incorporate the above constraints on the couplings. One useful way of achieving this is through the Casas-Ibarra (CI) parametrisation \cite{Casas:2001sr} extended to the radiative seesaw model \cite{Toma:2013zsa} which enables us to express the Yukawa coupling matrix as 
\begin{equation}
Y = U D_\nu^{1/2}R^{\dagger} \Lambda^{1/2}
\label{Yuk_CI}
\end{equation}
where, $U$ is the usual Pontecorvo-Maki-Nakagawa-Sakata (PMNS) mixing matrix, $D_\nu$ is the diagonal light neutrino mass matrix, R is an arbitrary complex orthogonal matrix satisfying $R R^T = 1$ and $\Lambda$ is a diagonal matrix with elements
\begin{equation}
\Lambda_i = \frac{2\pi^2}{\lambda_5}\zeta_i \frac{2 M_i}{v^2}
\end{equation}
\begin{equation}
\text{and } \zeta_i = \bigg(\frac{M_i^2}{8(M_{H^0}^2 - M_{A^0}^2)} \bigg[L_i(M_{H^0}^2)-L_i(M_{A^0}^2)\bigg]\bigg)^{-1},
\end{equation}
where $L_i(m^2)$ is the mass function defined as 
\begin{equation}
L_i(m^2) = \frac{m^2}{m^2-M_i^2} \text{ ln }\frac{m^2}{M_i^2}.
\end{equation}
Note that we are working in a basis where the charged lepton mass matrix is not diagonal. The PMNS mixing matrix can be parametrised as 
	\begin{equation}\label{eq8}
	U=U_{\text{PMNS}}=\left[\begin{array}{ccc}
	c_{12}c_{13}&s_{12}c_{13}&s_{13}e^{-i\delta}\\
	-c_{23}s_{12}-s_{23}s_{13}c_{12}e^{i\delta}& c_{23}c_{12}-s_{23}s_{13}s_{12}e^{i\delta}&s_{23}c_{13}\\
	s_{23}s_{12}-c_{23}s_{13}c_{12}e^{i\delta}&-s_{23}c_{12}-c_{23}s_{13}s_{12}e^{i\delta}&c_{23}c_{13}
	\end{array}\right]P,
	\end{equation}
	where $c_{ij} = \cos{\theta_{ij}}, \; s_{ij} = \sin{\theta_{ij}}$ and $\delta$ is the leptonic Dirac CP phase. The diagonal matrix $P=\text{diag}(1, e^{i\alpha}, e^{i(\beta+\delta)})$  contains the Majorana CP phases $\alpha, \beta$ that appears when $\nu$ is Majorana and are not constrained by neutrino oscillation data but has to be probed by alternative experiments. This leptonic mixing matrix is related to the diagonalising matrices of charged lepton and neutrino mass matrices as $U = V^{\dagger}_L U_{\nu}$ and as mentioned above, $V_L$ is not a unit matrix in our model. It consists of a rotation in ($1-3$) plane which can be parametrised as 
\begin{equation}
V_{L}= \begin{pmatrix}
c^l_{13} & 0 & s^l_{13}e^{-i\delta_{l}} \\
0 & 1 &  0  \\
-s^l_{13}e^{i \delta_{l}} & 0 & c^l_{13} \\ 
\end{pmatrix},
\end{equation}\\	
where $c^l_{13} = \cos{\theta^l_{13}}$, $s^l_{13} = \sin{\theta^l_{13}}$ and $\delta_l$ is an arbitrary phase which we assume to be zero for simplicity. Using this and the above parametric form of PMNS mixing matrix $U$, one can parametrise $U_{\nu}$ which can then be used to parametrise the light neutrino mass matrix as 
\begin{equation}\label{eq6}
	M_\nu= U_{\nu}{M_\nu}^{(\rm diag)} U^T_{\nu}.
	\end{equation}
In the above expression for $M_{\nu}$, the diagonal light neutrino mass matrix is denoted by ${M_\nu}^{(\rm diag)}= \textrm{diag}(m_1,m_2,m_3)$ where the light neutrino masses can follow either normal ordering (NO) or inverted ordering (IO). For NO, the three neutrino mass eigenvalues can be written as 
$$M^{\text{diag}}_{\nu}
= \text{diag}(m_1, \sqrt{m^2_1+\Delta m_{21}^2}, \sqrt{m_1^2+\Delta m_{31}^2})$$ while for IO, they can be written as 
$$M^{\text{diag}}_{\nu} = \text{diag}(\sqrt{m_3^2+\Delta m_{23}^2-\Delta m_{21}^2}, \sqrt{m_3^2+\Delta m_{23}^2}, m_3)$$ 
Structure of this parametric form of light neutrino mass matrix can now be compared with the structure of light neutrino mass matrix predicted by the model. Note that the model not only predicts a specific structure of right handed neutrino mass matrix given by eq.~\eqref{mrtm1}, but also predicts the Dirac Yukawa coupling matrix to have a similar structure 
\begin{equation}
Y = \begin{pmatrix}
Y_{11} & 0 & Y_{13}\\
0 & Y_{22} & 0 \\
Y_{31} & 0 & Y_{33}
\end{pmatrix}.
\label{dytm1}
\end{equation}
Using the formula for light neutrino masses given in eq.~\eqref{numasstm1}, it can be shown that the above mentioned textures of Dirac Yukawa coupling matrix $Y$ and right handed neutrino mass matrix $M_R$ lead to a very specific structure of light neutrino mass matrix with two independent zeros namely, $(M_{\nu})_{e \mu} = (M_{\nu})_{\mu e}=0, (M_{\nu})_{\mu \tau} = (M_{\nu})_{\tau \mu} =0$ where the equality $(M_{\nu})_{\alpha \beta}=(M_{\nu})_{\beta \alpha}$ results due to Majorana nature of light neutrinos giving rise to a complex symmetric structure of mass matrix.

We numerically solve these two texture zero complex equations in order to evaluate the unknowns namely, the lightest neutrino mass $m_1$ (NO), $m_3$ (IO), leptonic Dirac CP phase $\delta$ as well as two Majorana CP phases $\alpha, \beta$. The additional rotation angle in charged lepton sector $\theta^l_{13}$ is considered as a free parameter which can lie anywhere in $(0, \pi/2)$. The other known parameters namely, three mixing angles, two mass squared differences are varied in $3\sigma$ range \cite{Esteban:2018azc}. We find that these textures in light neutrino mass matrix predict a large value of the lightest neutrino mass, which is in tension with Planck 2018 bound on sum of absolute neutrino masses $\sum \limits_i m_i \leq 0.12$ eV \cite{Aghanim:2018eyx} as well as bounds on absolute neutrino mass scale from laboratory based experiments like KATRIN \cite{Aker:2019uuj}. Even if we consider a non-zero CP phase in charged lepton correction matrix $V_L$, this conclusion does not change. This is not surprising, given the fact that almost all possible two-zero textures in diagonal charged lepton basis are ruled out by latest experimental data \cite{Borgohain:2020now}.

One possible way to make it consistent with neutrino data without changing the model significantly is to change the $U(1)_X$ charge of the singlet scalar $\phi_2$ from 4 to 1. This results in a right handed neutrino mass matrix having only one zero at $(22)$ entry. While the lightest eigenstate of singlet fermion mass matrix can still be a DM candidate, no zeros appear in the light neutrino mass matrix even with the same Dirac Yukawa \eqref{dytm1}. Such a general structure of light neutrino mass matrix can be fitted with light neutrino data as there are sufficient free parameters, unlike in the previous case with two texture zeros. It is very unlikely that such a setup will change our DM and flavour physics results significantly. In the following subsection, we have added a discussion on this modified scenario.

\subsubsection{Modified Setup for Toy Model I}

As mentioned in the previous section, the light neutrino mass matrix that we obtain in this scenario violates the Planck 2018 bound on the sum of absolute neutrino masses. We also identified that a possible way out of this issue is by choosing the $U(1)_X$ charge of the singlet scalar $\phi_2$ to be 1 instead of 4. In this subsection, we will briefly point out the changes that will occur in our theoretical setup and how it might affect the other observables. First of all, the Yukawa interactions given in eq.~\eqref{eq:LYuk1} will be modified as given below in eq.~\eqref{eq:yukawamod}.
\begin{equation}
-\mathcal{L}_Y \supset \sum \limits_{i,j} Y_{ij} \bar{L}_i \tilde{H_2} N_j + Y_{22} \bar{L}_2 \tilde{H_2} N_2 + \sum_{i,j = (1,3)} Y^\varphi_{ij} \bar{N}_i^c N_j \varphi_1 + \sum_{i=1,3} Y^\varphi_{i2} \bar{N}_i^c N_2 \varphi_2.
\label{eq:yukawamod}
\end{equation}
Note that the first three terms of the Yukawa Lagrangian remain unchanged, however, the interaction term involving $N_2$ and $\phi_2$ has changed. Also, there will be a little change in the scalar potential, the trilinear term in eq.~\eqref{scalarpot1} now becomes $\left\{\delta \hspace{0.1cm} \varphi_2 \varphi_2 \varphi_1^\dag  + \text{h.c.}\right\}$. Hence, the pseudoscalar mass, which primarily depended on this trilinear term, modifies to $M_{A_2} = \big(-\frac{v_2^2 \delta}{\sqrt{2} v_1 s_\gamma^2}\big)^{1/2}$. Recall that the gauge boson mass $M_X$ mass and gauge coupling $g_X$ are related to the singlet vevs (eq.~\eqref{eq:GBMass1}). With the change in the $U(1)_X$ charge of $\varphi_2$, the aboove relation changes to
\begin{equation}
M_X^2 = \frac{1}{4} {g^\prime}^2 v^2 \epsilon^2   +  g_{X}^2 (4 v_1^2 + v_2^2),
\label{MX-mod}
\end{equation}
and for $\epsilon << 1$, we obtain $M_{Z^\prime} \simeq M_X = g_X \sqrt{(4 v_1^2 + v_2^2)}$. For simplicity, if we consider $v_1 = v_2$, then from eq.~\eqref{MX-mod}, $v_1 \approx 450$ GeV for $M_X = 1$ GeV and $g_X \simeq 0.001$. Therefore the masses of $s_1$ and $s_2$ will be restricted to be $\lsim 450$ GeV for the Yukawa and quartic couplings to remain perturbative. 

\begin{figure}[t]
	\centering
	\includegraphics[scale=0.6]{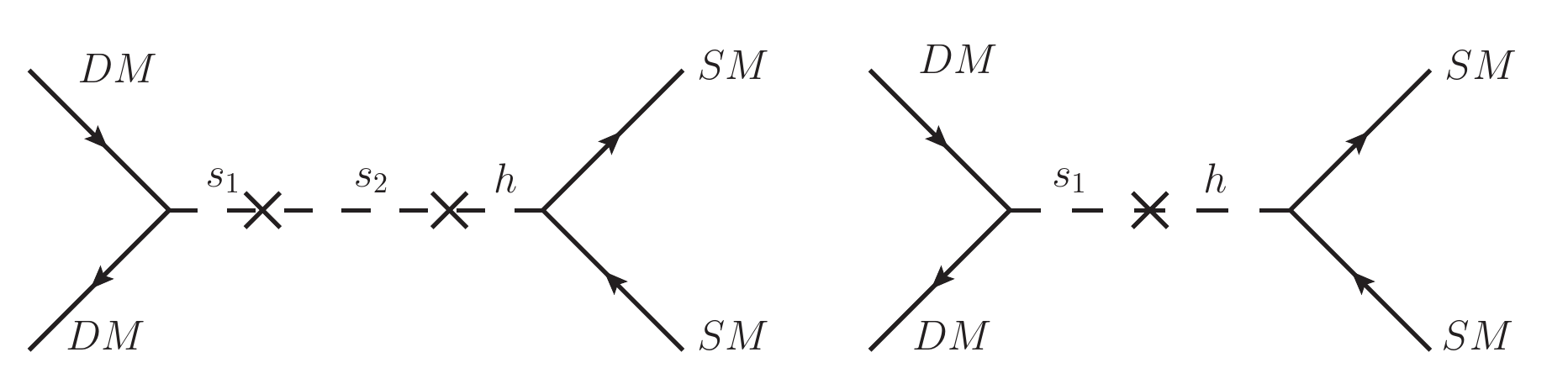}
	\caption{Dark matter annihilation diagrams.}
	\label{fig:dmann_mod}
\end{figure}

The analysis of relic abundance and the direct detection cross section will be in a similar line as discussed in subsection \ref{sec:toy1DM}. The annihilation via $Z^\prime$ remains the same. However, we can not consider a pure $N_2$ state as our DM candidate, since the the Yukawa Lagrangian does not have a Majorana mass term for $N_2$. Therefore, in principle, the lightest particle of $N_1$ and $N_3$ can be our DM candiate, and the dominating contributions will come from the annihilation diagrams shown in Fig.~\ref{fig:dmann_mod}. In such situation, as before, depending on the mass of $s_1$, the relic abundance will once again be satisfied near the resonances i.e. near $M_{DM} \sim M_{s_1}/2$. In the presence of $s$-channel annihilation, the role of co-annihilations are expected to be sub-dominant as in the previous setup.   

The possibility of mixing of the pure states $N_1$, $N_2$ and $N_3$ can be considered by rotating the interaction basis $N_i$ to a new basis $N_i^\prime$ by using a general unitary transformation as
\begin{equation}
\begin{pmatrix}
N_1 \\ N_2 \\ N_3
\end{pmatrix} = \mathcal{O}_{\nu_R} \begin{pmatrix}
N_1^\prime \\ N_2^\prime \\ N_3^\prime
\end{pmatrix}
\end{equation} 
which will result in a mass matrix of the form
\begin{equation}
\begin{pmatrix}
M_1^\prime \\ M_2^\prime \\ M_3^\prime
\end{pmatrix} = \mathcal{O}_{\nu_R} M_R \mathcal{O}_{\nu_R}^T.
\end{equation}
In the rotated basis, the lowest mass eigenstate can be considered as the DM candidate which will contribute via the annihilation diagram as given in Fig.~\ref{fig:dmann_mod}.

The Yukawa Lagrangian responsible for the RHN masses also gets modified such that the Majorana mass mixing matrix now becomes
\begin{equation}
M_R = \begin{pmatrix}
\frac{v_1}{\sqrt{2}}Y^\varphi_{11} & \frac{v_2}{\sqrt{2}}Y^\varphi_{12} &\frac{v_1}{\sqrt{2}}Y^\varphi_{13}\\
\frac{v_2}{\sqrt{2}}Y^\varphi_{21} & 0 & \frac{v_2}{\sqrt{2}}Y^\varphi_{23} \\
\frac{v_1}{\sqrt{2}}Y^\varphi_{31} & \frac{v_2}{\sqrt{2}}Y^\varphi_{32} & \frac{v_1}{\sqrt{2}}Y^\varphi_{33}
\end{pmatrix}.
\end{equation}
This is in contrast to the mass matrix we obtained before in eq.~\eqref{mrtm1}. Since the mixing angles ($s^\nu_{ij}$) of $\mathcal{O}_{\nu_R}$ are completely arbitrary, we have full freedom of choosing them in a way such that $M_2^\prime < M_1^\prime, M_3^\prime$ and the Yukawa couplings are also perturbative. 

It is important to note that the contributions to the other observables like anomalous magnetic moments, LFV decays and $R(D^{(*)})$ remain unaltered. We have already seen that a charged Higgs and RHN mediated diagram contributes to the magnetic moments of the leptons (cf. Fig.~\ref{fig:LFV_chrgdsc}). In the modified set-up, the changes occur in the Majorana- Yukawa interactions, which involves the coupling $Y_{ij}^{\phi}$, and they do not contribute to $\Delta a_{\mu,e}$, LFV decays or $R(D^{(*)})$.

\subsection{Toy Model II}
\subsubsection{Particle Content}
In this toy model, we have the same particle content as in the previous case, with the only difference being that all particles except $N_2$ are even under the discrete $\mathcal{Z}_2$ symmetry. This will once again prevent it from interacting directly with SM leptons. However, in this scenario, the neutrino mass generation mechanism will be different from the previous one. The particle content, along with their respective gauge quantum number and charges, has been described in Table~\ref{table3}. 

\begin{table}[t]
\begin{center}
\renewcommand{\arraystretch}{0.8}
\begin{tabular}{|c|c|c|c|}
\hline
Particles & $SU(3)_c \times SU(2)_L \times U(1)_Y$ & $U(1)_X $  & $\mathcal{Z}_2$ \\
\hline
$Q_L=\begin{pmatrix}u_{L}\\
d_{L}\end{pmatrix}$ & $(3, 2, \frac{1}{6})$ & 0  & +\\
$u_R$ & $(3, 1, \frac{2}{3})$ & 0 & + \\
$d_R$ & $(3, 1, -\frac{1}{3})$ & 0  & +\\
$L_1 = \begin{pmatrix}\nu_{e}\\
e\end{pmatrix}_L$ & $(1, 2, -\frac{1}{2})$ & $-1$ & + \\
$L_2 = \begin{pmatrix}\nu_{\mu}\\
\mu \end{pmatrix}_L$ & $(1, 2, -\frac{1}{2})$ & $2$ & + \\
$L_3 = \begin{pmatrix}\nu_{\tau}\\
\tau \end{pmatrix}_L$ & $(1, 2, -\frac{1}{2})$ & $-1$  & +\\
$e_R$ & $(1, 1, -1)$ & $-1$ & + \\
$\mu_R$ & $(1, 1, -1)$ & $2$ & + \\
$\tau_R$ & $(1, 1, -1)$ & $-1$ & + \\
$H_1$ & $(1,2,\frac{1}{2})$ & 0 & +\\
\hline
$N_{1R}$ & $(1, 1, 0)$ & $-1$ & + \\
$N_{2R}$ & $(1, 1, 0)$ & $2$ & - \\
$N_{3R}$ & $(1, 1, 0)$ & $-1$ & + \\
$H_2$ & $(1,2,\frac{1}{2})$ & $-3$ & +\\
$\varphi_1$ & $(1,1,0)$ & 2 & + \\
$\varphi_2$ & $(1,1,0)$ & 4 & + \\
\hline
\end{tabular}
\end{center}
\caption{Particle content for Toy model II.}
\label{table3}
\end{table}

\subsubsection{Lagrangian and Scalar Mass Spectrum}

In this scenario, the successful generation of charged lepton and light neutrino masses require $H_2$ to be charged under $U(1)_X$. The relevant Yukawa interactions are given by:
\begin{equation}
\begin{aligned}
-\mathcal{L}_Y & \supset \sum \limits_{i,j} Y_{ij}^\ell \bar{L}_i H_1 e_{jR} + \sum \limits_{j} Y_{jk}^\ell \bar{L}_j H_2 e_{kR} + \sum \limits_{i,j} Y_{ij}^\nu \bar{L}_i \tilde{H_1} N_{jR} + \sum \limits_{j} Y_{kj}^\nu \bar{L}_k \tilde{H_2} N_{jR} \\& + \sum \limits_{i,j} Y^{\varphi}_{ij} \bar{N_{iR}^c} N_{jR} \varphi_1 + Y^{\varphi}_{kk} \bar{N_{kR}^c} N_{kR} \varphi^{\dagger}_2
\end{aligned}
\label{LYuk2}
\end{equation}
where both $i$ and $j$ can take values $(1,3)$ and $k=2$. Thus only the second generation of lepton doublet couples to $N_{1,3}$ via the second Higgs doublet $H_2$. The scalar Lagrangian will be similar to the one defined in eq.~\eqref{eq:massbosons} with the scalar potential as given below:
\begin{equation}
\begin{aligned}
V(H_1,H_2, \varphi_1, \varphi_2) &= \mu_1^2|H_1|^2 +\mu_2^2|H_2|^2 + \mu_3^2|\varphi_1|^2 + \mu_4^2|\varphi_2|^2 +\frac{\lambda_{H_1}}{2}|H_1|^4+\frac{\lambda_{H_2}}{2}|H_2|^4 \\& + \frac{\lambda_{\varphi_1}}{2}|\varphi_1|^4 +\frac{\lambda_{\varphi_2}}{2}|\varphi_2|^4 + \lambda_{1}|H_1|^2|H_2|^2 + \lambda_{2}(H_1^\dag H_2)(H_2^\dag H_1) \\& + \lambda_3 (\varphi_1^\dag \varphi_1)(\varphi_2^\dag \varphi_2) + \lambda_{4} |H_1|^2 |\varphi_1|^2 +\lambda_{5} |H_1|^2 |\varphi_2|^2 + \lambda_6 |H_2|^2 |\varphi_1|^2 \\& + \lambda_7 |H_2|^2 |\varphi_2|^2 + \bigg\{\delta \hspace{0.1cm} \varphi_1 \varphi_1 \varphi_2^\dag  + \text{h.c.}\bigg\} + \frac{c}{\Lambda^2}\bigg\{(H_1^\dag H_2)^2 (\varphi_1 \varphi_2)+\text{h.c.} \bigg\}.
\end{aligned}
\label{eq:scalarpot2}
\end{equation}

In this case, all the scalars acquire a vev and are given by :
\begin{equation}
H_1=\begin{pmatrix} w^\pm \\  \frac{ v +h^{'} + i z^{'}}{\sqrt 2} \end{pmatrix} , \hspace{0.1cm} H_2=\begin{pmatrix} h^\pm\\ \frac{u+{H^0}^{'}+i{A^0}^{'}}{\sqrt 2} \end{pmatrix},  \hspace{0.1cm}
\varphi_1 = \bigg(\frac{v_1 + s_1^{'} + i A_1^{'}}{\sqrt{2}}\bigg), \hspace{0.1cm}  \varphi_2 = \bigg(\frac{v_2 + s_2^{'} + i A_2^{'}}{\sqrt{2}}\bigg)
\end{equation}
 Under such a scenario, electroweak symmetry breaking of the scalars require $\mu_i^2 < 0 \hspace{0.1cm}(i=1,2,3,4)$ and the minimization conditions are given by:
\begin{eqnarray}
\mu_1^2 &=& -\frac{1}{2}\bigg(u^2(\lambda v_1 v_2 + \lambda_1 + \lambda_2) + \lambda_4 v_1^2 +\lambda_5 v_2^2 + \lambda_{H_1} v^2\bigg), \nonumber \\
\mu_2^2 &=& -\frac{1}{2}\bigg(v^2(\lambda v_1 v_2 + \lambda_1 + \lambda_2) + \lambda_6 v_1^2 +\lambda_7 v_2^2 + \lambda_{H_2} u^2 \bigg), \nonumber \\
\mu_3^2 &=& -\frac{1}{4v_1}\bigg(u^2 v^2 v_2 \lambda + 2v_1(2\sqrt{2}v_2 \delta + \lambda_3 v_2^2 + \lambda_4 v^2 +\lambda_6 u^2) + 2\lambda_{\varphi_1}v_1^3 \bigg), \nonumber \\
\mu_4^2 &=& -\frac{1}{4v_2}\bigg(u^2 v^2 v_1 \lambda + 2v_1^2(\sqrt{2}v_2 \delta + \lambda_3 v_2) + 2v_2(\lambda_5 v^2 +\lambda_7 u^2 + \lambda_{\varphi_2}v_2^2) \bigg),
\label{eq:MinCond2}
\end{eqnarray}
where $\lambda = \frac{c}{\Lambda^2}$; the usefulness of this term will be discussed later in this subsection. The covariant derivative can be defined in the same way as in the previous case eq.~\eqref{eq:CovDer}. From the kinetic part of the scalar Lagrangian, we obtain the mass of the W-boson as :
\begin{equation}
M_W^2 = \frac{1}{4}g^2 (u^2 +v^2).
\label{Wmass2}
\end{equation}
One can rewrite the mass of W as $M_W^2 = \frac{1}{4}g^2 v_H^2$ where, $v_H^2 = (u^2 + v^2) = (246)^2$ GeV$^2$. We also express the ratio of the two vevs as $\frac{v}{u} = \text{tan }\beta$. The neutral gauge bosons ($W_\mu^3, B_\mu, X_\mu$) on the other hand mix and the mixing matrix is given by :
\begin{equation}
M_{GB}^2 = \begin{pmatrix}
\frac{1}{4}g^2 (u^2 + v^2) & -\frac{1}{4}g g'(u^2 + v^2) & -\frac{3}{2}g g_X u^2 \\
-\frac{1}{4}g g'(u^2 + v^2) & \frac{1}{4} {g'}^2 (u^2 + v^2) & \frac{3}{2}g^{'} g_X u^2  \\
-\frac{3}{2}g g_X u^2 & \frac{3}{2}g' g_X u^2 & g_X^2 \bigg(9 u^2 +4(v_1^2 + 4v_2^2)\bigg).
\end{pmatrix}
\label{GBMassMat2}
\end{equation}
After the usual Weinberg rotation as given in eq.~\eqref{Weinberg_Rot}, we obtain the masses of the physical neutral gauge bosons as :
\begin{eqnarray}
M_\gamma^2 &=& 0, \\
M_Z^2 &=& M_{Z^0}^2 = \frac{1}{4 C_W^2}g^2 v_H^2, \\
M_{Z^\prime}^2 &=& M_X^2 - \left(\frac{\Delta^4}{M_{Z^0}^2}\right),
\end{eqnarray}
where $M_X^2 = g_X^2(4(v_1^2 + 4v_2^2) + 9u^2) - 3 g' g_X u^2 \epsilon + \mathcal{O}(\epsilon^2)$ and $\Delta^2 = \frac{g}{4 C_W^2}\left(6 g_X u^2 - g' v_H^2 \epsilon\right)$.
One can immediately see that in the limit $\epsilon << 1$, $M_{Z^\prime}^2 = M_X^2 \simeq g_X^2 (9u^2 + 4(v_1^2 + 4v_2^2) + 9u^2 )$.

In this model, none of the scalars are $\mathcal{Z}_2$ odd, therefore, in principle, both the CP even and CP odd neutral components mix to give two $(4\times 4 )$ mixing mass matrices; one for $(h^{'}, s_1^{'}, s_2^{'}, {H^0}^{'})$ and the other for $(z^{'}, A_1^{'}, A_2^{'}, {A^0}^{'})$ as given below in eqs.~\eqref{eq:CPevenMassMat2} and \eqref{eq:CPoddMassMat2}, respectively. We also have a $(2\times2)$ mixing matrix for the charged scalars $(w^\pm, h^\pm)$ as given in eq.~\eqref{eq:ChargedMassMat}. 
\begin{equation}
\footnotesize
M_{sc}^2 = \begin{pmatrix}
\lambda_{H_1} v^2 & \frac{1}{2}\lambda u^2 v v_2 +\lambda_4 v v_1 & \frac{1}{2}\lambda u^2 v v_1 +\lambda_5 v v_2  & uv (\lambda v_1 v_2 + \lambda_1 + \lambda_2) \\
\frac{1}{2}\lambda u^2 v v_2 +\lambda_4 v v_1 & -\frac{\lambda u^2 v^2 v_2 - 4 \lambda_{\varphi_1} v_1^3}{4 v_1}  & \sqrt{2}v_1 \delta + \frac{1}{4}\lambda u^2 v^2 + \lambda_3 v_1 v_2  & u(\frac{1}{2}\lambda v^2 v_2 + \lambda_6 v_1) \\
\frac{1}{2}\lambda u^2 v v_1 +\lambda_5 v v_2 & \sqrt{2}v_1 \delta + \frac{1}{4}\lambda u^2 v^2 + \lambda_3 v_1 v_2 & -\frac{2\sqrt{2} v_1^2 \delta + \lambda u^2 v^2 v_1 - 4 \lambda_{\varphi_2} v_2^3}{4v_2} & u(\frac{1}{2}\lambda v^2 v_1 + \lambda_7 v_2) \\
uv (\lambda v_1 v_2 + \lambda_1 + \lambda_2) & u(\frac{1}{2}\lambda v^2 v_2 + \lambda_6 v_1) & u(\frac{1}{2}\lambda v^2 v_1 + \lambda_7 v_2) & \lambda_{H_2} u^2 
\end{pmatrix}
\label{eq:CPevenMassMat2}
\end{equation}

\begin{equation}
\footnotesize
M_{pseudo}^2 = \begin{pmatrix}
-\lambda u^2 v_1 v_2 & \frac{1}{2}\lambda u^2 v v_2 &   \frac{1}{2}\lambda u^2 v v_1 & \lambda u v v_1 v_2 \\
\frac{1}{2}\lambda u^2 v v_2 & -\frac{v_2(8\sqrt{2} v_1 \delta + \lambda u^2 v^2)}{4v_1} & \sqrt{2}v_1 \delta - \frac{1}{4} \lambda u^2 v^2 & -\frac{1}{2}\lambda u v^2 v_2 \\
\frac{1}{2}\lambda u^2 v v_1 & \sqrt{2}v_1 \delta - \frac{1}{4} \lambda u^2 v^2 & -\frac{v_1(2\sqrt{2} v_1 \delta + \lambda u^2 v^2)}{4v_2} & -\frac{1}{2}\lambda u v^2 v_1 \\
\lambda u v v_1 v_2 & -\frac{1}{2}\lambda u v^2 v_2 & -\frac{1}{2}\lambda u v^2 v_1 & -\lambda v^2 v_1 v_2
\end{pmatrix}
\label{eq:CPoddMassMat2}
\end{equation}

\begin{equation}
\footnotesize
M_{ch}^2 = \frac{(v_1 v_2 \lambda + \lambda_2)}{2} \begin{pmatrix}
- u^2 & \frac{1}{2} uv  \\ \frac{1}{2} uv  & - u^2 \end{pmatrix}
\label{eq:ChargedMassMat}
\end{equation}
We therefore require two $(4\times 4 )$ rotation matrices (cf. Appendix \ref{RotMat4}) to diagonalize the CP even and CP odd Higgs which we denote by $\mathcal{R}_\alpha$ and $\mathcal{R}_\theta$ respectively (as shown in eq.~\eqref{HiggsRot}) and an orthogonal rotation by angle $\gamma$ for the charged scalars.

\begin{equation}
\begin{pmatrix} h\\ s_1 \\s_2 \\ H^0\end{pmatrix} = \mathcal{R}_\alpha^T \begin{pmatrix}
h^{'} \\ s_1^{'} \\ s_2^{'} \\ {H^0}^{'}\end{pmatrix}, \begin{pmatrix}
G_z\\ G_{z'} \\A_2 \\ A^0\end{pmatrix} = \mathcal{R}_\theta^T \begin{pmatrix}
z^{'}\\ A_1^{'}\\ A_2^{'}\\ {A^0}^{'}
\end{pmatrix} \text{and, } \begin{pmatrix} G^\pm\\ H^\pm \end{pmatrix} = \mathcal{R}_\gamma^T \begin{pmatrix}
w^\pm\\ h^\pm \end{pmatrix}
\label{HiggsRot}
\end{equation}
where $G_z, G_{z'}$ and $G^\pm$ are the massless Goldstones corresponding to the physical vector bosons $Z, Z^\prime$ and $W$ respectively. Notice that in the scalar potential in eq.~\eqref{eq:scalarpot2}, we have added a higher dimensional symmetry breaking term proportional to $\lambda (=\frac{c}{\Lambda^2})$ apart from the trilinear term. The relevance of this term can be easily understood from the scalar mass matrix given in eq.~\eqref{eq:CPoddMassMat2}. In this matrix, the elements of the first and fourth rows and columns are proportional to $\lambda$; hence, if we set $\lambda = 0$, the resulting mass matrix will be a $2\times 2$ matrix with determinant zero, which results in zero-mass pseudo-scalar fields (not allowed). Also, in absence of this term, the $U(1)_X$ symmetry can be broken by the vev of $H_2$ alone, and we don't need the additional singlet scalars.

We denote the angles in $\mathcal{R}_{\alpha(\theta)}$ by $\alpha_{ij}(\theta_{ij})$. Thus, we have many unconstrained terms in the rotation matrices $\mathcal{R}_{\alpha,\theta}$ with at least 6 mixing angles in each, and so we make the following assumptions to simplify the analysis:

\begin{enumerate}
\item[(i)] The mixing angles of $h$ with the singlet scalars are $\alpha_{12} \equiv \alpha_2$ and $\alpha_{13} \equiv \alpha_3$, respectively. Also, we have not considered very large mixing scenarios.
\item[(ii)] For simplicity, the mixing angles of $H^0$ with the singlet scalars are set to zero, i.e  $\alpha_{24} = \alpha_{34} \equiv 0$. Also, the possibility of mixing between the two singlet scalars has been neglected, i.e $\alpha_{23} \simeq 0$.
\item[(iii)] We denote the mixing of the $h$ and $H^0$ by $\alpha_{14} \equiv \alpha$.
\end{enumerate}
A similar approximation is also considered for the rotation matrix $\mathcal{R}_{\theta}$. This helps us to eliminate some of the mixing angles for each of the matrices. Therefore, we are left with the following free parameters: 
\begin{equation}
\text{tan}\beta, v_1, M_h, M_{s_1,s_2,H^0}, M_{A^0},M_{A_2},M_{H^\pm}, \text{Mixing angles }(\alpha, \alpha_2, \alpha_3, \theta, \theta_2, \theta_3, \gamma).
\end{equation}
The couplings expressed in terms of masses and mixing angles can be found in Appendix \ref{sec:Const2}. These model parameters are constrained from both theoretical requirements of unitarity, vacuum stability, perturbativity etc. and experimental data on electroweak observables, Higgs decays and so on. We have assumed small values of $\alpha_2$ and $\alpha_3$ so that we can utilize the existing bound on the parameters like $\alpha$ and $\beta$ of a two Higgs doublet model (2HDM) scenario with and without an additional singlet. For recent analyses of extended 2HDM see \cite{Chen:2013jvg, Drozd:2014yla,Muhlleitner:2016mzt,vonBuddenbrock:2018xar,Arhrib:2018qmw}. It has been shown that large singlet doublet admixture is allowed by the LEP and LHC data \cite{Muhlleitner:2016mzt}. However, a large admixture does not allow a large value for $|\cos(\beta-\alpha)|$ \cite{Chen:2013jvg}. In our analysis, we have considered the scenarios with $\text{tan } \beta \leq 5$ and $|\text{cos}(\beta -\alpha)| < 0.1$, also, we have assumed  $\text{sin }\alpha_2 \le 0.1$ and $\text{sin }\alpha_3 \le 0.1$. 

We identify $h$ to be the 125 GeV Higgs boson discovered at the LHC and restrict the parameters in the following range :
\begin{equation}
\begin{aligned}
	& 1 < \text{tan }\beta < 5, \quad \text{cos }(\beta - \alpha) \in [-0.1,0.1], \quad v_1 = 450 \text{ GeV}, \\& M_{H^0} \in [1,100] \text{ GeV}, \quad M_{s_1} \in [150,450]\text{ GeV}, \quad M_{s_2} \in [10,100] \text{ GeV}, \\& M_{H^\pm} \in [100,300] \text{ GeV}, \quad  M_{A^0} \in [100,300] \text{ GeV}, \quad  M_{A_2} \in [100,300] \text{ GeV}, \\& \text{sin }\alpha_{2,3} \in [-0.1,0.1], \quad \text{sin }\theta \in [-0.1,0.1], \quad \text{sin }\theta_{2,3} \in [-0.1,0.1], \quad \text{sin }\gamma \in [-0.1,0.1].
	\end{aligned}
	\end{equation}
	For the above range of masses, the scale $\Lambda \sim (300 - 600)$ GeV for $c = -1$. Note that $\tan\beta > 1$ allows only the scenario $M_h > M_{H^0}$ otherwise $\lambda_{H_2}$ will pick up a very large value. One can have the scenario $M_h < M_{H^0}$ when $\tan\beta < 1$, however, these choices will lead to the large values of $\lambda_{H_1}$, and at the same time top-Yukawa $y_{t} >> 1$. 

\subsubsection{DM Phenomenology}

\begin{figure}[t]
	\centering
	\subfigure[]
	{\includegraphics[scale=0.45]{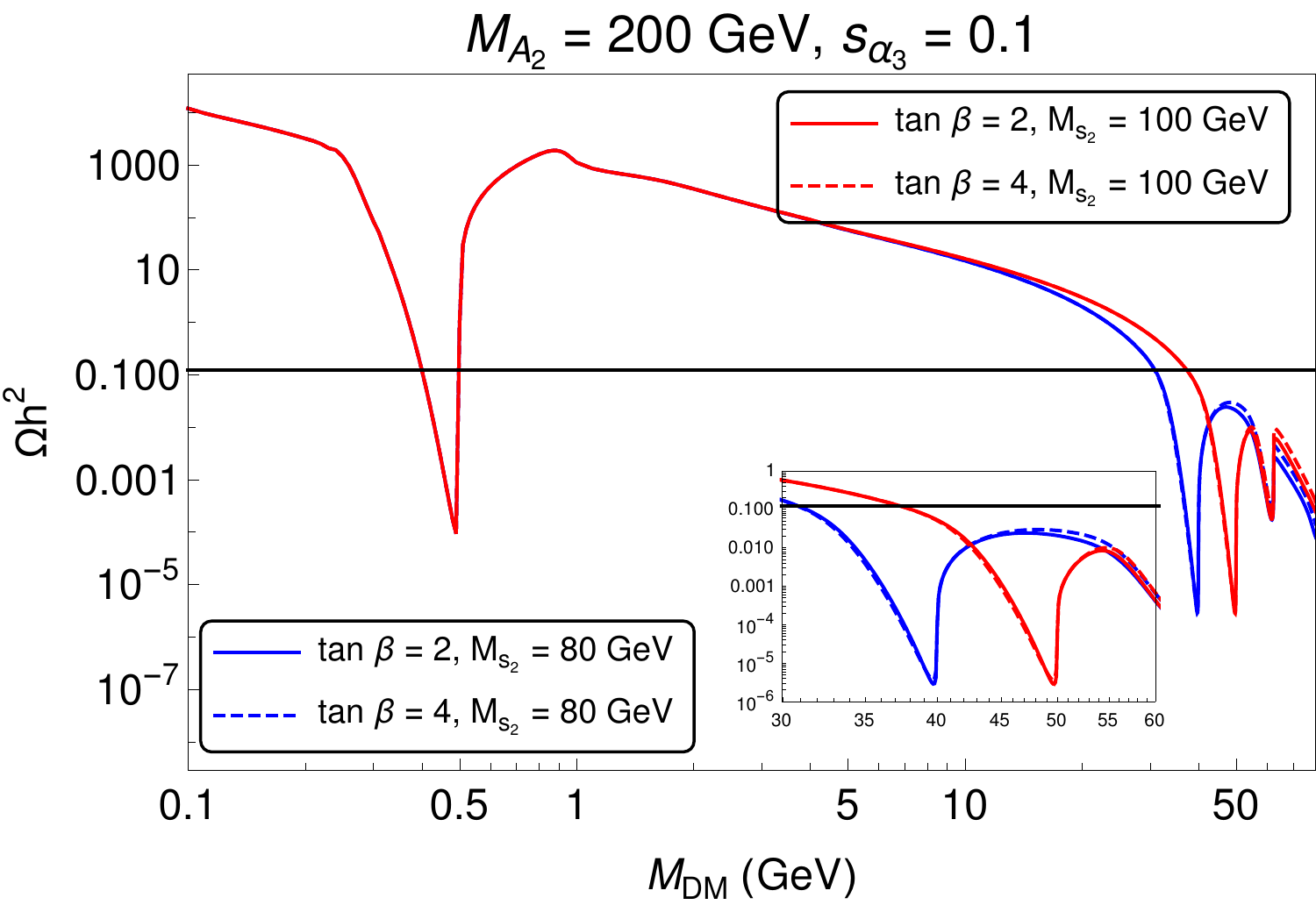}
		\label{fig:relic1}}
	\subfigure[]
	{\includegraphics[scale=0.39]{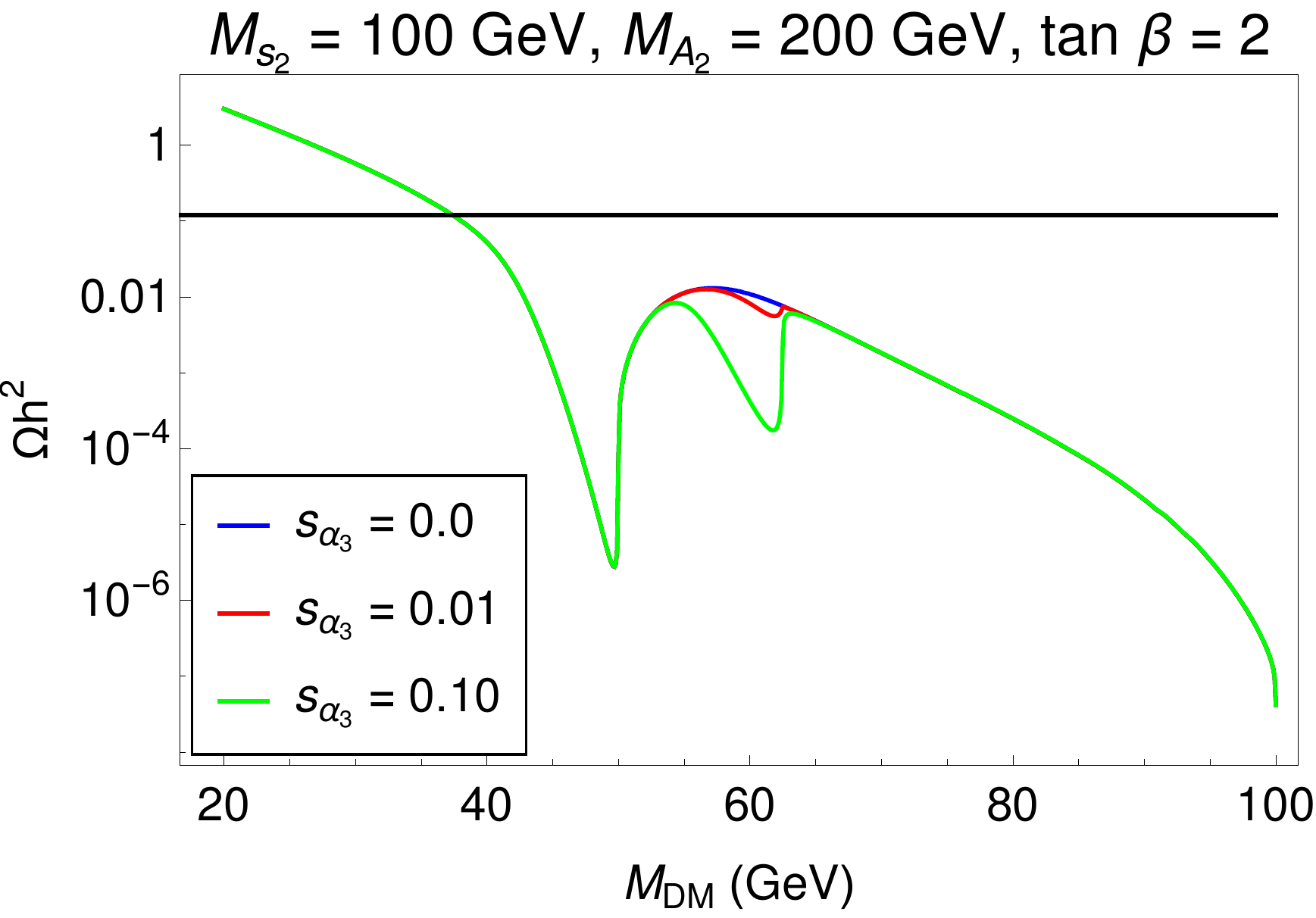}
		\label{fig:relic2}}
	\caption{(a) The variation of relic abundance with the dark matter mass for different values of $\text{tan }\beta$ and the mass $M_{s_2}$. In the inset we have zoomed into the annihilation peaks of the DM for $30 < M_{DM} (\text{GeV})< 60$. The black solid line denotes the Planck observed relic of DM. (b) Same as in Fig.~\ref{fig:relic1} for different values of sine of the mixing angle $\alpha_3$.}
	\label{fig:relic_vs_DM}
\end{figure}

\begin{figure}[htp!]
	\centering
	\subfigure[]
	{\includegraphics[scale=0.45]{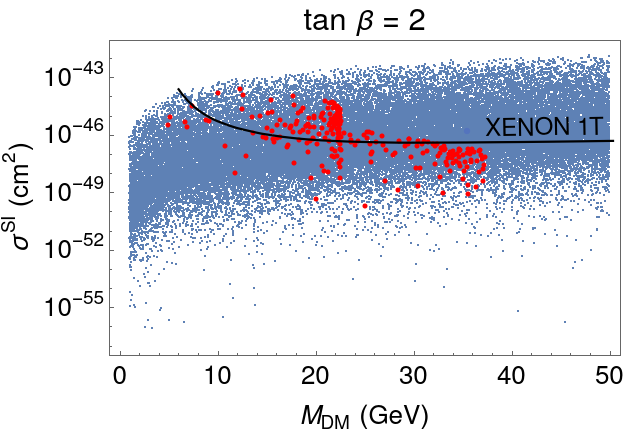}
		\label{fig:ddvsmass}}~
	\subfigure[]
	{\includegraphics[scale=0.35]{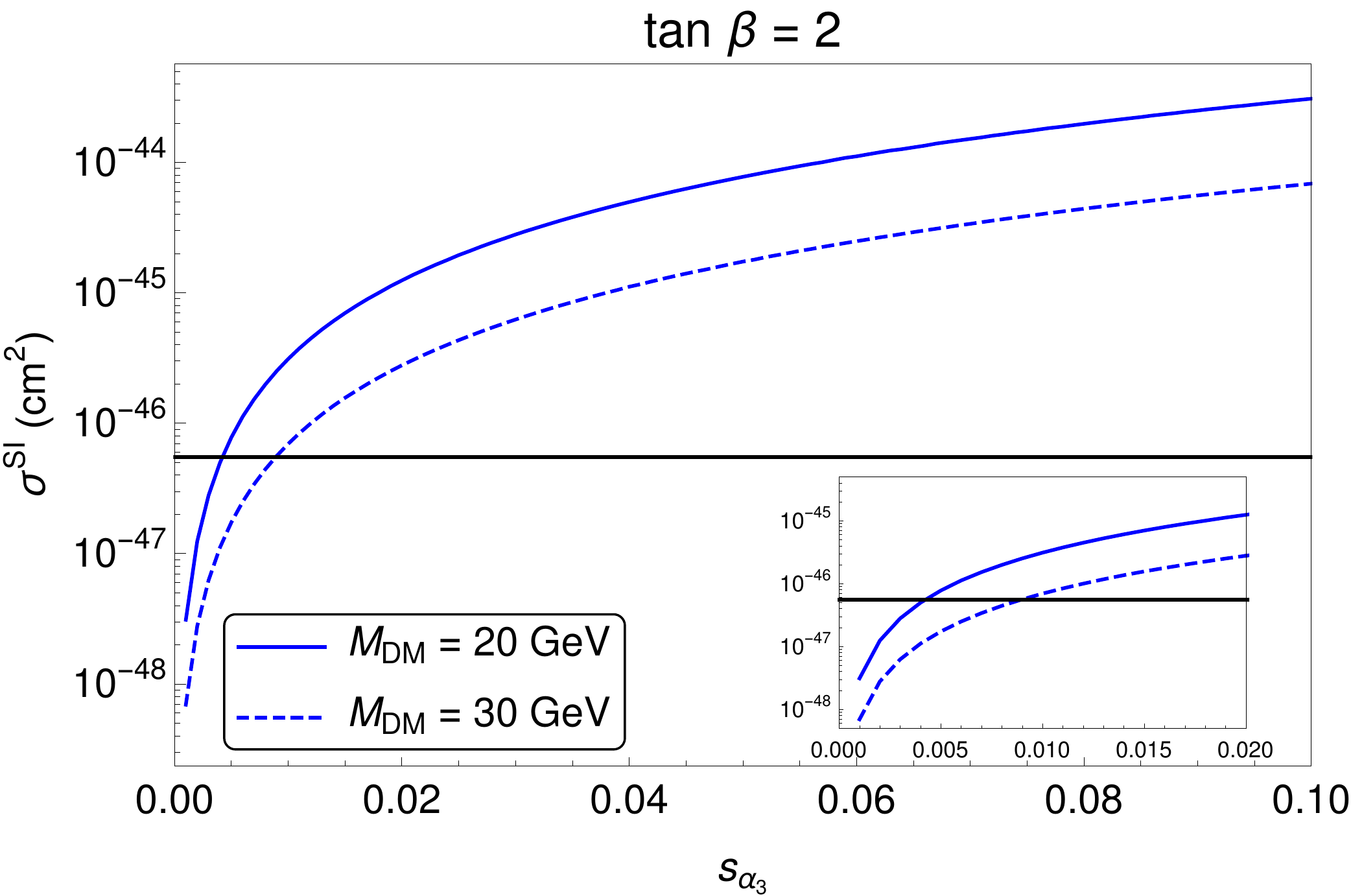}
		\label{fig:ddvsangle}}
	\caption{(a) The bounds on $M_{DM}$ from relic and the allowed limit on DM direct detection cross section ($\sigma^{SI}$). (b) The variation of $\sigma^{SI}$ with $s_{\alpha_3}$ for different allowed values of $M_{DM}$.}
	\label{fig:DDvsDM}
\end{figure}

In this case, at the leading order, the contributions to the relic abundance and the direct detection cross-section will come from similar type of annihilation diagrams, as shown in Fig.~\ref{fig:dmann}. Hence the true relic abundance is expected to be satisfied only around the resonances of the different scalar and vector mediators. There will be no coannihilations in this case. Apart from $M_{Z'}$ and $g_{Z'}$, the other model parameters which will have a dominant role in DM searches are given by $M_{s_2}$, $s_{\alpha_3}$ and $Y_{22}^{\varphi} \approx \frac{\sqrt{2} M_{N_2}}{v_2}$. The other parameters which will have a subdominant role are given by $M_{s_1}$, $M_{A_2}$, $\tan\beta$ and $s_{\alpha_2}$. Therefore, we have fixed their values at $M_{s_1} = 400$ GeV, $M_{A_2} = 200$ GeV and $s_{\alpha_2}=0.1$, respectively. In Fig.~\ref{fig:relic1}, we have shown the variation of the dark matter relic abundance with DM mass for two different values of $\text{tan }\beta$. The nature of the curve is similar to the one observed in our toy model 1 (see Fig.~\ref{fig:relicY22}). When the DM mass is in the sub-GeV range, the $Z^{'}$ mediated annihilation will be dominant similar to the previous case. As expected, the current bound on relic density will be satisfied at the DM masses close to the value $M_{Z'}/2$, and at a value $M_{DM} < M_{s_2}/2$. There are different peaks for $M_{DM} > M_{s_2}/2$ which correspond to the different resonance annihilation of the DM through the Higgs portal. In all the resonances for $M_{DM} > M_{s_2}/2$, the relic is much below the present observed abundance. The allowed values of DM mass are mostly limited in the sub-GeV to less than $50$ GeV mass. Note that the relic is almost insensitive to the value of $\tan\beta$. Also, as shown in Fig.~\ref{fig:relic2} the sine of mixing angle $\alpha_3$ does not have an impact on the allowed regions of $M_{DM}$. Although, we have chosen very small values of $s_{\alpha_3}$, the situation will not change even for larger values of $s_{\alpha_3}$. 

In Fig.~\ref{fig:ddvsmass}, we have shown the regions of $M_{DM}$ allowed by relic density bound and the current experimental limit on the DM direct detection cross section $\sigma^{SI}$ from XENON 1T. To generate this plot we consider $\tan\beta = 2$, and the values of the other relevant parameters are the following:  $0 < s_{\alpha_3} < 0.01$, and $ 10 \leq M_{s_2} \leq 100$ GeV. All the other relatively less relevant parameters are fixed at the values as mentioned above. The maximum value of $M_{DM}$ allowed by the data on the relic and $\sigma^{SI}$ is $\sim 40$ GeV. Note from Fig.~\ref{fig:ddvsangle} that the current limit on $\sigma^{SI}$ put stringent bound on $s_{\alpha_3}$. For example, for $M_{DM} \approx 30$ GeV the allowed value of $s_{\alpha_3}$ can not be larger than $0.01$. Here, we have shown the plot for $\tan\beta = 2$; however, as shown above, the results will be similar for other allowed values of $\tan\beta$. Like in the case of toy model-I, the contributions to spin-dependent direct detection cross sections are negligibly small in this model as well.

\begin{figure}[t]
	\centering
	\includegraphics[scale=0.48]{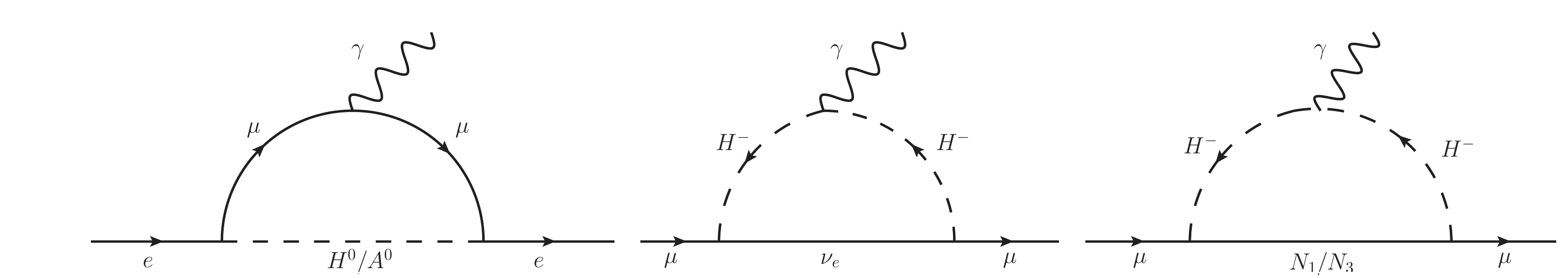}
	\caption{The diagrams which will contribute to muon and electron magnetic moments.}
	\label{fig:MagMomDiag}
\end{figure}

\subsubsection{Electron Anomalous Magnetic Moment and LFV}
	
\paragraph{Magnetic Moments :-}
In this model, apart from the contribution from a $U(1)_X$ gauge particle as has been discussed in sub-section \ref{sec:muong2x}, the contributions to the muon and electron magnetic moment will come from the respective diagrams shown in Fig.~\ref{fig:MagMomDiag}. The contributions from these diagrams from left to the right, respectively, are summarised in the following equations:
    
\begin{equation}
\Delta a_e^{(H)} = \frac{m_e^2}{8\pi^2 M_H^2}\frac{|Y^\ell_{12}|^2}{12},~~~~  \text{with $H\equiv (H^0,A^0)$},  
\label{eq:delaetoy2}
\end{equation}
\begin{equation}
\Delta a_\mu^{\nu} = -\frac{m_\mu^2}{8\pi^2 M_{H^\pm}^2}\frac{|Y^\ell_{32}|^2}{12},
\label{eq:delamutoy2nu}
\end{equation}
\begin{equation}
\Delta a_\mu^{N} = -\frac{m_\mu^2 |R_2|^2}{8\pi^2 M_{H^\pm}^2} \int_0^1 dx \frac{x^2(1-x)}{x + (1-x)\frac{M_N^2}{M_{H^\pm}^2}}.
\label{eq:delamutoy2N}
\end{equation}
Here, we have defined $|R_2|^2 \equiv \left((Y^\nu_{21})^2 + (Y^\nu_{23})^2 \right)$ in the same way as we defined $R_1, R_3$ in the previous toy model. To do so, we have assumed the same masses for $N_1$ and $N_3$. Note that the contributions in $\Delta a_e^{(H)}$ is sensitive to the Yukawa coupling $|Y^\ell_{12}|$, and the contributions in $\Delta a_{\mu}$ are coming from the diagrams with $\nu_{\tau}$ and $N_1/N_3$ in the loop, respectively.  

\begin{figure}[t]
	\centering
	\subfigure[]
	{\includegraphics[scale=0.45]{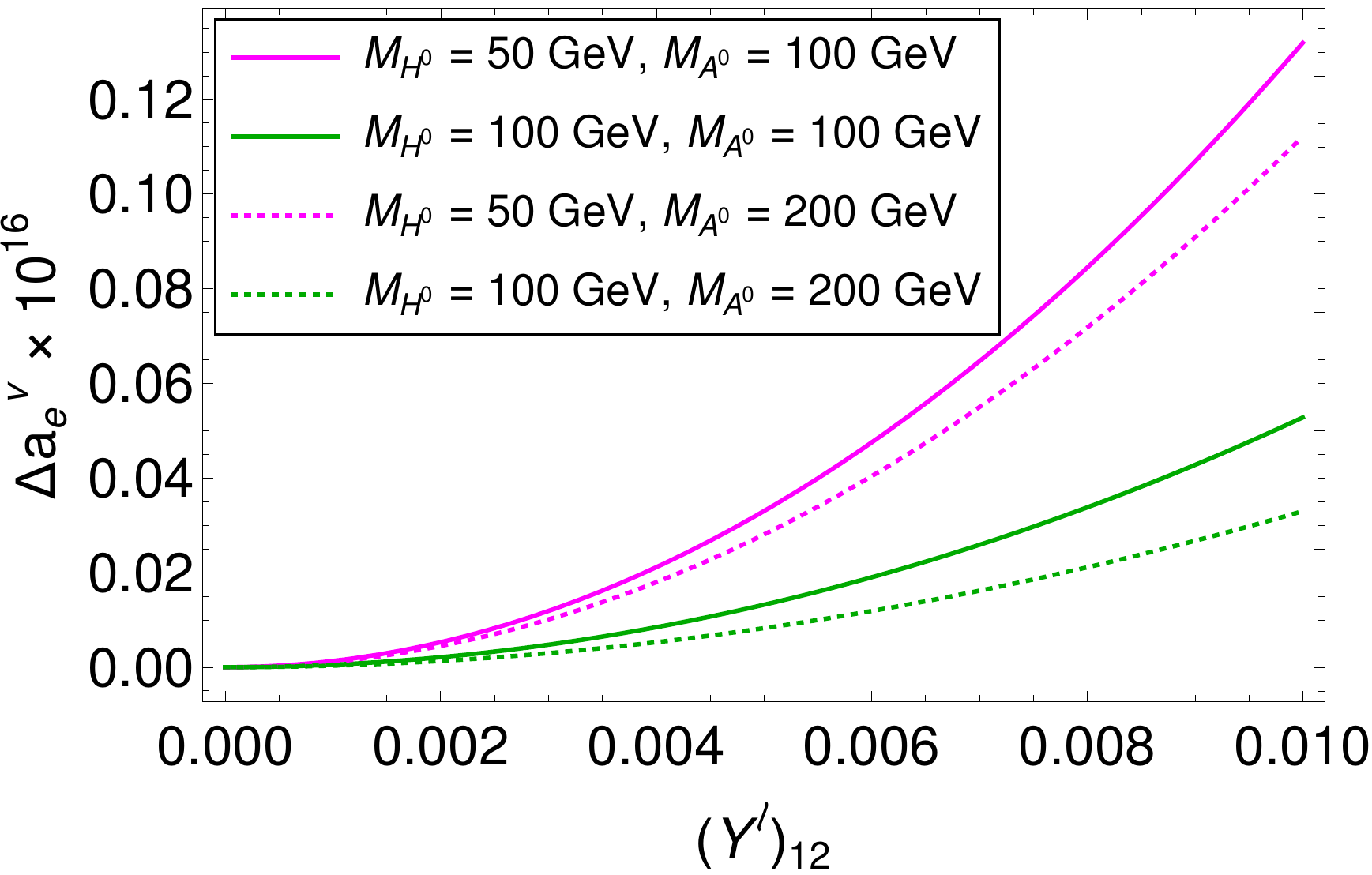}\label{fig:delae_toy2}}~
	\subfigure[]
		{\includegraphics[scale=0.45]{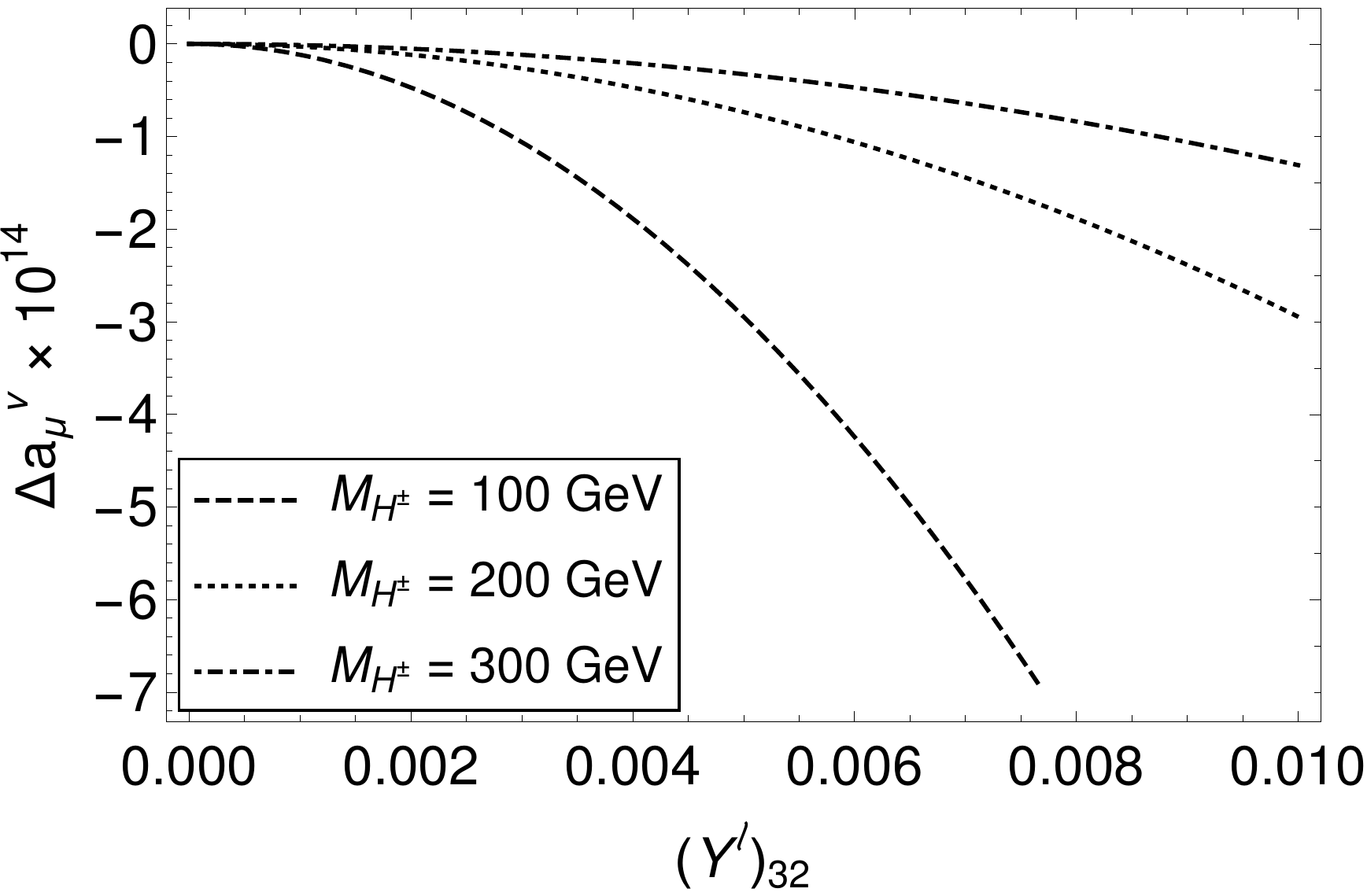}\label{fig:delamu_nu_toy2}}
	\caption{(a) Variation of $\Delta a_{e}^\nu$ with the relevant Yukawa coupling $Y^\ell_{12}$ for different values of $M_{H^0}$ and $M_{A^0}$. (b) Dependencies of $\Delta a_{\mu}^\nu$ with $Y^{\ell}_{32}$. }
	\label{fig:magnetic_toy2}
\end{figure}

\begin{figure}[t]
	\centering	
	\subfigure[]
	{\includegraphics[scale=0.36]{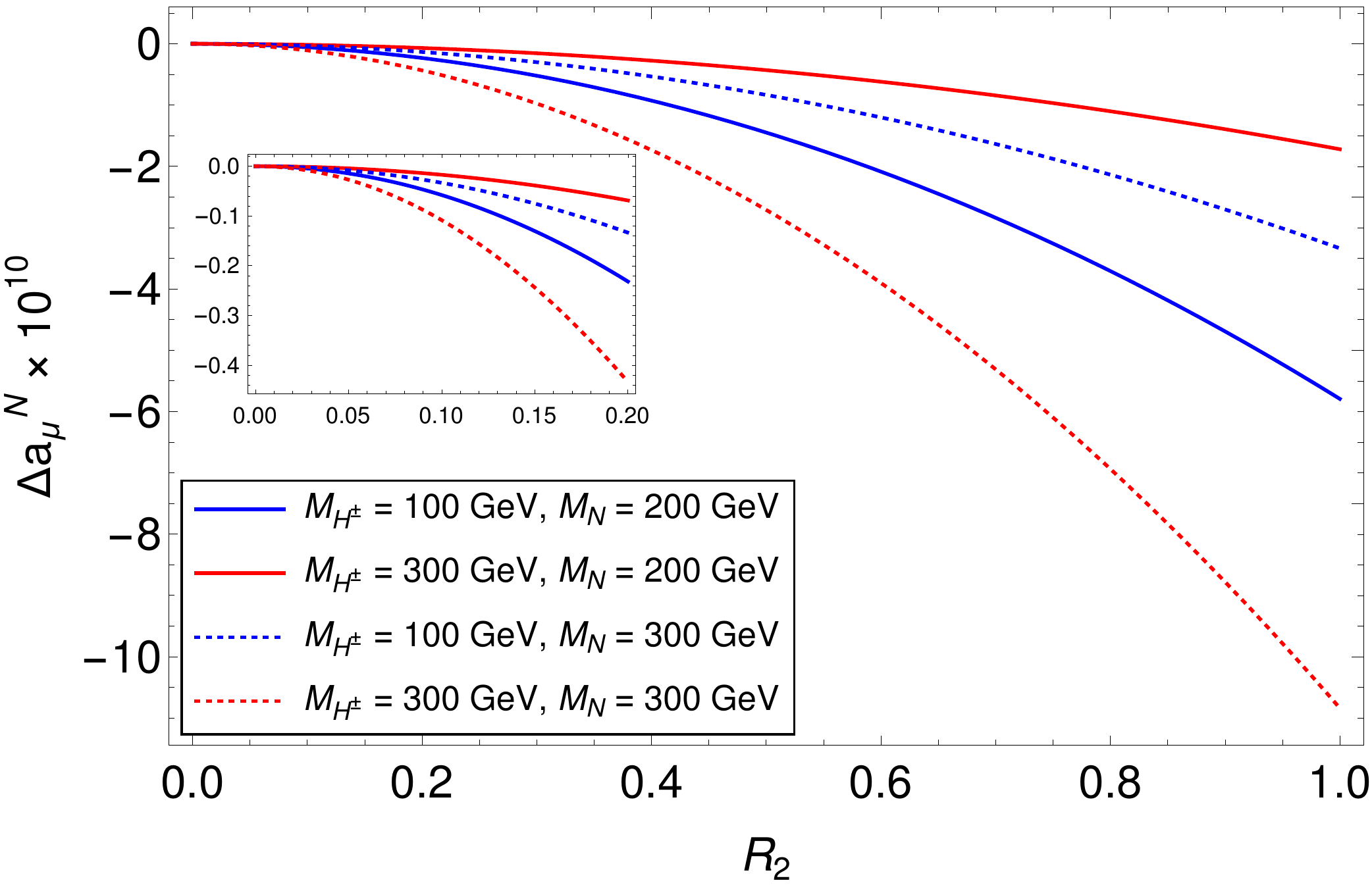}\label{fig:delamu_N_toy2}}~
	\subfigure[]
	{\includegraphics[scale=0.43]{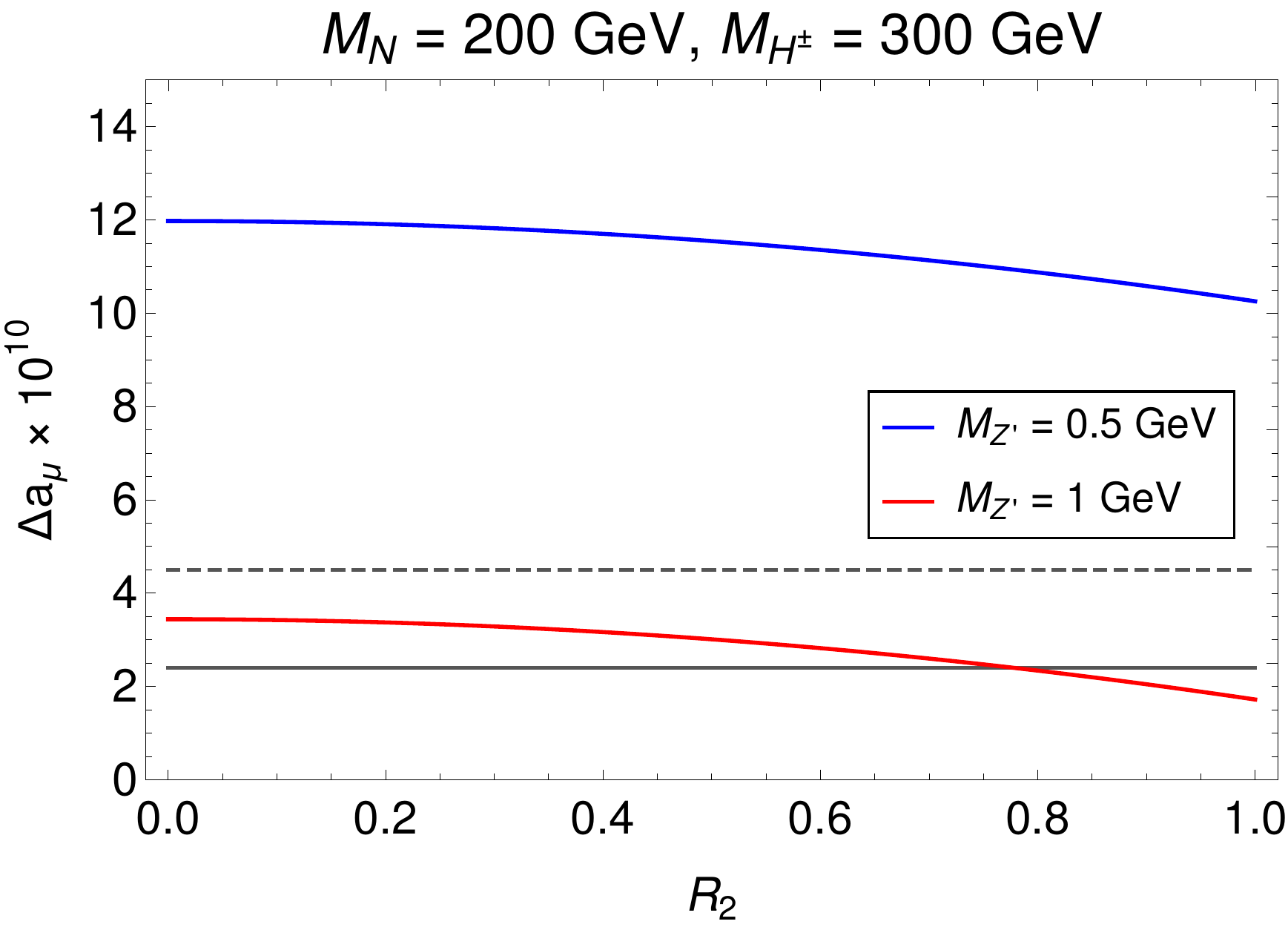}\label{fig:delamu_Tot_toy2}}
	\caption{(a) The variation of $\Delta a_{\mu}$ with $R_2 \equiv \sqrt{\left({Y^\nu_{21}}^2 + {Y^\nu_{23}}^2 \right)}$ for different values of $M_{N_1}=M_{N_3}=M_{N}$ and $M_{H^+}$. (b) The total contributions in $\Delta a_{\mu}$ from the diagrams in Fig.~\ref{fig:MagMomDiag} and \ref{fig:g2}. The grey horizontal line is the allowed 3-$\sigma$ lower limit \cite{pdg2018}, while the dashed grey line represents the 3-$\sigma$ lower limit of a very recent estimate \cite{Aoyama:2020ynm}. }
	\label{fig:muong2_toy2}
\end{figure}
The variations of $\Delta a_e$ with $|Y^\ell_{12}|$ for different values of $M_{H^0}$ and $M_{A^0}$ are shown in Fig.~\ref{fig:delae_toy2}. Note that the contribution to $\Delta a_{e}$ is highly suppressed and the values like $Y^{\ell}_{12} \gsim 0.01$ are allowed. As can be seen from Fig.~\ref{fig:delamu_nu_toy2}, the contribution to $\Delta a_{\mu}$ from the diagram with $\nu_{\tau}$ in the loop is highly suppressed. In Fig.~\ref{fig:muong2_toy2}, we have shown the variation of $\Delta a^{\mu}$ with $|R_2|$ for different values of $M_{H^0}$ and $M_N$. Note that the contributions in $\Delta a_{\mu}$ from the diagram with right-handed neutrinos are significant and have negative values. We have already shown earlier that the contribution from the diagram with $X$ with a mass $M_{X} \approx 0.5$ GeV can accommodate the current discrepancy in $\Delta a_{\mu}$. The large negative contribution from diagrams with $N_1$ or $N_3$ (in the loop) will reduce the value of $\Delta a_{\mu}$ obtained from a diagram with $X$. However, note that, for lower values of $R_2 $ i.e. $R_2 \lsim 0.5$, the effects are not that significant. Therefore, to explain $\Delta a_{\mu}$ we can restrict $R_2$ to a low value. In Fig.~\ref{fig:delamu_Tot_toy2}, we have shown the variation of the total contribution to $\Delta a_{\mu}$ with $R_2$ for the ultimate choices of the other relevant parameters.         
	
\begin{figure}[t]
	\centering
	\subfigure[]
	{\includegraphics[scale=0.8]{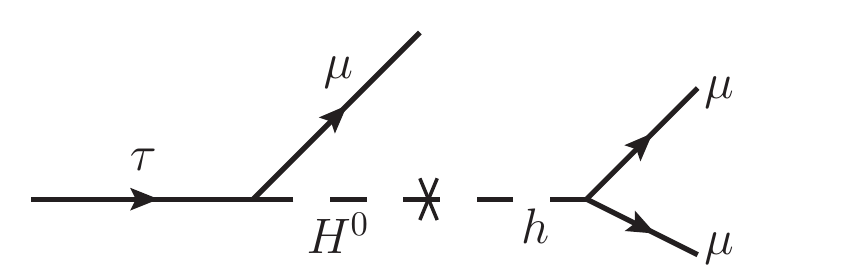}\label{fig:diagtau3mu}}~
	\subfigure[]
	{\includegraphics[scale=0.35]{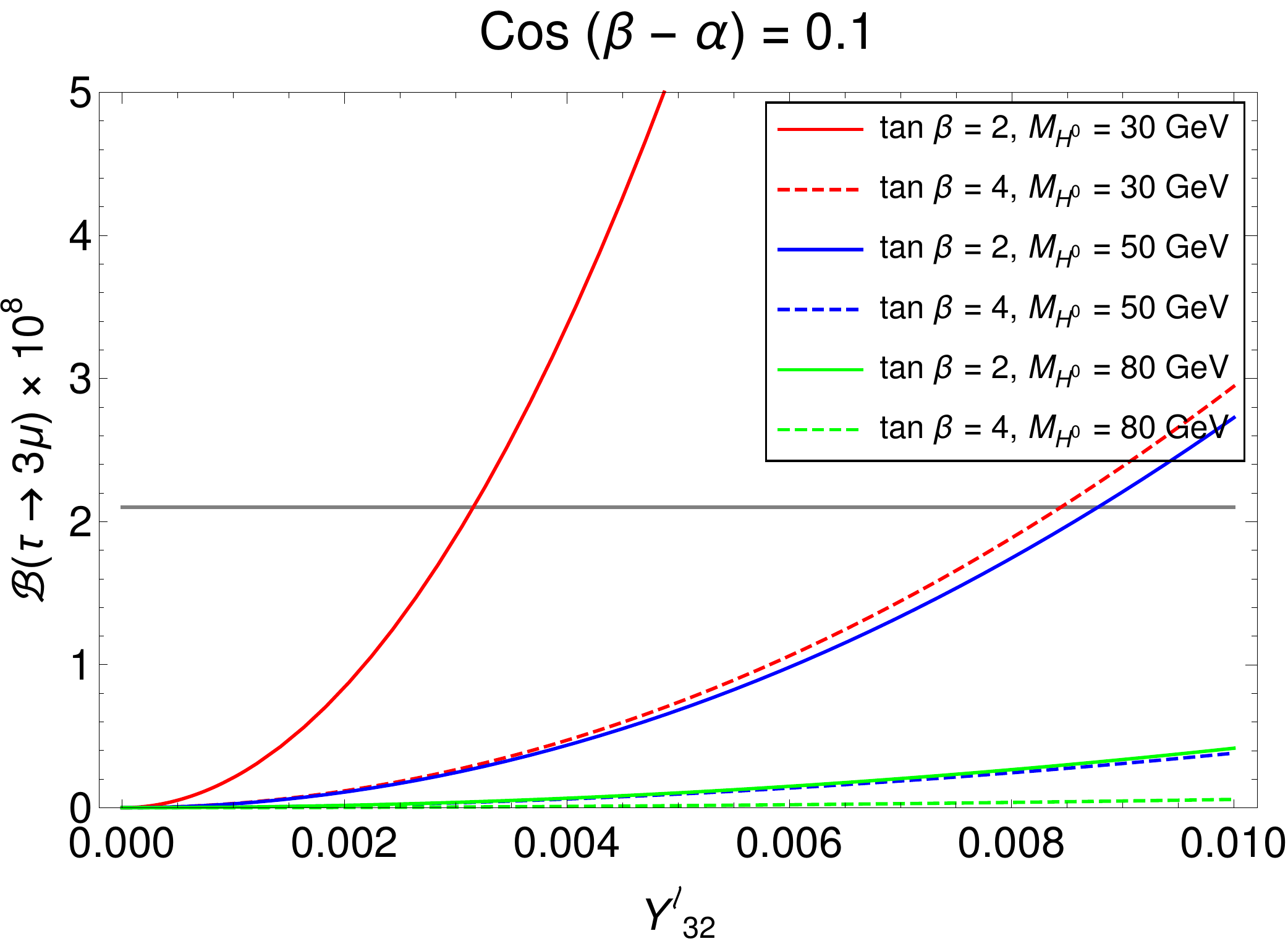}\label{fig:tau3muplot}}
	\caption{(a) Higgs mediated diagram contributing to $\tau \to 3 \mu$ process. (b) Variation of $\mathcal{B}_\tau$ with the coupling $Y^\ell_{32}$ for different values of $\text{tan}\beta$ and $M_{H^0}$. The constraint $\text{cos}(\beta-\alpha) = 0.1$ has been used while calculating the branching fraction. The gray line denotes the experimental upper bound on the branching fraction.}
	\label{fig:Tau3Mu}
\end{figure}

\paragraph{Lepton Flavour Violation :-} In our second model, there won't be any contribution to the processes like $\tau\to \mu\gamma$, $\mu\to e \gamma$ or $\tau\to e \gamma$. However, from eq.~\eqref{LYuk2}, one can see that the charge lepton mass matrix is not diagonal and is given by
\begin{equation}
	M_\ell = \begin{pmatrix}
	\frac{v}{\sqrt{2}}Y^\ell_{11} & \frac{u}{\sqrt{2}}Y^\ell_{12} & \frac{v}{\sqrt{2}}Y^\ell_{13} \\
	0 & \frac{v}{\sqrt{2}}Y^\ell_{22} & 0 \\
	\frac{v}{\sqrt{2}}Y^\ell_{31} & \frac{u}{\sqrt{2}}Y^\ell_{32} & \frac{v}{\sqrt{2}}Y^\ell_{33}
	\end{pmatrix}.
	\label{MLepton}
	\end{equation} 
	Due to the presence of off-diagonal terms in the charged lepton mass matrix, we have a contribution to the lepton flavour violating decay $\tau \to 3\mu$ as shown in Fig.~\ref{fig:diagtau3mu}. Although the contribution is mixing suppressed, the stringent limit on the branching fraction will put a direct constraint on the Yukawa coupling $Y^\ell_{32}$ since the process occurs at tree level. The upper bound on the branching fraction from the Belle Collaboration \cite{Hayasaka:2010np} is 
	\begin{equation}
	\mathcal{B}_\tau < 2.1 \times 10^{-8}
	\end{equation}
	at 90\% CL. The amplitude for the process can be written in the form 
	\begin{equation}
	\mathcal{M}_{\tau} = g_{LL}^s (\bar{\mu}_L \mu_R)(\bar{\mu}_R \tau_L)
	\end{equation}
	where
	\begin{equation}
	g_{LL}^s = \bigg(\frac{{Y^\ell_{32}}^{*}m_\mu s_\alpha}{v M_{H^0}^2}\bigg)
	\end{equation}
	and the branching fraction is given by\cite{Calcuttawala_2018}
	\begin{equation}
	\mathcal{B}_\tau= \left(\frac{T_\tau m_\tau^5|g_{LL}^s|^2}{128\times 48\pi^3}\right) \int_0^1 dx \int_0^1 d(\text{cos}\theta) \left[3x^2-2x^3+x^2\text{cos}\theta -2x^3 \text{cos}\theta \right]
	\end{equation}
	where $T_\tau$ is the lifetime of the $\tau$ lepton, $x = 2E_{\bar{\mu}}/m_\tau$ is the reduced energy of the antimuon, and $\theta$ is angle between the polarization of the $\tau$ and the momentum of the antimuon.
	 
In Fig.~\ref{fig:tau3muplot}, we show the variation of the branching fraction of $\tau \to 3 \mu$ with $Y^\ell_{32}$ for three different values of $M_{H^0}$, and in each of these cases, we have chosen two different values of $\tan\beta$. It is evident from the plot that experimental upper limit on $\mathcal{B}(\tau \to 3 \mu)$ restricts the allowed regions of $Y^\ell_{32}$ and $M_{H^0}$, and the preferable choice is $Y^\ell_{32} \lsim 0.005$ for $M_{H^0} \gsim 50$ GeV. The decay width is also sensitive to the value of $\tan\beta$. Therefore, for all practical purposes it is convenient to set $Y^\ell_{32}$ to a very small value, say $0.001$, in order to evade this strong bound.

\begin{figure}[t]
		\centering
		\subfigure[]
		{\includegraphics[scale=0.45]{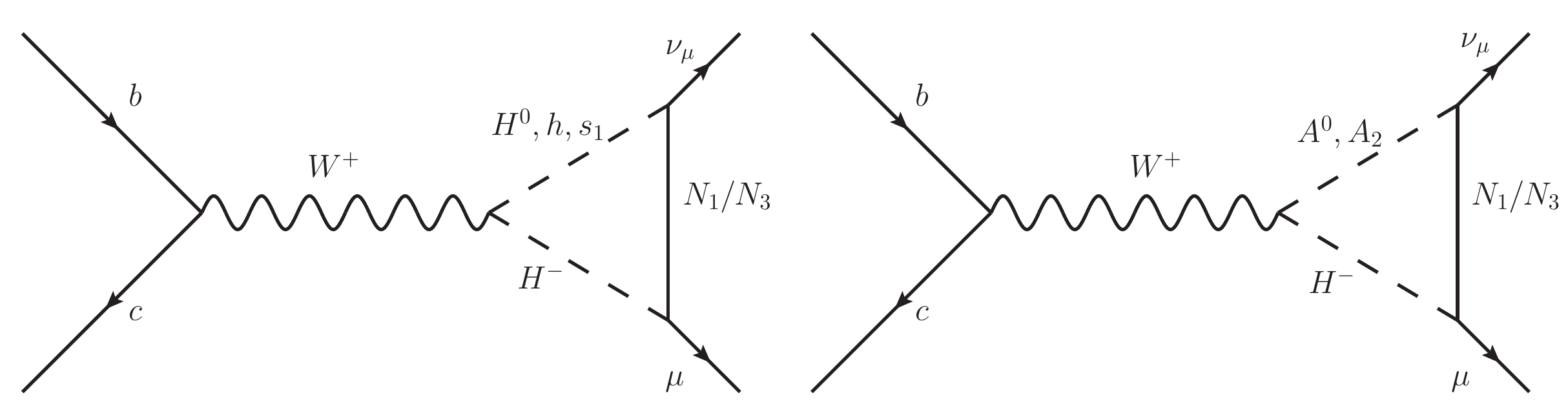}
			\label{fig:btocmu}}\\
		\centering
		\subfigure[]
		{\includegraphics[scale=0.5]{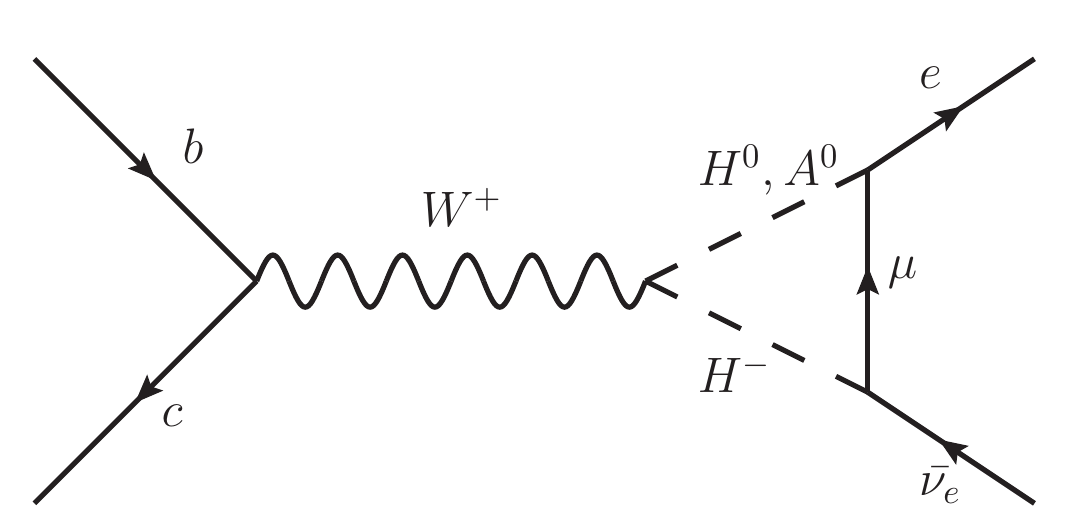}\label{fig:btoce}}~
		\subfigure[]
		{\includegraphics[scale=0.5]{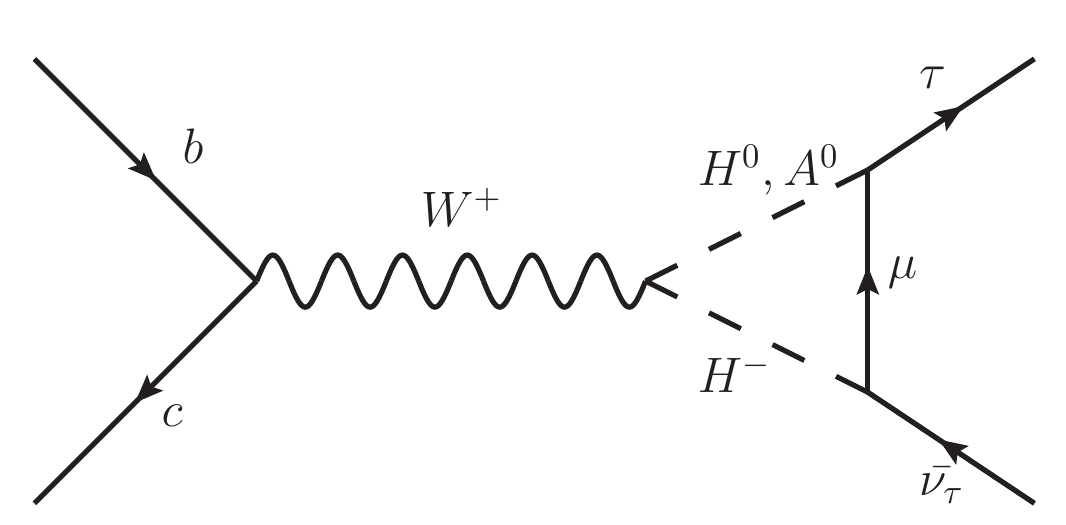}
			\label{fig:btoctau}}
		\caption{Diagrams contributing to $b \to c \ell \bar\nu_\ell$ decays ($\ell = e,\mu,\tau$).}
		\label{fig:btoc}
	\end{figure}
	
	\subsubsection{Additional contribution to $b \to c \ell \bar{\nu_\ell}$}
		
\begin{figure}[t]
\centering
			\subfigure[]
			{\includegraphics[scale=0.4]{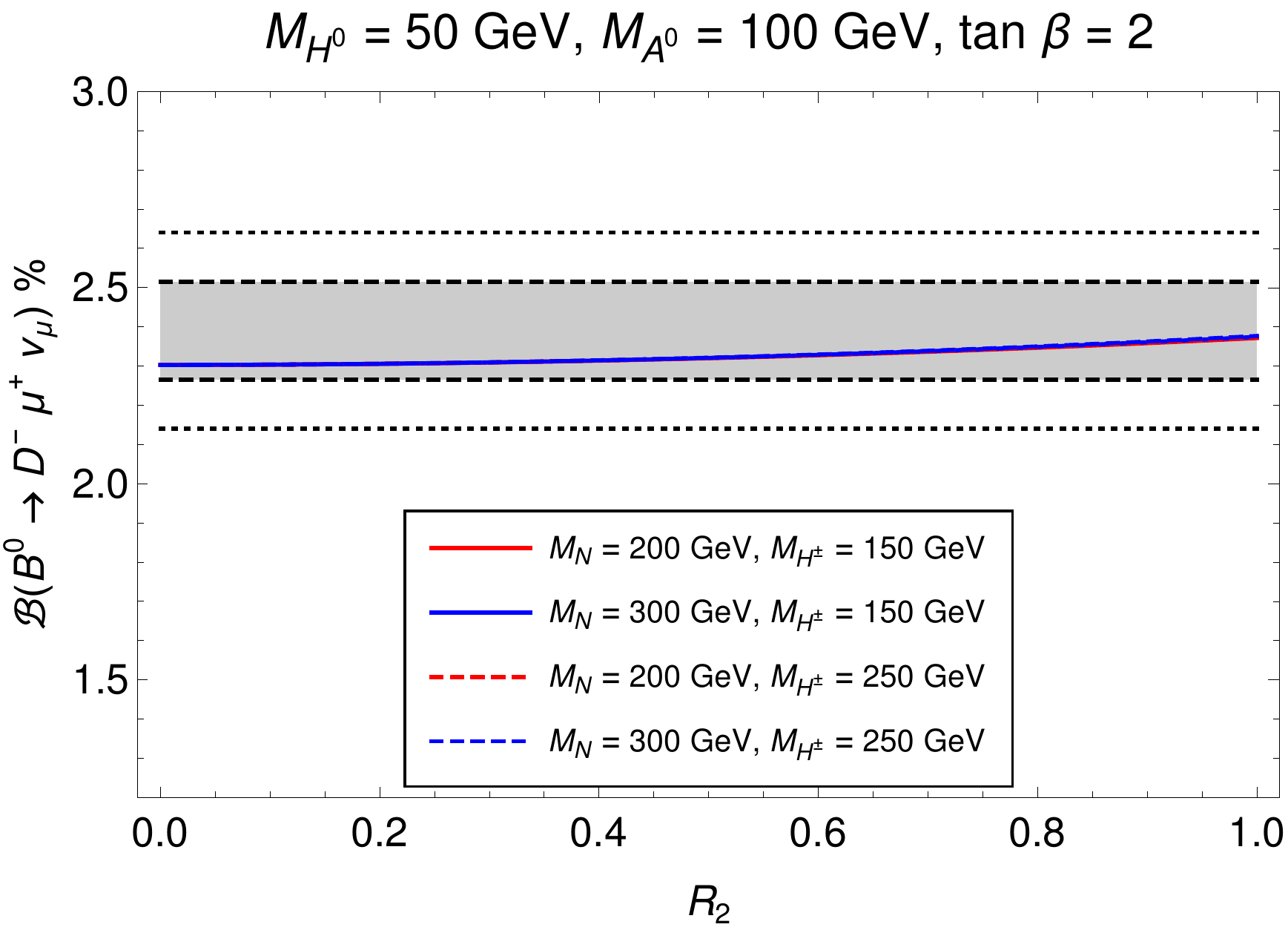}
				\label{BtoD_toy2}}
			\subfigure[]
			{\includegraphics[scale=0.4]{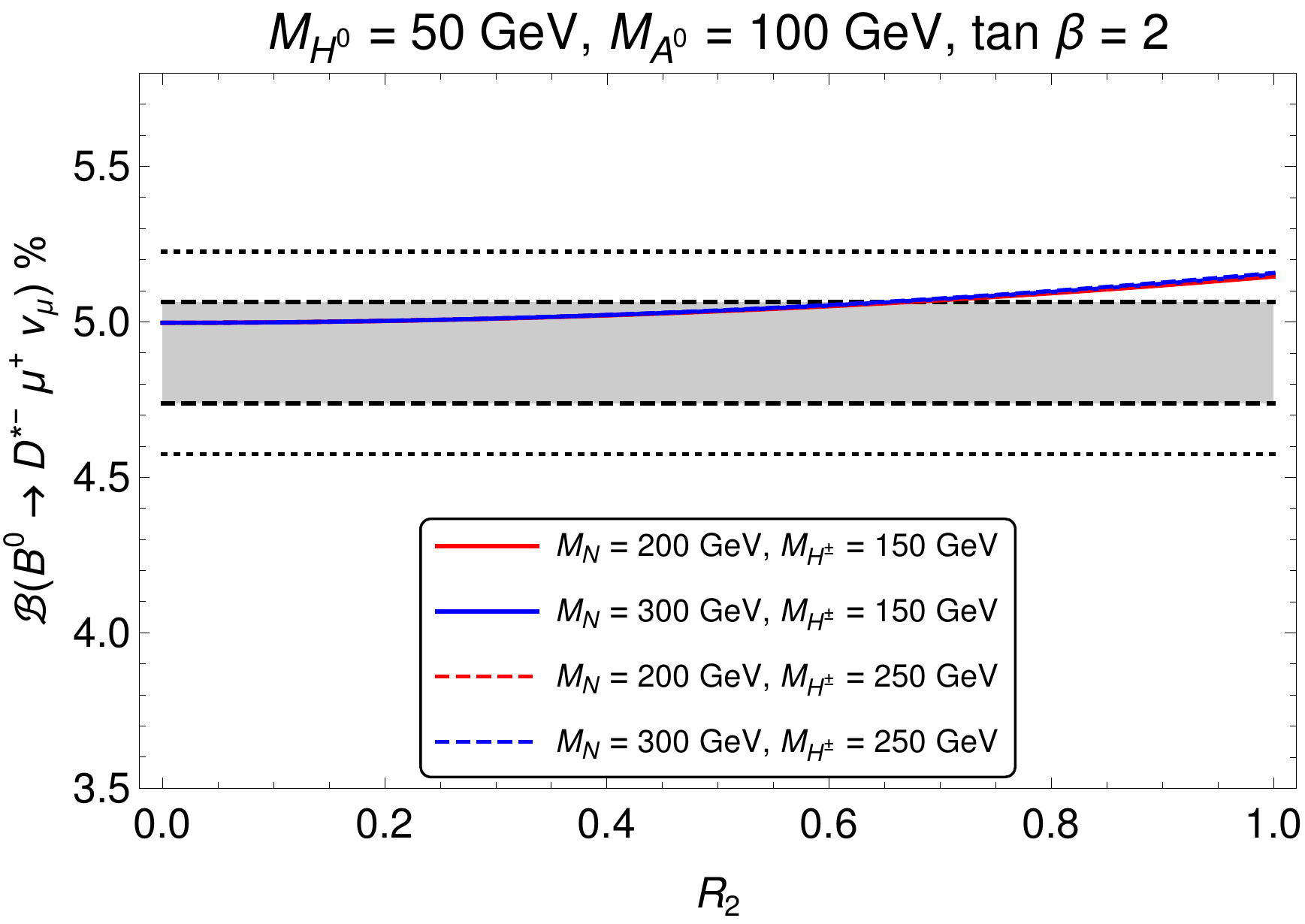}
				\label{BtoDst_toy2}}
			\caption{Variation of the branching fraction of ${\bar B}\to D \mu \bar\nu_\mu$ (Fig.~\ref{BtoD_toy2}) and ${\bar B}\to D^* \mu \bar\nu_\mu$ (Fig.~\ref{BtoDst_toy2}) with $R_2$ for two different values of $M_N$ and $M_{H^\pm}$ each. The other relevant parameters are kept fixed as shown in the plot labels. In both the plots, the gray shaded band is the measured branching fraction of the $B$-decays in their $1\sigma$ CL respectively while the dotted lines denote the $2\sigma$ allowed band.}
			\label{R2_BDBDst}
		\end{figure}
		\begin{figure}[t]
			\subfigure[]
			{\includegraphics[scale=0.37]{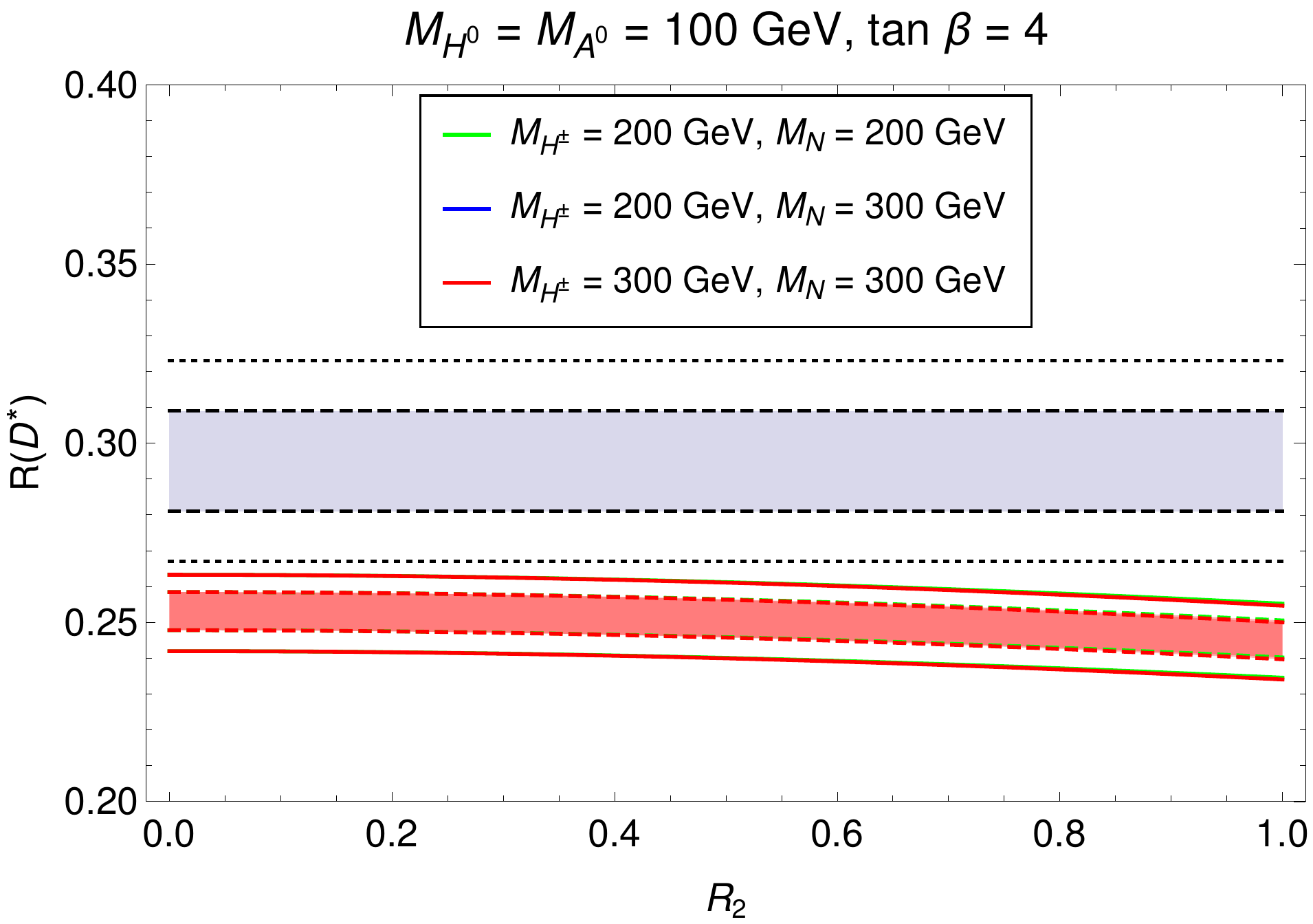}
				\label{fig:rdst}}
			\subfigure[]
			{\includegraphics[scale=0.37]{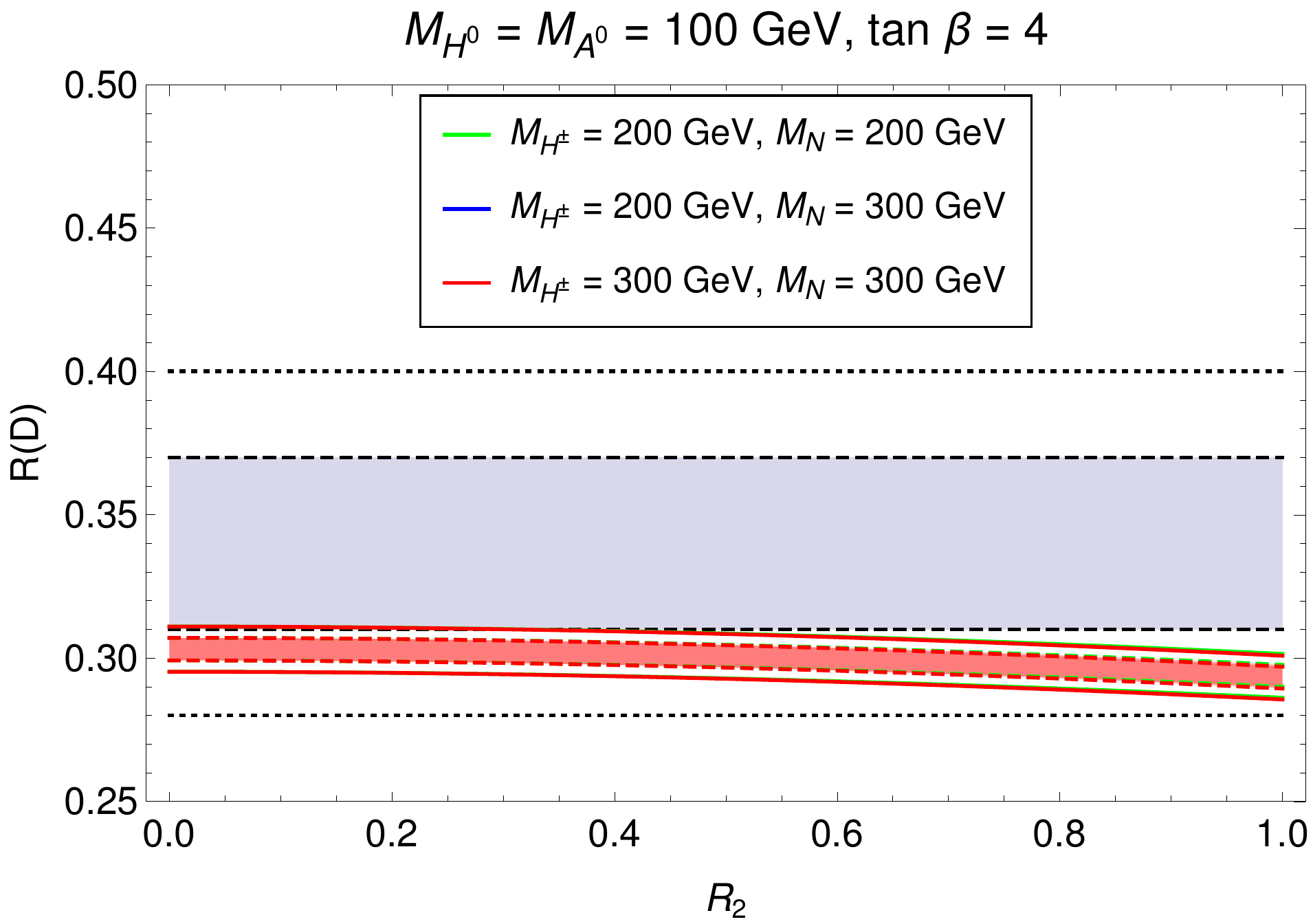}
				\label{fig:rd}}
			\caption{Variation of $R(D^*)$ (left) and $R(D)$ (right) with $R_2$ for different values of $M_N$ and $M_{H^\pm}$. The shaded gray region is the experimental $1\sigma$ allowed region while the dotted lines denote the same in the $2\sigma$ CI. The shaded and unshaded colored bands (red,blue,green) signify the theoretical uncertainty in the $1\sigma$ and $2\sigma$ CI respectively.}
			\label{fig:R2_rd}
		\end{figure}
	
	In this case, the diagrams that will contribute to $b \to c \ell\bar{\nu_{\ell}} $ (with $\ell = e, \mu$ and $\tau$) decays are given in Figs.~\ref{fig:btoc}. The diagrams in Fig.~\ref{fig:btocmu} will contribute to $b\to c \mu\bar\nu_{\mu}$ decays, whereas those in Figs.~\ref{fig:btoce} and \ref{fig:btoctau} will contribute to $b\to c e\bar\nu_{e}$ and $b\to c \tau\bar\nu_{\tau}$ decays respectively. The resulting Wilson coefficient contributing to $b\to c \mu \bar\nu_\mu$ can be written as:

\begin{equation}
\mathcal{C}_{V_1}^{(H)}= \frac{{Y^{\nu}_{2j}}^2}{32 \pi^2} g_{HH^-W^+} \int_0^1 dx \int_0^{1-x} \text{ln } \Delta_{WHH}^j,~~~~~ \text{with $H \equiv (H^0, h, s_1, A^0, A_2)$},
\label{eq:CV1toy2}
\end{equation}
	where the coupling $g_{HH^-W^+}$ between the charged Higgs, W-boson and the different neutral scalars are given by :
	\begin{equation}
	\begin{aligned}
	g_{H^0H^-W^+} &= \frac{g}{2}(c_\gamma \mathcal{R}_{\alpha_{44}} + s_\gamma \mathcal{R}_{\alpha_{14}}), ~~~~~~~
	g_{hH^-W^+} = \frac{g}{2}(c_\gamma \mathcal{R}_{\alpha_{41}} + s_\gamma \mathcal{R}_{\alpha_{11}}), \\
	g_{s_1H^-W^+} &= \frac{g}{2}(c_\gamma \mathcal{R}_{\alpha_{42}} + s_\gamma \mathcal{R}_{\alpha_{12}}), ~~~~~~~
	g_{A^0H^-W^+} = \frac{g}{2}(c_\gamma \mathcal{R}_{\theta_{44}} + s_\gamma \mathcal{R}_{\theta_{14}}), \\
	g_{A_2H^-W^+} &= \frac{g}{2}(c_\gamma \mathcal{R}_{\theta_{43}} + s_\gamma \mathcal{R}_{\theta_{13}}), 
	\end{aligned}
	\end{equation}
	where,  $\mathcal{R}_{\alpha_{ij}}$ is the $(ij)$ element of the rotation matrix given that diagonalises the CP even mass matrix given in Appendix \ref{RotMat4}, and similarly for $\mathcal{R}_{\theta_{ij}}$. The factor $\Delta_{WHH}^j$ is given in eq.~\eqref{WHH}. However, it is evident from the Lagrangian that the dominant contributions would come from the diagrams containing $H^0$ and $A^0$ in the loop since their coupling with the neutrinos are not mixing suppressed unlike the other Higgses.
	
The contributions to $b\to c e \bar\nu_e$ and $b \to c \tau\bar\nu_{\tau}$ from the diagrams as mentioned above are negligibly small. The contributions come from the second term of the Yukawa Lagrangian in eq.~\eqref{LYuk2}. The Wilson coefficients in these cases are sensitive to the off-diagonal Yukawas of the charged lepton mass mixing matrix. Also, we have already obtained a direct bound on the coupling $Y^\ell_{32}$ from the branching fraction of $\tau \to 3\mu$ in the previous section and fixed it to $0.001$.  For such a small Yukawa, the resulting WC will have a value of order $\mathcal{O}(10^{-9})$ which is negligibly small and hence it can safely be neglected. Note that even if we choose $Y^\ell_{32} \approx 0.01$, the contributions to the WC will be of order $\mathcal{O}(10^{-7})$. Similar conclusion holds for the electron final state. Therefore, we will only focus on the contribution to the muon channel.

If we assume, for simplicity, the masses of $N_1$ and $N_3$ are equal, then the Wilson coefficient will be proportional to the parameter $R_2$ which we have already defined in the previous sub-section. We can then constrain the parameter space of $R_2$ and $M_N \equiv M_{N_1} = M_{N_3}$ from the data on $R(D), R(D^*)$ and ${\bar B} \to D^{(*)} \mu \bar\nu_\mu$. In Fig.~\ref{R2_BDBDst}, we show the variation of the branching fractions of ${\bar B} \to D^{(*)} \mu \bar\nu_\mu$ with $R_2$ for two different values of $M_N$ and $M_{H^\pm}$ each, keeping the masses of the neutral scalars $H^0,A^0$ fixed. Also we keep $\text{tan }\beta$ fixed at 2 and choose $\alpha$ such that $\text{cos}(\beta - \alpha) = 0.1$. We see that even for a large ($\sim \mathcal{O}(1)$) $R_2$, the branching fraction of ${\bar B} \to D^{(*)} \mu \bar\nu_\mu$ remains within the $2\sigma$ experimental range. Also, we have noted that the change in the branching fraction with the mass of the charged scalar or the heavy neutrino is insignificant (the legends overlap in the figure). We have also studied the impact on $R(D^*)$ and $R(D)$, and the results are shown in Figs.~\ref{fig:rdst} and \ref{fig:rd}, respectively. Note that it is hard to explain $R(D^*)$ even if we take both the theory and the measured errors within their 2$\sigma$ confidence interval (CI)\footnote{It was discussed in refs. \cite{Jaiswal:2017rve,Jaiswal:2020wer} that at the moment, the predictions of $R(D^*)$ depend too much on the experimental results on $B \to D^* \mu\nu_{\mu}$ and $B \to D^* e \nu_{e}$ decay, and with the changes in the data the predictions are changing. For a prediction independent of any experimental inputs, we have to wait till the inputs from the lattice at non-zero recoil angle of the outgoing meson are available.}. However, we can conveniently explain the observation in $R(D)$ even if we consider the data and theory at their 1$\sigma$ CI. Also, even though large values of $R_2$ is allowed by the branching fractions to the muon mode, the data on $R(D^{(*)})$ restricts the value of $R_2$ to $\lsim 0.5$ for the entire range of RHN or charged scalar mass.

\subsubsection{Neutrino Mass Generation}
In this case the minimal seesaw mechanism will help us give rise to the neutrino mass. From the Yukawa interactions in eq.~\eqref{LYuk2}, one can obtain the Dirac neutrino mass matrix $M_D$ and the Majorana mass matrix $M_R$ to be :
\begin{equation}
M_D = \begin{pmatrix}
\frac{v}{\sqrt{2}}Y^\nu_{11} & 0& \frac{v}{\sqrt{2}}Y^\nu_{13} \\
\frac{u}{\sqrt{2}}Y^\nu_{21} & 0&  \frac{u}{\sqrt{2}} Y^\nu_{23}\\
\frac{v}{\sqrt{2}}Y^\nu_{31} & 0& \frac{v}{\sqrt{2}}Y^\nu_{33}
\end{pmatrix}, \hspace{0.2cm} M_R = \begin{pmatrix}
\frac{v_1}{\sqrt{2}}Y^\varphi_{11} & 0 &\frac{v_1}{\sqrt{2}}Y^\varphi_{13}\\
0 & \frac{v_2}{\sqrt{2}} Y^\varphi_{22} & 0 \\
\frac{v_1}{\sqrt{2}}Y^\varphi_{31} & 0 & \frac{v_1}{\sqrt{2}}Y^\varphi_{33}
\end{pmatrix}.
\label{MDMR}
\end{equation}
Thus the light active neutrino masses can be obtained from the seesaw formula given by:
\begin{equation}
m_\nu = -\bigg(M_D^T M_R^{-1} M_D\bigg).
\end{equation}
By using the structure of $M_D, M_R$ in the type I seesaw \cite{Minkowski:1977sc, GellMann:1980vs, Mohapatra:1979ia, Schechter:1980gr} formula for light neutrino masses mentioned above, we find a general structure of light neutrino mass matrix without any textures unlike that found in toy model I. However, the light neutrino mass matrix has rank 2 predicting the lightest neutrino mass to be vanishing. While neutrino oscillation experiments can not constrain such a scenario, other experiments like neutrinoless double beta decay which is sensitive to absolute neutrino mass scale can shed more light on such scenario in future.

\section{KOTO Anomaly}
\label{sec6}

Very recently, the KOTO experiment at J-PARC reported an excess of events over the SM expectation for the rare decay process $K_L \to \pi^0 \nu \bar{\nu}$ \cite{KOTOTalk}. They have reported four candidate events in the relevant signal region, whereas the SM expectation is only $0.1 \pm 0.02$ events. The corresponding measured value is given by 
\begin{equation}
\mathcal{B}(K_L \to \pi^0 \nu \bar{\nu})_{\text{KOTO}} = 2.1 ^{+2.0(+4.1)}_{-1.1(-1.7)} \times 10^{-9},
\end{equation}
where the quoted errors are given at $68\%$ and $95\%$ (within the parenthesis) CI, respectively. The SM prediction, on the other hand, is \cite{Cirigliano:2011ny, Buras:2015qea}
\begin{equation}
\mathcal{B}(K_L \to \pi^0 \nu \bar{\nu})_{\text{SM}} = (3.0 \pm 0.3) \times 10^{-11},
\end{equation}
which is about two orders of magnitude smaller than the one measured at KOTO. An upper bound on the branching fraction of $K^+ \to \pi^+ \nu \bar{\nu}$ decay is reported by the NA62 Collaboration \cite{CortinaGil:2020vlo, NA62Talk}:
\begin{equation}
\mathcal{B}(K^+ \to \pi^+ \nu \bar{\nu})_{\text{NA62}} < 1.85 (2.44) \times 10^{-10}.
\end{equation}
Here again, we have quoted the $90\% (95\%)$ confidence level (CL) limit and the measured value is consistent with the respective SM prediction of $(9.11 \pm 0.72) \times 10^{-11}$. The explanation of these excess events in $K_L \to \pi^0 \nu \bar{\nu}$ require new contribution in $d \to s + invisible$ decays beyond the SM. However, the same NP will contribute to $K^+ \to \pi^+ \nu \bar{\nu}$ decay as well. Also, the above mentioned branching fractions follow a model-independent bound \cite{Grossman:1997sk},
\begin{equation}
\mathcal{B}(K_L \to \pi^0 \nu \bar{\nu}) \lsim 4.3 \times \mathcal{B}(K^+ \to \pi^+ \nu \bar{\nu}),
\label{eq:GNbounds}
\end{equation}
which is totally based on isospin symmetry. From the measured values, it looks difficult to explain both the branching fractions simultaneously. However, it is important to note that the experimental measurement at NA62 \cite{CortinaGil:2020vlo} excludes the following kinematic regions of the missing mass : $0.01 < M_{\text{miss}}^2 < 0.026$ ${\text{GeV}}^2/c^2$ and $M_{\text{miss}}^2 > 0.068$ ${\text{GeV}}^2/c^2$, respectively. Hence, for the missing masses within these kinematic regions, one can avoid the bound given in eq.~\eqref{eq:GNbounds}. Following this, there are different NP explanations available in the literature, for example, see \cite{Fuyuto:2014cya,Kitahara:2019lws,Egana-Ugrinovic:2019wzj,Dev:2019hho,Jho:2020jsa,Gori:2020xvq,Liu:2020qgx,Datta:2020auq,Dutta:2020scq}. 

\begin{figure}[t]
	\centering
	\subfigure[]
	{\includegraphics[scale=0.4]{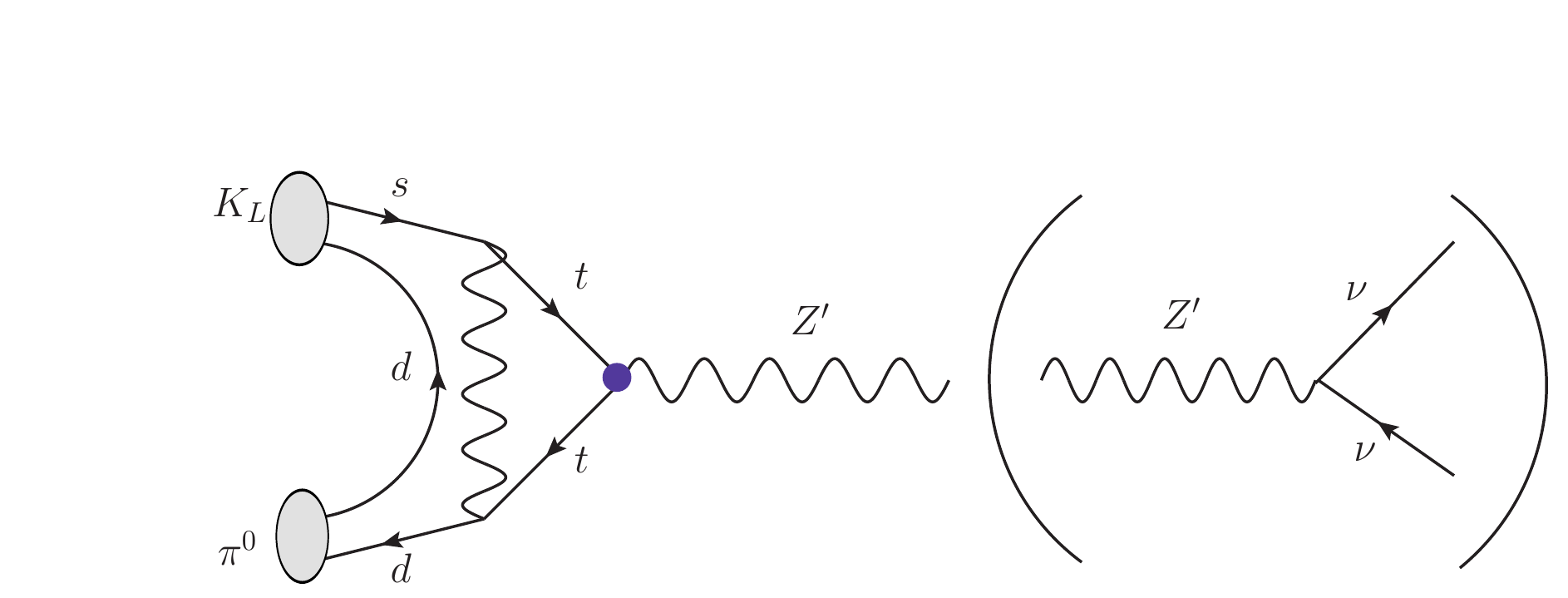}
		\label{fig:kotodiag}}~~~~~~~~~~~~
	\subfigure[]
	{\includegraphics[scale=0.52]{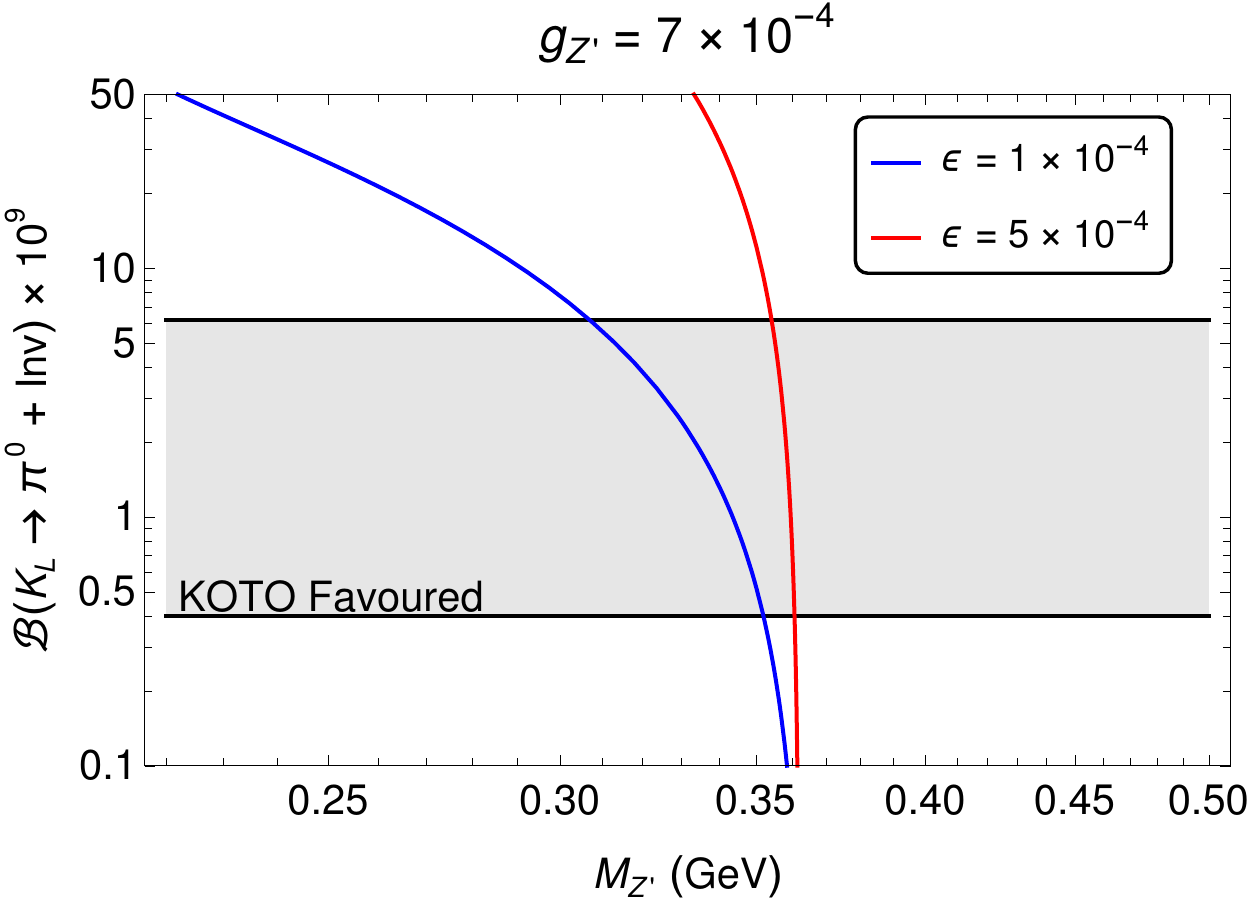}
		\label{fig:koto}}
	\caption{(a) The diagram that will contribute to $K_L \to \pi^0 \nu\nu$ decays. (b) Variation of the branching fraction of $K_L \to \pi^0 + \text{Inv}$ with the mass $M_{Z'}$ for different values of the kinetic mixing parameter $\epsilon$. The grey band represents the meaured value of $\mathcal{B}(K_L \to \pi^0 \nu \bar{\nu})$.  }
	\label{fig:KtoPi}
\end{figure}
The presence of a low scale $U(1)_X$ gauge boson in our model allows us to study the prospect of explaining the KOTO excess via $K_L \to \pi^0 Z' (Z' \to \nu \bar{\nu})$ decays which is possible in both of our toy models. The relevant Feynman diagram is given in Fig.~\ref{fig:kotodiag}. The corresponding branching fraction can be expressed as 
\begin{equation}
\mathcal{B}(K_L \to \pi^0 \nu \bar{\nu}) = \mathcal{B}(K_L \to \pi^0 Z') \times \mathcal{B} (Z' \to \nu\bar{\nu}),    
\end{equation}
i.e via the resonance production of $Z'$ and then by a subsequent $Z' \to \nu\bar{\nu}$ decay. The branching fraction of the rare $K_L \to \pi^0 Z'$ decay is given by \cite{Jho:2020jsa}
\begin{equation}
\begin{aligned}
& \mathcal{B}(K_L \to \pi^0 Z') = \frac{m_{K_L}^3}{\Gamma_{K_L}} \frac{(\text{Im } g_{dsZ'}^{\text{eff}})^2}{64\pi M_{Z'}^2} \left[\lambda\left(1, \frac{m_{\pi^0}^2}{m_{K_L}^2}, \frac{M_{X}^2}{m_{K_L}^2}\right)\right]^{3/2}\left[f_+^{K_L \pi^0}(M^2_{Z'})\right]^2. 
\end{aligned}
\end{equation}
The effective vertex for the interaction $[{\bar d_L}\gamma_{\mu}s_L] Z'$ is given by 
\begin{equation}
g_{dsZ'}^{\text{eff}} = \frac{V_{ts}V_{td}^* e^3 \epsilon}{16\pi^2 C_W S_W^3} \left(1-\frac{4}{3}S_W^2\right) C(x_t),
\end{equation}
which is obtained after calculating the loop diagram given in Fig.~\ref{fig:kotodiag}. Here, $f_+^{K_L\pi^0}$ is the meson decay form factor, $\lambda(x,y,z) = x^2 + y^2 + z^2 - 2 (xy + yz + zx)$, and $C(x_t)$ can be found in eq.~\eqref{Cxt}. Note that in our model $Z'$ plays the role of a mediator of DM ($N_2$) interactions, and the bound on the relic abundance is satisfied at the DM mass $M_{N_2} \approx M_{Z'}/2$. Since in our model $Z'$ primarily decays to $\mu^+\mu^-$, $e^+ e^-$ and $\nu\bar\nu$, therefore the resonance production of $Z'$ and its subsequent on-shell decay to $N_2 N_2$ will be kinematically forbidden. As we have seen earlier the other low energy data like $\Delta a_{\mu}$, data in $b\to s \mu\mu, b \to see$ etc. preferred a mass region $ 0.22 \lsim M_{Z'} \lsim$ 1 (in GeV) for a given coupling $ 0.0005 \lsim g_{Z'} \lsim 0.001$, for example see Fig.~\ref{fig:Mxgx}. In Fig.~\ref{fig:koto}, we have shown the variation of $\mathcal{B}(K_L \to \pi^0 \nu \bar{\nu})$ with $M_{Z'}$ for different values of the mixing parameter $\epsilon$. Note that the coupling $g_{Z'}$ cancels in the ratio $\frac{\Gamma(Z'\to \nu\bar{\nu})}{\Gamma_{Z'}}$, hence the branching fraction is insensitive to the variation of $g_{Z'}$. As can be seen from the plot, for mass range $0.30 < M_{Z'}  < 0.35$ GeV, we are able to explain the observed branching excess in $\mathcal{B}(K_L \to \pi^0 \nu \bar{\nu})$. Note that in this region we can avoid the model-independent bound given in eq.~\eqref{eq:GNbounds}. Though we have discussed the phenomenology of our toy models with $M_{Z'}= 1$ GeV, the observations made are equally valid for the mass window $ 0.3 \lsim M_{Z'} (\text{GeV}) < 1$.

\section{Possible Collider Signatures}
\label{sec7}
\subsection{Higgs Invisible decays}
\begin{figure}[htp!]
		\centering
		\includegraphics[scale=0.6]{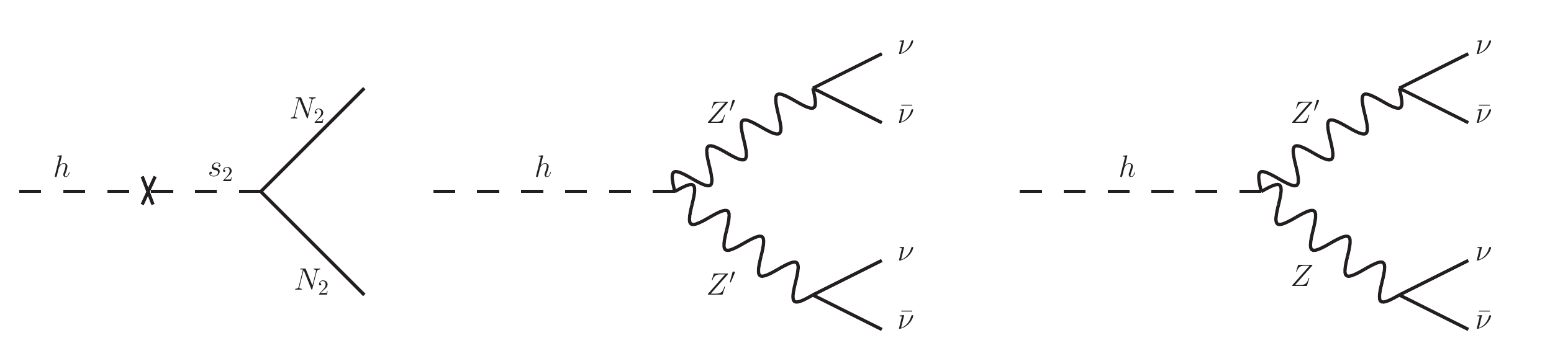}
		\caption{Invisible Higgs decay channels.}
		\label{fig:HiggsInvDiag}
	\end{figure}

In any NP model it is quite an exciting prospect to look into the non-standard or undetected  decays of the SM Higgs as a complementary search for BSM particles. The toy models that we discussed above constitute of a dark matter particle that couple to the SM Higgs boson through its mixing with singlet scalars. Also, there is a viable parameter space in both the models where the dark matter mass is lighter than the Higgs. Under such a scenario, it would be a useful exercise to study the model contribution to Higgs invisible decay and use the available data on it to further constrain the model parameters. All the relevant diagrams contributing to the Higgs invisible decay are shown in Fig.~\ref{fig:HiggsInvDiag}. From the figure it is almost clear that the dominating invisible decay of Higgs would be to the DM $N_2$ for $M_{N_2} < M_h/2$ and the decay to $4\nu$ via gauge bosons would in general be suppressed compared to this tree level decay. However, since the new gauge boson in our model, $Z^\prime$, is light (sub-GeV), the contribution mediated by $Z^\prime$ could also be quite significant. The third decay, involving one heavy boson $Z$ and a light $Z^\prime$, is in general suppressed by the very small value of the gauge mixing parameter $\epsilon$ ($\sim 10^{-4}$) that we considered. Hence, one can safely neglect the contribution from this decay mode. Therefore we will only consider the first two decay modes in our calculation.
	
Both ATLAS and CMS have looked into such invisibly decaying Higgs mainly through its inclusive production in the vector boson fusion mode, as well as in the associated production of a Higgs with a $Z$ boson. The constraint on the Higgs invisible decay branching fraction from the ATLAS experiment at LHC is \cite{Aaboud:2019rtt}
	
	\begin{equation}
	\mathcal{B}(h \to \text{Invisible}) = \frac{\Gamma(h \to \text{Invisible})}{\Gamma(h \to SM) + \Gamma(h \to \text{Invisible}) } \leq 26\%
	\end{equation}
	while the recent ATLAS announcement \cite{ATLAS:2020cjb} puts a more stringent constraint at $13\%$. The Higgs decay to SM particles is known to be around $4$ MeV. In the following two subsections, we will discuss the impact of this upper limit on the model parameters, in particular the mixing angles. 
	
	\subsubsection{Toy Model 1}
	
\begin{figure}[t]
		\centering
		\includegraphics[scale=0.5]{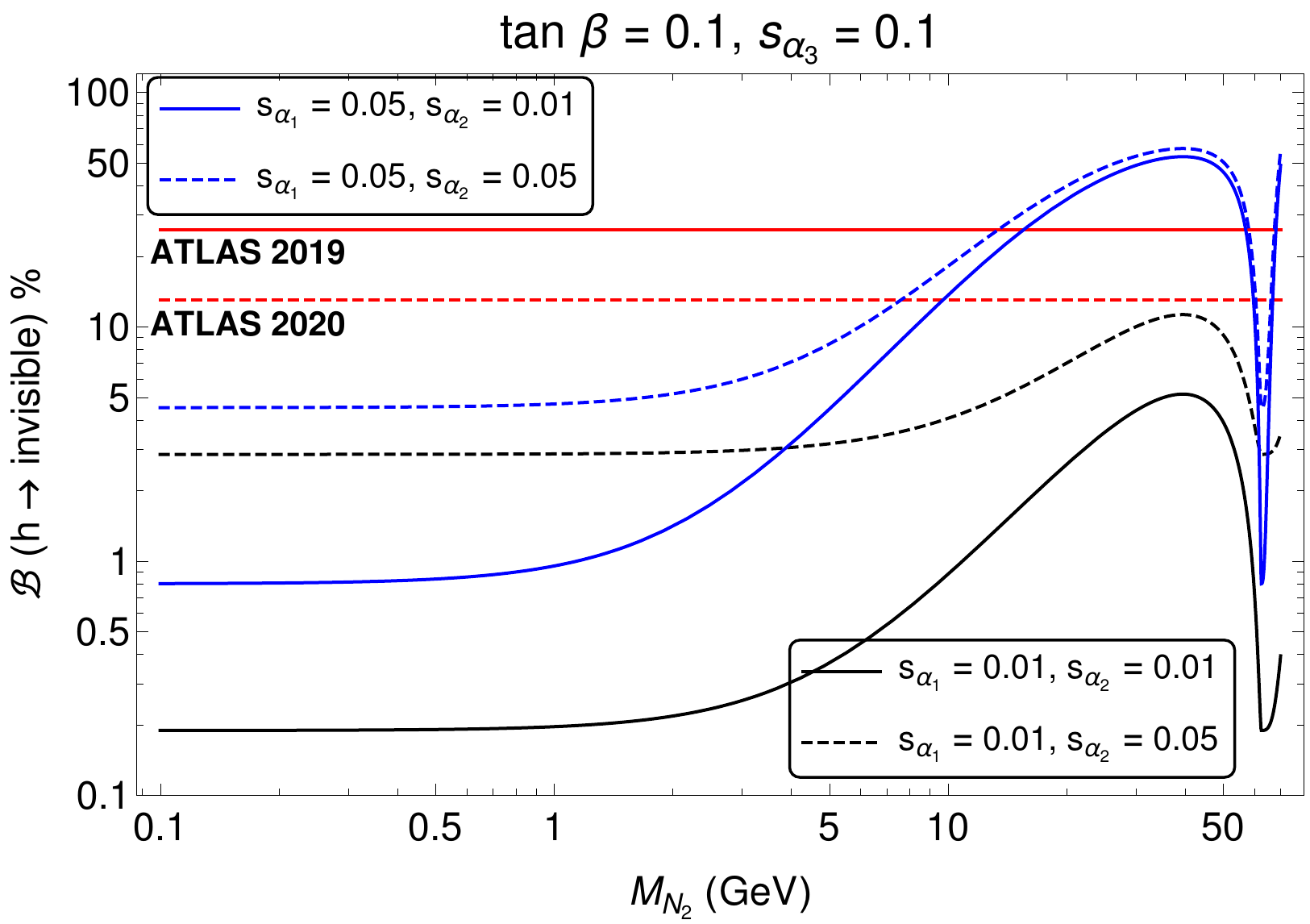}~~~~~~~
		\includegraphics[scale=0.6]{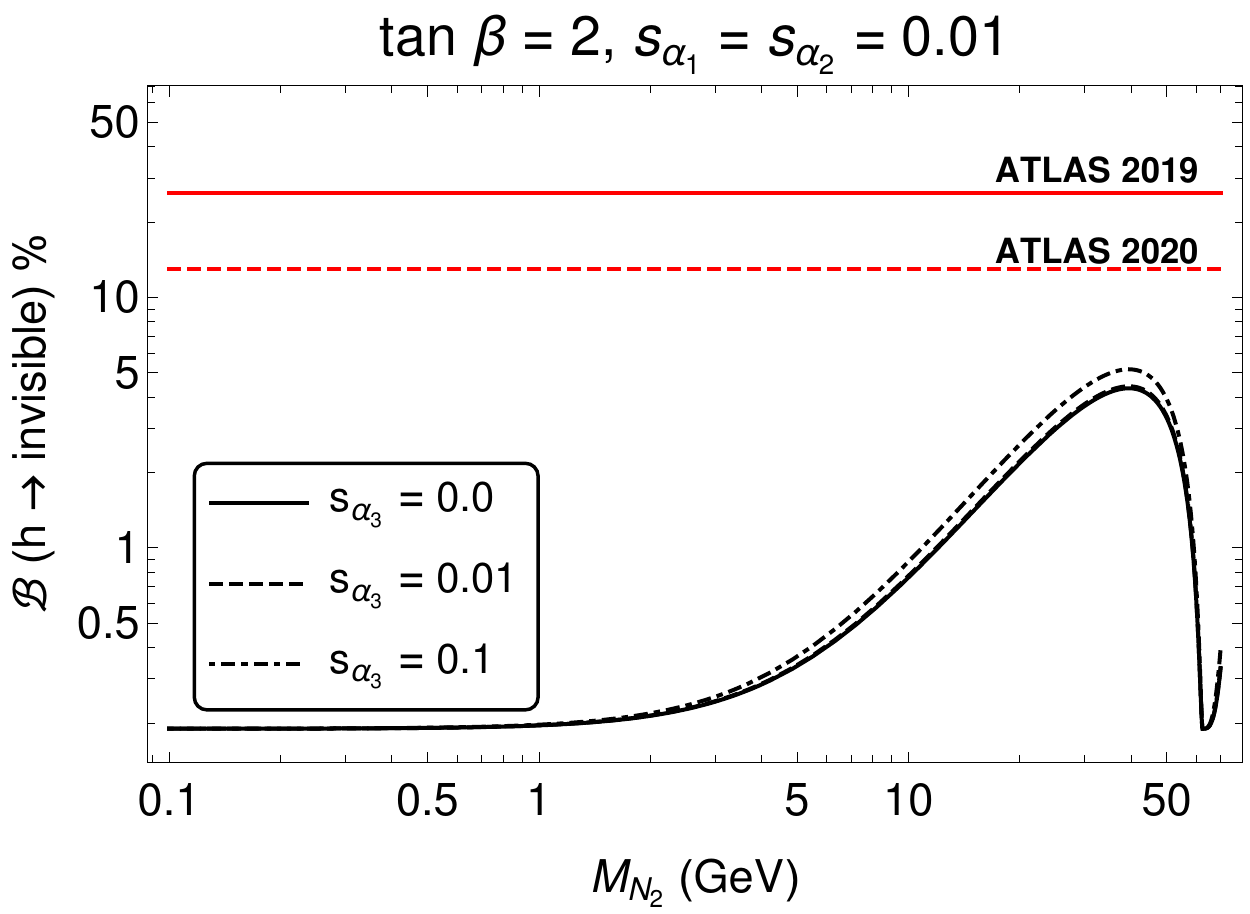}
		\caption{Higgs invisible branching fraction as a function of $M_{N_2}$ for two different values of $s_{\alpha_1}$ and $s_{\alpha_2}$ each is shown. The red solid line is the upper bound from ATLAS 2019 \cite{Aaboud:2019rtt} while the red dashed line is the upper bound  from their recent announcement in April 2020\cite{ATLAS:2020cjb}.}
		\label{fig:HiggsInv_Toy1}
	\end{figure}

	The invisible decay width of Higgs to dark matter is given by :
	\begin{equation}
	\Gamma(h \to N_2 N_2) = \frac{1}{8\pi} \frac{M_{N_2}^2}{v_2^2}M_h \bigg(1- \frac{4 M_{N_2}^2}{M_h^2}\bigg)^{3/2} (s_{\alpha_2} + c_{\alpha_2}^2 s_{\alpha_1}s_{\alpha_3})^2.
	\label{eq:hn1n2toy1}
	\end{equation}
	On the other hand, the decay of Higgs to SM neutrinos via $X X$, or equivalently via $Z^\prime Z^\prime$ is given by :
	\begin{equation}
	\Gamma(h \to Z' Z' \to 4\nu_\ell) = \frac{1}{8\pi}\frac{g_{hZ^\prime Z^\prime}^2}{M_h}\left(1-\frac{4 M_{Z^\prime}^2}{M_h^2}\right)^{1/2}\left(3+\frac{M_h^4}{4 M_{Z^\prime}^4}  - \frac{M_h^2}{ M_{Z^\prime}^2}\right) \times \sum_\ell [\mathcal{B}(Z^\prime \to 2\nu_\ell)]^2
	\end{equation}
	where $g_{hZ^\prime Z^\prime} = 8g_X^2 (c_{\alpha_2} s_{\alpha_1} v_1 + 4 s_{\alpha_2}v_2)$ is the effective coupling of SM Higgs with $Z^\prime$ via mixing with the singlet scalars. For our model, $\sum^{}_\ell \left[\mathcal{B}(Z^\prime \to 2\nu_\ell)\right]^2 \approx 0.14$ for $M_{Z^\prime} = 1$ GeV and $g_{Z'} = 10^{-3}$. Therefore, the total invisible decay width of Higgs is given by the sum of the decay widths as mentioned above. In Fig.~\ref{fig:HiggsInv_Toy1}, we have shown the variation of the Higgs invisible decay with the mass of dark matter for different values of the mixing angles. Note that only small mixing like $s_{\alpha_2} = 0.01$, $s_{\alpha_1}=0.01$ are allowed by the current limit for the entire mass range of $N_2$. However, $s_{\alpha_3}$ could be as large as $0.1$.  
	
		\begin{figure}[t]
			\centering
			\includegraphics[scale=0.65]{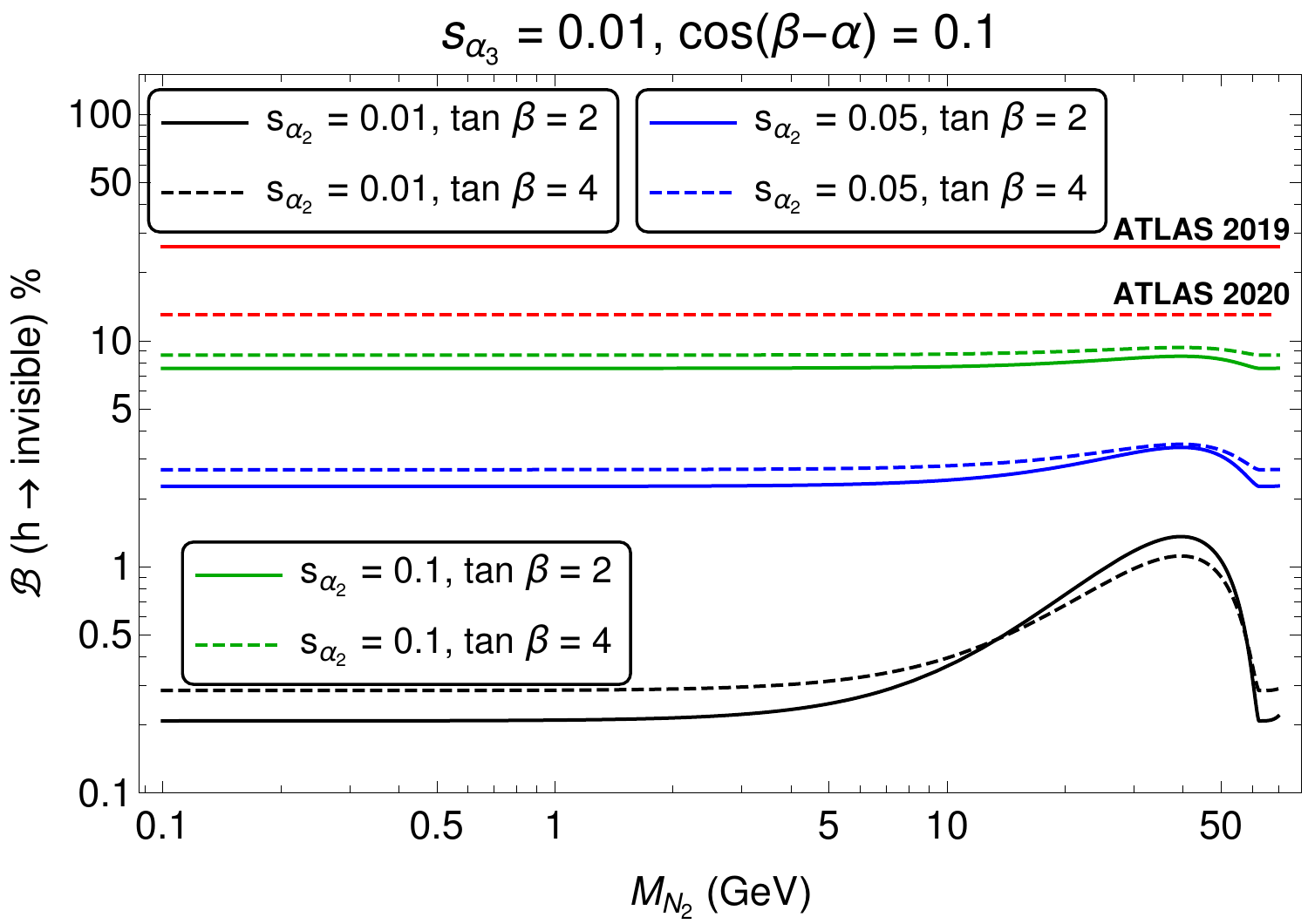}
			\caption{The variation of Higgs invisible branching fraction with the DM mass for different values of the other relevant parameters in toy model 2. The red solid and dashed lines have a similar description as given in Fig.~\ref{fig:HiggsInv_Toy1}.}
			\label{fig:HiggsInv_Toy2}
		\end{figure}
		
	\subsubsection{Toy Model 2}
	In our second model, the contribution to the Higgs invisible decay width is given by
	\begin{equation}
	\Gamma(h \to N_2 N_2) = \frac{1}{8\pi} \frac{M_{N_2}^2}{v_2^2}M_h \bigg(1- \frac{4 M_{N_2}^2}{M_h^2}\bigg)^{3/2} \mathcal{R}_{31}^2.
	\end{equation}
	Also, the decay of Higgs to SM neutrinos via $Z^\prime Z^\prime$ is given by :
	\begin{equation}
	\Gamma(h \to Z' Z' \to 4\nu_\ell) = \frac{1}{8\pi}\frac{g_{hZ^\prime Z^\prime}^2}{M_h}\left(1-\frac{4 M_{Z^\prime}^2}{M_h^2}\right)^{1/2}\left(3+\frac{M_h^4}{4 M_{Z^\prime}^4}  - \frac{M_h^2}{ M_{Z^\prime}^2}\right) \times \sum_\ell [\mathcal{B}(Z^\prime \to 2\nu_\ell)]^2
	\end{equation}
	where $g_{hZ^\prime Z^\prime} = 8g_X^2 (\mathcal{R}_{12} v_1 + 4 \mathcal{R}_{13} v_2)$ is the effective coupling of SM Higgs with $Z^\prime$ via mixing with the singlet scalars. Again, the sum of squares of branching fraction, $\sum^{}_\ell \left[\mathcal{B}(Z^\prime \to 2\nu_\ell)\right]^2 \approx 0.14$ for $M_{Z^\prime} = 1$ GeV as mentioned before. In Fig.~\ref{fig:HiggsInv_Toy2}, we have shown the dependencies of the $\mathcal{B}(h \to \text{invisible})$ with the DM mass and other relevant parameters in toy model 2, like sine of the mixing angles and $\tan\beta$. Note that our chosen benchmark values like $s_{\alpha_2}=0.01$, $s_{\alpha_3}=0.01$ and $\tan\beta = 2$ or $4$ are allowed by the current bound on $\mathcal{B}(h \to \text{invisible})$.

\subsection{LFV decays of Higgs}
\begin{figure}[t]
	\centering
	\subfigure[]
	{\includegraphics[scale=0.6]{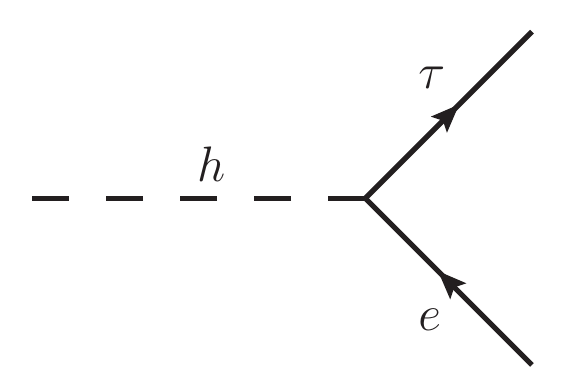}\label{fig:lfvhtaue} }
	\subfigure[]
	{\includegraphics[scale=0.6]{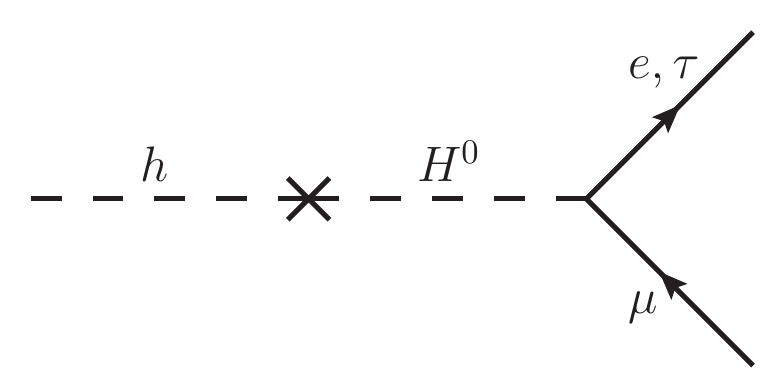}\label{fig:lfvhtaumu}}
	\caption{Lepton Flavour Violating decays of Higgs boson.}
	\label{fig:LFV_Higgs}
\end{figure}

 It is evident from the Yukawa interactions in eqs. \eqref{eq:LYuk1} and \eqref{LYuk2} that there exists lepton flavour violating decays of the Higgs ($h$) for both the Toy models. However, there are notable differences between the allowed LFV channels in the two models. The $U(1)_X$ charge assignments of the charged leptons are such that in both the models we will get the $h \to \tau e$ decay, for example, see Fig.~\ref{fig:lfvhtaue}. However, only Toy Model II contributes to LFV $h \to \mu\tau$ and $h \to \mu e$ decays via the mixing of the $h$ with the $H^0$ as shown in the Fig.~\ref{fig:lfvhtaumu}.
 
\begin{figure}[t]
	\centering
	\subfigure[]
	{\includegraphics[scale=0.55]{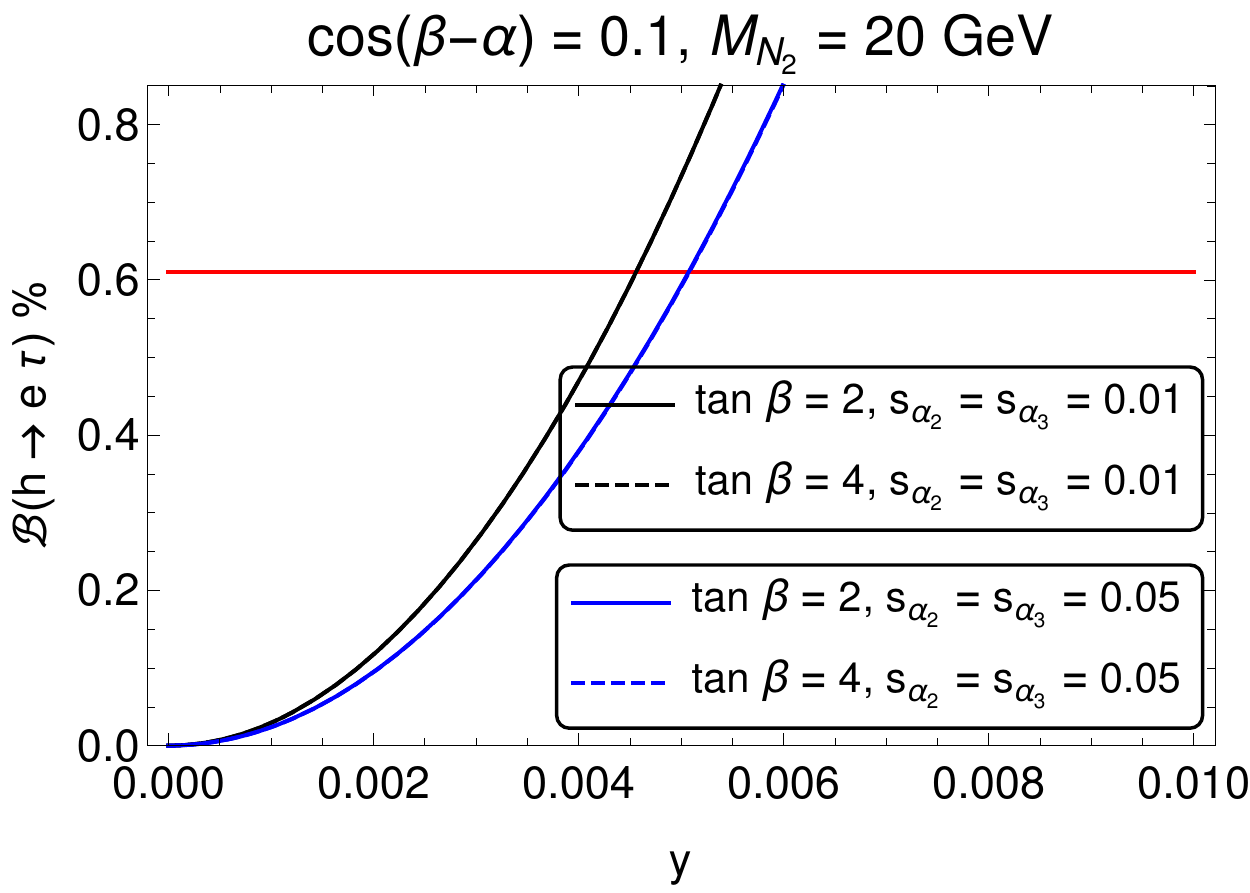}
		\label{fig:LFV_Higgs_E}}
	\subfigure[]
	{\includegraphics[scale=0.55]{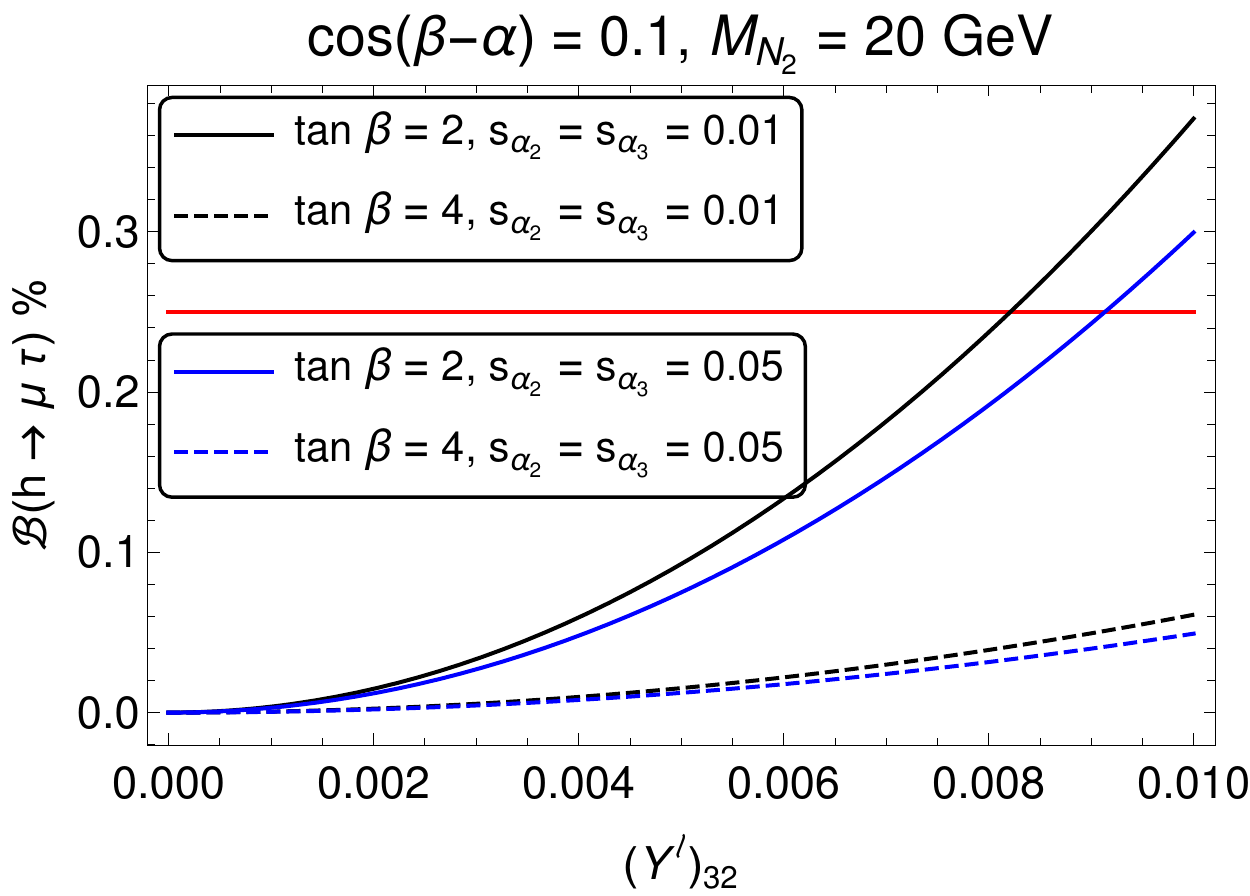}
		\label{fig:LFV_Higgs_Mu}}
	\caption{(a)Variation of $\mathcal{B}(h \to e \tau)$ with off-diagonal Yukawa coupling $y$ for two different values of tan $\beta$ and other mixing angles. The red line denotes the upper limit on the branching ratio. (b) Similar variation for $\mathcal{B}(h \to \mu \tau)$ is shown.}
	\label{fig:LFV_Higgs_plot}
\end{figure}

So far no excess have been observed in these channels at the LHC searches and the most recent upper limits on the lepton flavour violating branching fractions of the Higgs boson by CMS \cite{Sirunyan:2017xzt} reads
\begin{equation}
\begin{aligned}
&\mathcal{B}(h \to e \tau) < 0.61\%, \\
&\mathcal{B}(h \to \mu \tau) < 0.25\%.
\end{aligned}
\end{equation}
These limits will be helpful to constrain the lepton flavour violating Higgs couplings $Y^\ell_{ij}$ where $(i,j = 1,3)$ \& $i \neq j$. The general expression for the branching fraction for the LFV Higgs decay is given by
\begin{equation}
\begin{aligned}
\mathcal{B}(h \to \ell_i \ell_j) &= \left(\frac{|y|^2}{16\pi M_h \Gamma_h^{\text{tot}}}\right)\left[\left(1-\left(\frac{m_{\ell_i} + m_{\ell_j}}{M_h}\right)^2\right)\left(1-\left(\frac{m_{\ell_i} - m_{\ell_j}}{M_h}\right)^2\right)\right]^{1/2}(M_h^2 - m_{\ell_i}^2 - m_{\ell_j}^2)\\
&\approx \left(\frac{|y|^2 M_h}{16\pi \Gamma_h^{\text{tot}}}\right),~~~\text{for }~~ m_{\ell_i},m_{\ell_j} << M_h.
\end{aligned}
\end{equation}
Here, $\Gamma_h^{\text{tot}} = \Gamma(h \to SM) + \Gamma(h \to \text{Invisible})$ and $y$ denotes the effective LFV coupling. In most of the allowed parameter spaces, we can expect $\Gamma(h \to \text{Invisible}) <<  \Gamma(h \to SM)$; however, there are regions where it might be relevant to consider. In both the models, for $h \to \tau e $ decays the effective LFV coupling is given by $y \equiv \sqrt{{Y^\ell_{13}}^2 + {Y^\ell_{31}}^2}$. As mentioned earlier, there won't be any contribution to $h \to \mu\tau$ or $h\to \mu e$ decays in Toy Model I. In Toy Model II, the expression for the braching fraction for $h \to \mu\tau$ decay is given by 
\begin{equation}
\mathcal{B}(h \to \mu \tau) = \left(\frac{s_\alpha^2 |Y^\ell_{32}|^2}{2}\right)\left(\frac{M_h}{16\pi \Gamma_h^{\text{tot}}}\right),
\label{eq:brhtaumu}
\end{equation}
which in the limit $\alpha \to 0$, gives us the corresponding expression for $h \to e \tau$ decay. We will obtain the expression for $\mathcal{B}(h \to \mu e)$ after replacing $Y^\ell_{32}$ by $Y^\ell_{12}$ in eq.~\eqref{eq:brhtaumu}.  

In Fig.~\ref{fig:LFV_Higgs_E} we have shown the variation of $\mathcal{B}(h \to e \tau)$ with the effective coupling $y$. Since the contribution will be similar for both the models, we have not shown it separately for the two. It can be clearly understood from the plot that the coupling cannot be larger than $\sim 0.005$ irrespective of the value of $\tan\beta$ or other angles. There is very little dependence on  $s_{\alpha_2}$ or $s_{\alpha_3}$ which is coming from the contributions in $\Gamma(h \to \text{Invisible})$ (see eq.~\eqref{eq:hn1n2toy1}) in the denominator. For illustrative purpose, we have shown the variation for $M_{N_2} = 20$ GeV; however, we have checked that the variation does not change significantly on changing the DM mass. 

In Fig.~\ref{fig:LFV_Higgs_Mu} we show a similar variation of the branching ratio to the $\mu \tau$ mode with $Y^\ell_{32}$ for two different values of tan $\beta$ and other mixing angles. As mentioned earlier, this decay mode is specific for Toy Model II only. Since this process is mixing induced, both $\tan\beta$ and $Y^\ell_{32}$ are tightly constrained from the data. As expected, the branching fraction is sensitive to both the mixing parameters $\beta$ and $\alpha$. Note that for $\tan\beta = 2$, the allowed values of $Y^\ell_{32}$ is $Y^\ell_{32} < 0.01$. However, for $\tan\beta > 2 $, more higher values of the Yukawa coupling are allowed. In general, higher values of the $\tan\beta$ prefers higher values of the Yukawa coupling. This is expected since the constraint $|\text{cos}(\beta - \alpha)| = 0.1$ implies smaller $\text{sin }\alpha$ for large tan $\beta$. Once again the conclusions are not affected significantly by the DM mass. Also, we have noted that for values like $\tan\beta = 2$ and $s_{\alpha_2} = s_{\alpha_3} = 0.01$, the branching fraction $\mathcal{B}(h \to \mu e) \le 0.4\%$ for $Y^\ell_{12} \le 0.01$.

\begin{table}[h!!!]
\footnotesize
	\centering
	\renewcommand{\arraystretch}{1.2}
	\begin{tabular}{|c|c|}
		\hline
		$\ell^+ \ell^- + \slashed{E_T}$  & $4\ell$ \\
		\hline
		~~$pp \to Z \to Z \hspace{0.1cm}s_2 \to \ell^+ \ell^- \slashed{E_T} $ (Toy Model-I and II) ~~& $pp \to \ell^+ \ell^- \gamma \to 4 \ell$  \\
		~~$pp \to Z \to Z' \hspace{0.1cm}s_2 \to \ell^+ \ell^- \slashed{E_T} $ (Toy Model-I and II) ~~  & $pp \to \ell^+ \ell^- Z' \to 4 \ell$   \\
		~~$pp \to h \to h \hspace{0.1cm}s_2 \to \ell^+ \ell^- \slashed{E_T} $ (Toy Model-I and II) ~~ & $pp \to Z' Z' \to 4 \ell$  \\
		~~$pp \to H^+ H^- \to \mu^+ \mu^- \slashed{E_T}$ (Only in Toy Model-I) ~~& \\
		\hline
	\end{tabular}
	\caption{Collider signatures resulting in dilepton and 4-lepton final states for both Toy Models.}
	\label{tab:Coll}
\end{table}

\begin{table}[!h]
\footnotesize
	\centering
	\renewcommand{\arraystretch}{1.2}
	\begin{tabular}{|c|c|c|}
		\hline
		Sl.No. & Benchmark point (BP) &  $\sigma(pp \to H^+ H^- \to \mu^+ \mu^- \slashed{E_T})$ (fb)\\
		\hline
		(A) & $M_{H^\pm} = 500$ GeV, $M_{N_2} = 20$ GeV, $Y_{22} = 0.1$ & 2.45  \\
		(B) & $M_{H^\pm} = 500$ GeV, $M_{N_2} = 40$ GeV, $Y_{22} = 0.1$ & 2.39  \\
		(C) & $M_{H^\pm} = 500$ GeV, $M_{N_2} = 20$ GeV, $Y_{22} = 0.2$ & 39.2   \\
		(D) & $M_{H^\pm} = 800$ GeV, $M_{N_2} = 20$ GeV, $Y_{22} = 0.1$ & 0.58 \\
		\hline
	\end{tabular}
	\caption{Production cross-section of the dimuon $+ \slashed{E_T}$ final state generated from intermediate inert charged Higgses. This production is exclusively for Toy model I.}
	\label{tab:DiMuon}
\end{table}

\begin{table}[!h]
\footnotesize
	\centering
	\renewcommand{\arraystretch}{1.4}
	\begin{tabular}{|c|c|c|c|c|c|}
		\hline
		\multirow{2}{*}{Sl.No.} & \multirow{2}{*}{Benchmark point (BP)} & \multicolumn{2}{|c|}{$\sigma^{2\ell + \slashed{E_T}}$ (fb)} & \multicolumn{2}{|c|}{$\sigma^{4\ell}$ (fb)} \\
		\cline{3-6}
		& & Toy I & Toy II & Toy I &  Toy II\\
		\hline
		1. & $M_{N_2} = 20$ GeV, $M_{Z^\prime} = 1$ GeV & 0.10 & 0.074 & \multirow{2}{*}{25.54} & \multirow{2}{*}{25.54} \\
		2. & $M_{N_2} = 40$ GeV, $M_{Z^\prime} = 1$ GeV & 0.07 & 0.066 & &  \\
		\hline
		3. & $M_{N_2} = 20$ GeV, $M_{Z^\prime} = 0.5$ GeV & 0.10 & 0.074 & \multirow{2}{*}{25.57} & \multirow{2}{*}{25.57} \\
		4. & $M_{N_2} = 40$ GeV, $M_{Z^\prime} = 0.5$ GeV & 0.07 & 0.065 & &\\
		\hline
	\end{tabular}
	\caption{Production cross-section of the dilepton $+ \slashed{E_T}$ ($\sigma^{2\ell + \slashed{E_T}}$) and $4\ell$ ($\sigma^{4\ell}$) final states for some specific benchmark points of the two toy models. The intermediate channels that lead to such final states are listed in Table.~\ref{tab:Coll}. Please note that the cross-section for the dilepton $+ \slashed{E_T}$ channel for Toy model I quoted here excludes the contribution from the charged Higgs mediated diagram (which we have separately shown in Table.~\ref{tab:DiMuon}).}
	\label{tab:CrossSec}
\end{table}

\begin{figure}[t]

		\centering
	\centering
	\includegraphics[scale=0.6]{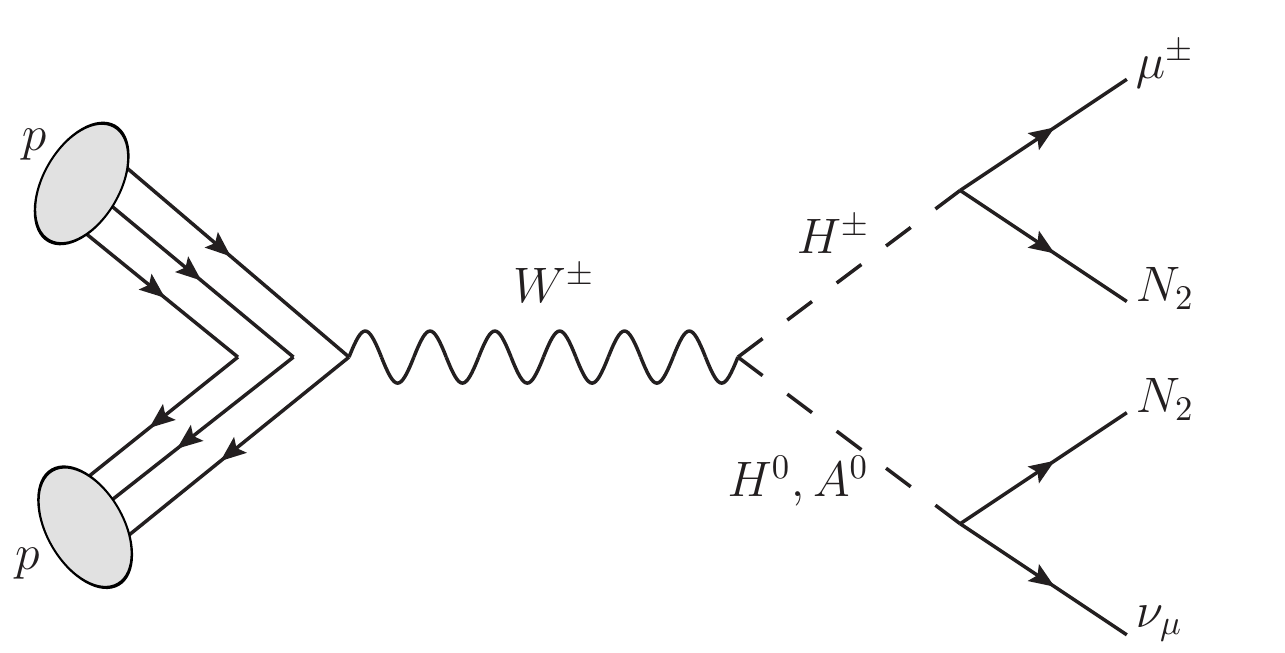}
	\caption{Feynman diagram for the production of $\mu^\pm + \slashed{E_T}$ final state at LHC (Toy Model I).}
	\label{fig:monoMuon}
\end{figure}

\begin{table}[h]
\footnotesize
	\centering
	\renewcommand{\arraystretch}{1.4}
	\begin{tabular}{|c|c|c|}
		\hline
		Sl.No. & Benchmark point (BP) &  $\sigma$($\mu^\pm + \slashed{E_T}$) (fb)\\
		\hline
		1. & $M_{H^\pm} = 500$ GeV, $M_{H^0} = 200$ GeV, $\lambda_5 = 0.01$, $M_{N_2} = 20$ GeV, $Y_{22} = 0.1$ & $0.0044$  \\
		2. & $M_{H^\pm} = 500$ GeV, $M_{H^0} = 200$ GeV, $\lambda_5 = 0.01$, $M_{N_2} = 40$ GeV, $Y_{22} = 0.1$ & $0.0041$  \\
		3. & $M_{H^\pm} = 500$ GeV, $M_{H^0} = 200$ GeV, $\lambda_5 = 0.01$, $M_{N_2} = 20$ GeV, $Y_{22} = 0.2$ & $0.071$  \\
		4. & $M_{H^\pm} = 500$ GeV, $M_{H^0} = 200$ GeV, $\lambda_5 = 0.01$, $M_{N_2} = 20$ GeV, $Y_{22} = 0.5$ & $2.77$  \\
		5. & $M_{H^\pm} = 200$ GeV, $M_{H^0} = 200$ GeV, $\lambda_5 = 0.01$, $M_{N_2} = 20$ GeV, $Y_{22} = 0.2$ & $0.263$ \\
		\hline
	\end{tabular}
	\caption{Production cross-section of the monomuon $+ \slashed{E_T}$ final state in Toy Model I.}
	\label{tab:MonoMuon}
\end{table}

\begin{table}[!h]
\footnotesize
	\centering
	\renewcommand{\arraystretch}{1.3}
	\begin{tabular}{|c|c|c|c|}
		\hline
		No.& Benchmark Point & $\sigma(\tau^+ \mu^- \mu^- e^+) + \sigma(\tau^- \mu^+ \mu^+ e^-)$  & $\sigma(\tau \tau \mu e)$ (fb) \\
		& & (in fb)  &  \\
		\hline
		1. & $M_{H^0} = 50$ GeV, $M_{A^0} = 100$ GeV, $Y^\ell_{12} = Y^\ell_{32} = 0.005$ & 1.106 & 0.026\\
		2. & $M_{H^0} = 100$ GeV, $M_{A^0} = 100$ GeV, $Y^\ell_{12} = Y^\ell_{32} = 0.005$ & 0.008 & $1.85\times 10^{-4}$ \\
		\hline
	\end{tabular}
	\caption{Possible signature of Toy Model II with the corresponding production cross-sections.}
	\label{tab:sigtoy2}
\end{table}

\subsection{Other Possible Signatures}
In this subsection, we would like to briefly discuss some other possible collider signatures of our Toy models at the LHC, in addition to the specific ones mentioned above. Here, we will only mention a few exciting channels for the search at the LHC; a detailed analysis is beyond the scope of this paper. It is quite evident that the most notable collider signals would be the multi-lepton final states with or without an associated missing energy ($\slashed{E_T}$). More importantly, the dilepton ($2\ell$) + $\slashed{E_T}$ channel which can be probed with excellent precision in the high luminosity colliders in the near future, has the potential to discriminate the two Toy models. A few of the dominating production channels are mentioned in Table \ref{tab:Coll}. Note that the search for $4\ell$ final states will not be a unique test of our toy models. This is because, in all the cases, the primary decay channels are via the production of $Z'$ which is very common in NP models with an additional $U(1)_X$ gauge bosons. The details of the collider searches for such an extension with a sub-GeV $M_{Z'}$ can be seen from \cite{Deppisch:2019ldi}. However, the search of $(2\ell + \slashed{E_T} )$ could be helpful to probe Toy model I since the process is also mediated by the production and decay of the new scalars. In Table \ref{tab:Coll}, we have mentioned only the dominating production channels. We have noted that for the allowed values of the model parameters, as discussed earlier, the production cross-section  $\sigma^{2\ell + \slashed{E_T}}$ in Toy Model-I is much larger than that in Toy Model-II. This is due to the presence of an additional channel $pp \to H^+ H^- \to \mu^+ \mu^- \slashed{E_T}$ in Toy Model-I. The corresponding production cross-section of this specific channel for some benchmark scenarios are given in Table \ref{tab:DiMuon}. Here, the events are generated in \texttt{MADGRAPH} \cite{Alwall:2014hca} at $\sqrt{s} = 14$ TeV. Depending on the charged Higgs mass and associated coupling $Y_{22}$, the cross-section can be quite large. For the rest of three channels as mentioned in Table \ref{tab:Coll}, the estimated production cross-sections for a few benchmark scenarios are given in Table \ref{tab:CrossSec}. As one can see, the production cross-sections are very small, which is expected since the diagrams mentioned here are mostly mixing induced which we have assumed to be small. Note that the cross-sections are insensitive to the mass of $Z'$. We have checked that in both the toy models, among these three channels the dominating contribution will come from $pp \to Z \to Z \hspace{0.1cm}s_2 \to \ell^+ \ell^- \slashed{E_T} $. 
Therefore, this channel will not be helpful to discriminate the signatures of Toy Model-I from that of Toy Model-II. However, as one can see that $\sigma(pp \to H^+ H^- \to \mu^+ \mu^- \slashed{E_T})$ is much larger than the production cross-sections for the rest of three channels, therefore, at the colliders a dedicated search for $\mu^+ \mu^- + \slashed{E_T}$ signature could be helpful to probe Toy Model-I. Please note than in order to obtain long-lived charged scalars whose decay length ($c\tau$) $ \gsim 0.1$ mm, we need Yukawa coupling of the order of $\sim (10^{-6} - 10^{-5})$. This will give very low production cross-section of the final state and therefore techniques like displaced muon and kink vertex will not be applicable. So we do not discuss it further.

Another interesting collider signature could be the production of ($\mu^\pm + \slashed{E_T}$) which is an exclusive feature of Toy Model I and can be a smoking gun signal. It can be observed at a $pp$ collider like the LHC where the intermediate particles leading to such a final state are the inert charged Higgs ($H^\pm$) and inert neutral scalars ($H^0,A^0$) as shown in Fig.~\ref{fig:monoMuon}. The readers may recall that the inert Higgs couples to muon along with the dark matter. Due to electroweak interaction, it is possible to have sizeable production of $H^+$-$H^0$ which then decay to give a mono-muon plus missing energy final state. There are no other contributing diagrams to this muon specific signal. The other mono-lepton channels (say the mono-electon for example) will be kinematically suppressed due to the associated heavy neutrinos in the final state. The mono-muon signal is cleaner than the mono-jet searches and therefore it is possible to tag the muon. The major background is the $W(\ell \nu)$ process but one can expect a clean signal away from the $W$-boson mass window. The other minor backgrounds include $t, t\bar{t}, Z/\gamma* (\ell \ell), \gamma + \text{jets}$ and $VV$ (where $V$ stands for the SM vector bosons $W,Z$). As can be seen from Table.~\ref{tab:MonoMuon}, for a few suitable benchmark values of the scalar and DM masses and coupling $Y_{22}$, it is possible to obtain a few femtobarns of production cross-section. In Toy Model-II, the $U(1)_X$ charge of $H_2$ forbids its coupling with muon and $N_2$ simultaneously, instead it couples with the other RHNs ($N_1,N_3$). Also the scalars are in general lighter than $N_1, N_3$.
Therefore, once again, the $\mu^{\pm} + \slashed{E_T}$ production at LHC will be kinematically suppressed. The probable collider signatures of Toy Model-II will be the productions of $\tau^+ \mu^- \mu^- e^+$, $\tau^- \mu^+ \mu^+ e^-$ and $\tau \tau \mu e$ events at the LHC via the production and decay of $H^0 H^0$. This is possible only in Toy Model-II since in this model, $H^0$ takes part in LFV interactions, which is not allowed in Toy Model-I. In a few benchmark scenarios, the corresponding production cross-sections are given in Table \ref{tab:sigtoy2}. As expected, the cross-sections are highly sensitive to the mass of $H^0$. Note that for the above mentioned four-lepton final states the SM background will be highly suppressed. Therefore, a dedicated search of these four lepton states with specific flavour and charge could be useful to test our Toy Model-II.

\section{Summary}
\label{sec9}
We have extended the SM by an Abelian $U(1)_X$ gauge group which results in a massive gauge boson ($X$) that couples only to leptons and has a small kinetic mixing with the SM $Z$ boson. We have considered only the low masses of $X$ ($M_X \lsim 1$ GeV). In this kind of extension, we will get new contributions to flavour changing processes like $b\to s \ell^+\ell^-$ decays, and the new contribution will be in $\Delta C^{\ell}_9$ which is the WC of the operator $\mathcal{O}_9$. Here, $\mathcal{O}_9$ is a left-handed quark current operator with vector muon/electron coupling. Also, in this model, the contributions to such flavour changing processes will be in both the electron and muon final states. At the same time, we will get new contributions in anomalous magnetic moment of the muon. We use the present data on $R(K), R(K^{*})$, the ratio of branching fraction $\mathcal{B}(B^0 \to  K^{*0}\chi (\mu^+ \mu^-))/\mathcal{B}(B^0 \to K^{*0}\mu^+ \mu^-)$, $\mathcal{B}(B \to K^{(*)} e^+e^-)$, and muon anomalous magnetic moment to constrain $U(1)_X$ charges of the SM leptons. Also, the values $\Delta C^{\mu}_9$ and $\Delta C^{e}_9$ which are obtained from the analysis are consistent with the global fit results of the data in $b\to s \ell^+\ell^-$ decays including various angular observables. Additionally, we consider all upper bounds from different experimental data on such light Abelian gauge boson mass and its couplings.

Now charging the SM fermion under a generic $U(1)_X$ symmetry makes the theory anomalous. To get an anomaly-free renormalisable model, we have incorporated additional chiral fermions into the model. In order to fit our requirements with a minimal particle content, we have considered a scenario where the three generation of leptons having vector type $U(1)_X$ interactions have corresponding charges $(n_1,n_2,n_3) = (-1,2,-1)$ respectively. Such a choice is consistent with the data and also ensures anomaly cancellations after adding one right-handed neutrino per fermion generation having equal, and opposite $U(1)_X$ charges as that of SM lepton in that generation. Also, we have added additional singlet and doublet Higgs fields to get the desired mass spectrum. To prevent a direct coupling of the RHNs with the lepton doublets via SM Higgs, we impose a discrete $\mathcal{Z}_2$ symmetry on the particles in two different ways which lead to two distinct models and phenomenology. This kind of symmetry restrictions will provide a natural candidate for DM in our extended models. At the same time, the chosen particle content of the models can also generate light neutrino masses, in agreement with neutrino oscillation data.

We are able to successfully study the DM phenomenology which is almost similar for the two Toy models but they have different neutrino mass generation mechanisms. The scalar content is very rich with an additional scalar doublet and two scalar singlets apart from the usual SM Higgs doublet. However, the second scalar doublet has very distinct features and plays different roles in each of the two Toy models. The low gauge boson $X$ mass allows us to evade stringent constraints from LHC, while facing tight constraints from other low energy experiments. We ensure that our analysis is consistent with the LEP II bounds on $U(1)_X$ gauge boson mass and coupling and the bounds from other light boson search experiments. Few preliminary results have also been shown and discussed.

The two toy models lead to phenomenological implications that can be tested at the collider experiments. The promising channels are the 4-lepton final states, dilepton ($2\ell$) + $\slashed{E_T}$ etc. Also, in the low energy experiments the potential signatures may come from the FCNC processes, like $b \to s(d) + invisible$, $s \to d + invisible$, $c \to u + invisible$ which will lead to rare decays of $B_q$/$K$/$D$ mesons to a relatively lighter meson final state with invisible particles. For example, the recently observed excess of events in the $K_L \to \pi^0 + invisible$ decay at the KOTO experiment can be explained for a mass window $0.30 < M_{Z'} (M_X)  < 0.35$ GeV. For these values of $M_{Z'}$, all the other observables discussed in this article are consistent with the corresponding data.  Also, both the models will contribute to LFV $h \to \tau e$ decays.   

We have discussed some distinct features of both the models which could be helpful to discriminate the signatures of the two models at different experiments. At the LHC, the production of $\mu^\pm + \slashed{E_T}$ and lepton-specific $\mu^+\mu^- + \slashed{E_T}$ events could be the possible signatures of Toy Model-I which is not possible to get in Toy Model-II. On the other hand, the search for the specific multi-leptonic states like $\tau^+ \mu^- \mu^- e^+$, $\tau^- \mu^+ \mu^+ e^-$ and $\tau \tau \mu e$ could be useful to identify the potential signatures of Toy Model-II. In the context of Higgs LFV, Toy Model-II contributes to $h \to \tau\mu$ and $h \to \mu e $ decays while Toy Model-I does not. Similarly, there are a few examples of the potential observables in the low energy sector: Toy Model-I contributes significantly to the semileptonic or purely leptonic decays $B_q$/$K$/$D$ mesons via the following quark level transitions: $b\to c (u) \tau \bar\nu_{\tau}$, $s\to u \tau \bar\nu_{\tau}$, $c\to s \tau \bar\nu_{\tau}$. However, Toy Model-II does not have significant contributions to these decays with $\tau$ in the final states. In Toy Model-II, we do not have any contribution in $\tau\to e\gamma$, $\tau\to \mu\gamma$ and $\mu\to e\gamma$ LFV decays; however, this model contributes to $\tau \to 3\mu$ decay. On the other hand, Toy Model-I contributes only in $\tau \to e\gamma$, not in all the other LFV decays as considered above. More precise data from future experiments will be able to discriminate between such toy models while confirming or ruling out some part of the available parameter space.

\acknowledgments
We are grateful to Biplob Bhattacherjee for some illuminating discussion. LM would like to thank Aritra Biswas, Sneha Jaiswal, Dibyendu Nanda for useful discussions and in particular Sunando Kumar Patra for computational help.
This work of SN is supported by the Science and Engineering Research Board, Govt. of India, under the grant CRG/2018/001260. DB acknowledges the support from Early Career Research Award from the department of science and technology-science and engineering research board (DST-SERB), Government of India (reference number: ECR/2017/001873). 

\newpage
\appendix
\begin{appendices}
\section{Subleading Annihilation Diagrams}
\label{sec:SubDM}

\begin{figure*}[h!]
	\centering	
	\includegraphics[scale=0.4]{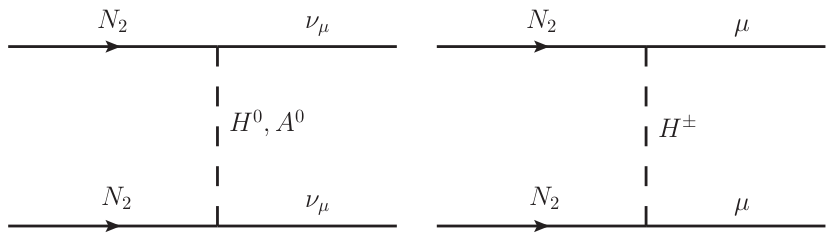}
	\caption{Subleading contributions to the relic.}
	\label{fig:subann}
\end{figure*}
\begin{figure}[htp!!]
	\centering
	\includegraphics[scale=0.7]{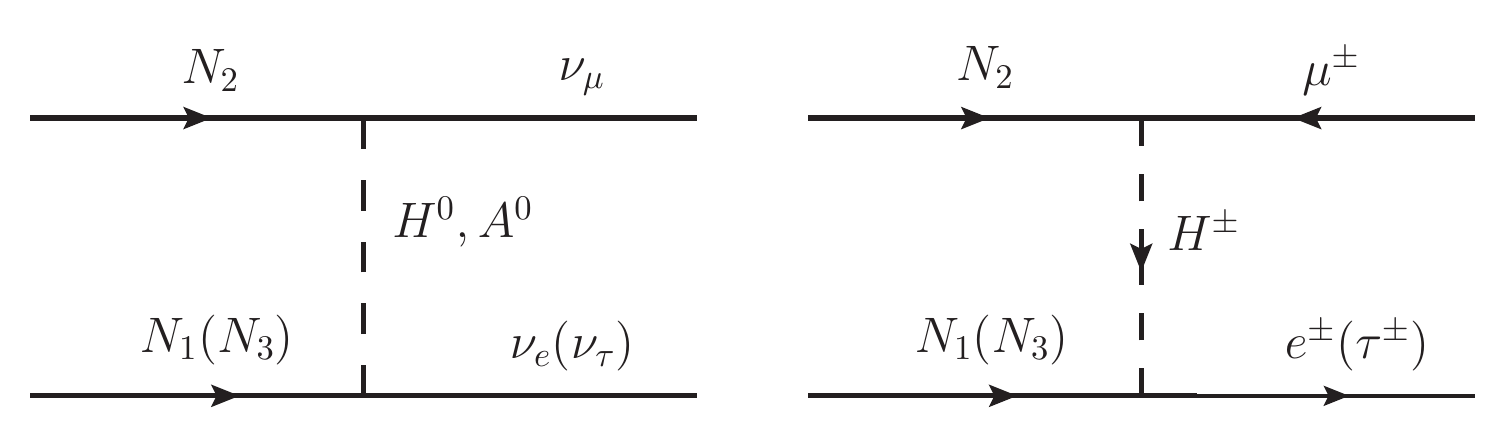}
	\includegraphics[scale=0.7]{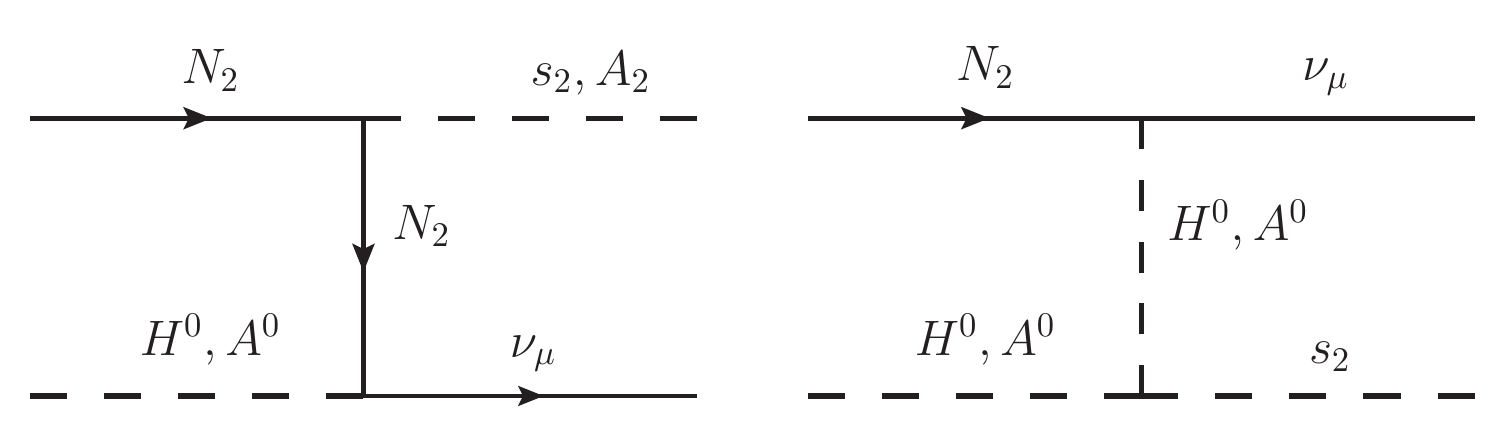}
	\includegraphics[scale=0.7]{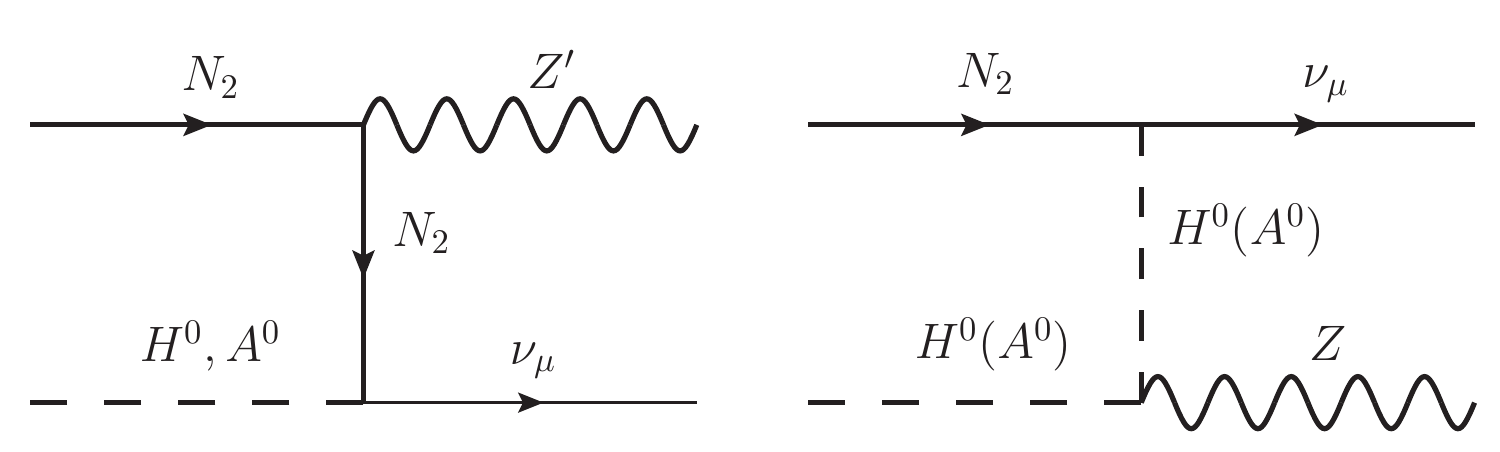}
	\caption{Subleading Co-annihilation diagrams.}
	\label{fig:subcoann}
\end{figure}

\section{Coupling Constants : Toy Model I}
\label{sec:Const1}
The coupling constants in terms of masses and mixing angles are given by :

\begin{equation*}
\begin{aligned}
\lambda_1 &= \frac{c_{\alpha_2}^2(c_{\alpha_1}^2 M_h^2 + M_{s_1}^2 s_{\alpha_1}^2) + M_{s_2}^2 s_{\alpha_2}^2}{v^2}, \\
\lambda_3 &= \frac{2(\lambda_L v^2 + M_{H^\pm}^2 - M_{H^0}^2)}{v^2}, \\
\lambda_4 &= \left(\frac{M_{H^0}^2 + M_{A^0}^2 -2 M_{H^\pm}^2}{v^2}\right),
\end{aligned}
\end{equation*}
\begin{equation}
\allowdisplaybreaks
\begin{aligned}
\lambda_5 &= \left(\frac{M_{H^0}^2 - M_{A^0}^2 }{v^2}\right), \\
\lambda_6 &= \frac{1}{v_1^2} \left(c_{\alpha_2}^2 M_{s_2}^2 s_{\alpha_3}^2 + M_h^2 (c_{\alpha_3}\hspace{0.1cm}  s_{\alpha_1} + c_{\alpha_1}\hspace{0.1cm}  s_{\alpha_2}\hspace{0.1cm} s_{\alpha_3})^2 +M_{s_1}^2 (c_{\alpha_1} \hspace{0.1cm} c_{\alpha_3} - c_{\alpha_1}\hspace{0.1cm}  s_{\alpha_2} \hspace{0.1cm} s_{\alpha_3})^2 \right), \\
\lambda_7 &= \frac{1}{v_2^2}\bigg( c_{\alpha_3}^2 (c_{\alpha_2}^2 M_{s_2}^2 + (c_{\alpha_1}^2 M_h^2 + M_{s_1}^2 s_{\alpha_1}^2) s_{\alpha_2}^2) + (c_{\alpha_1}^2 M_{s_1}^2 + M_h^2 s_{\alpha_1}^2)s_{\alpha_3}^2 \\ & + 2 c_{\alpha_1}\hspace{0.1cm}  c_{\alpha_3} (-M_h^2 + M_{s_1}^2) s_{\alpha_1} \hspace{0.1cm} s_{\alpha_2} \hspace{0.1cm} s_{\alpha_3}\bigg) - \frac{M_{A_2}^2 s_{\gamma}^2 v_1^2}{4v_2^4},\\
\lambda_8 &= \frac{1}{v_1 v_2}\bigg(c_{\alpha_1}^2\hspace{0.1cm}  c_{\alpha_3} (-M_{s_1}^2 + M_h^2 s_{\alpha_2}^2) s_{\alpha_3} + c_{\alpha_3}(c_{\alpha_2}^2 M_{s_2}^2 + s_{\alpha_1}^2 (-M_h^2 + M_{s_1}^2 s_{\alpha_2}^2))s_{\alpha_3} \\ & + c_{\alpha_1}\hspace{0.1cm} s_{\alpha_1}\hspace{0.1cm} s_{\alpha_2} (M_h^2 - M_{s_1}^2)(c_{\alpha_3}^2 - s_{\alpha_3}^2) \bigg)  + \frac{M_{A_2}^2 s_{\gamma}^2}{2v_2^2}, \\
\lambda_{\varphi_1} &= \frac{1}{v_1 v}\bigg(c_{\alpha_1}\hspace{0.1cm} c_{\alpha_2} \hspace{0.1cm} c_{\alpha_3} \hspace{0.1cm} s_{\alpha_1}(-M_h^2 + M_{s_1}^2) + c_{\alpha_2}\hspace{0.1cm} s_{\alpha_2} \hspace{0.1cm} s_{\alpha_3}(-c_{\alpha_1}^2 M_h^2 + M_{s_2}^2 + s_{\alpha_1}^2 M_{s_1}^2)\bigg), \\
\lambda_{\varphi_2} &= \frac{1}{v_2 v}\bigg(c_{\alpha_2} \hspace{0.1cm} c_{\alpha_3} \hspace{0.1cm} s_{\alpha_2}(- c_{\alpha_1}^2 M_h^2 + M_{s_2}^2 - s_{\alpha_1}^2 M_{s_1}^2) + c_{\alpha_1}\hspace{0.1cm} c_{\alpha_2} \hspace{0.1cm}s_{\alpha_1}\hspace{0.1cm} s_{\alpha_3}( M_h^2 - M_{s_1}^2)\bigg), \\
\delta &= -\bigg(\frac{M_{A_2}^2 s_{\gamma}^2}{2\sqrt{2} v_2}\bigg).
\end{aligned}
\end{equation}

\section{4 $\times$ 4 Rotation Matrix}
\label{RotMat4}
The components of a general $4\times 4$ real orthogonal matrix without phase are given by :

\begin{equation}
\begin{aligned}
\mathcal{R}_{11} &= c_{12} c_{13}c_{14}, \\
\mathcal{R}_{12} &= c_{13} c_{14} s_{12}, \\
\mathcal{R}_{13} &= c_{14} s_{13}, \\
\mathcal{R}_{14} &= s_{14}, \\ 
\mathcal{R}_{21} &= -c_{23} c_{24} s_{12} - c_{12} c_{24}s_{13} s_{23} -c_{12}c_{13}s_{14}s_{24}, \\
\mathcal{R}_{22} &= c_{12} c_{23} c_{24} - c_{24} s_{12} s_{13} s_{23} -c_{13} s_{12} s_{14} s_{24}, \\
\mathcal{R}_{23} &= c_{13} c_{24} s_{23} - s_{13} s_{14} s_{24}, \\
\mathcal{R}_{24} &= c_{14} s_{24},\\
\mathcal{R}_{31} &= -c_{12} c_{23} c_{34} s_{13} + c_{34} s_{12} s_{23} - c_{12} c_{13} c_{24} s_{14} s_{34} + c_{23} s_{12} s_{24} s_{34} + c_{12} s_{13} s_{23} s_{24} s_{34}, \\
\mathcal{R}_{32} &= -c_{12} c_{34} s_{23} + s_{12}(-c_{13} c_{24} s_{14} + s_{13} s_{23} s_{24})s_{34} - c_{23}(c_{34} s_{12} s_{13} + c_{14} s_{24} s_{34}),\\
\mathcal{R}_{33} &= c_{13} c_{23} c_{34} - c_{24} s_{13} s_{14} s_{34} - c_{13} s_{23} s_{24} s_{34},\\
\mathcal{R}_{34} &= c_{14} c_{24} s_{34},\\
\mathcal{R}_{41} &= -c_{12} c_{13} c_{24} c_{34} s_{14} + c_{12} s_{13}(c_{34} s_{23} s_{24} + c_{23} s_{34}) + s_{12}(c_{23} c_{34} s_{24} - s_{23} s_{34}),\\
\mathcal{R}_{42} &= -c_{13} c_{24} c_{34} s_{12} s_{14} + s_{12} s_{13} (c_{34} s_{23} s_{24} + c_{23} s_{34}) + c_{12} (-c_{23} c_{34} s_{24} + s_{23} s_{34}),
\end{aligned}
\end{equation}

\begin{equation*}
\begin{aligned}
\mathcal{R}_{43} &= -c_{24} c_{34} s_{13} s_{14} - c_{13}(c_{34} s_{23} s_{24} + c_{23} s_{34}), \\
\mathcal{R}_{44} &= c_{14} c_{24} c_{34}.
\end{aligned}
\end{equation*}
where $s_{ij} \equiv \text{sin } \alpha_{ij}$ and $c_{ij} \equiv \text{cos } \alpha_{ij}$.

\section{Coupling Constants : Toy Model II}
\label{sec:Const2}
The coupling constants in terms of masses and mixing angles are given by :

\begin{equation}
\allowdisplaybreaks
\begin{aligned}
\lambda_{H_1} &= \frac{c_\alpha^2 (c_{\alpha_3}^2 (c_{\alpha_2}^2 M_h^2 + M_{s_1}^2 s_{\alpha_2}^2) + M_{s_2}^2 s_{\alpha_3}^2) + M_{H^0}^2 s_\alpha^2}{v^2},\\
\lambda_{H_2} &= \frac{(s_{\alpha_2}^2 (c_{\alpha_3}^2 (c_{\alpha_2}^2 M_h^2 + M_{s_1}^2 s_{\alpha_2}^2) + M_{s_2}^2 s_{\alpha_3}^2) + M_{H^0}^2 c_\alpha^2 ) \text{tan}^2\beta}{v^2},\\
\lambda_{\varphi_1} &= -\frac{(c_\theta^2 M_{A_2}^2 s_{\theta_3}^2 + M_{A^0}^2 s_\theta^2)v^2}{4v_1^4} + \frac{c_{\alpha_2}^2 M_{s_1}^2 + M_h^2 s_{\alpha_2}^2}{v_1^2},\\
\lambda_{\varphi_2} &= -\frac{3(c_\theta^2 M_{A_2}^2 s_{\theta_3}^2 + M_{A^0}^2 s_\theta^2)v^2}{16v_2^4} + \frac{c_{\alpha_3}^2 M_{s_2}^2 + (M_h^2 c_{\alpha_2}^2 + M_{s_1}^2 s_{\alpha_2}^2)s_{\alpha_3}^2}{v_2^2}, \\
\delta &= \frac{(M_{A^0}^2 s_\theta^2 + c_\theta^2 M_{A_2}^2 s_{\theta_3}^2)v^2}{8\sqrt{2}v_1^2 v_2},\\
\lambda &= -\frac{(M_{A^0}^2 s_\theta^2 + c_\theta^2 M_{A_2}^2 s_{\theta_3}^2)\text{tan}^2\beta}{v^2 v_1 v_2},\\
\lambda_1 &= \frac{\text{tan}\beta (c_\alpha s_\alpha (M_{H^0}^2 - c_{\alpha_3}^2 (c_{\alpha_2}^2 M_h^2 + M_{s_1}^2 s_{\alpha_2}^2) -M_{s_2}^2 s_{\alpha_3}^2) + 2 M_{H^\pm}^2 s_\gamma^2 \text{ tan} \beta)}{v^2},\\
\lambda_2 &= \frac{(-2 M_{H^\pm}^2 s_\gamma^2 + M_{A^0}^2 s_\theta^2 + c_\theta^2 M_{A_2}^2 s_{\theta_3}^2) \text{tan}^2 \beta}{v^2},\\
\lambda_3 &= \frac{(M_{A^0}^2 s_\theta^2 + c_\theta^2 M_{A_2}^2 s_{\theta_3}^2)v^2 + 8 c_{\alpha_2}(M_h^2 - M_{s_1}^2) s_{\alpha_2}s_{\alpha_3}v_1 v_2}{8 v_1^2 v_2^2},\\
\lambda_4 &= \frac{(M_{A^0}^2 s_\theta^2 + c_\theta^2 M_{A_2}^2 s_{\theta_3}^2)v + 2 c_\alpha c_{\alpha_2} c_{\alpha_3}(-M_h^2 + M_{s_1}^2)s_{\alpha_2}v_1}{2 v v_1^2},\\
\lambda_5 &= \frac{(M_{A^0}^2 s_\theta^2 + c_\theta^2 M_{A_2}^2 s_{\theta_3}^2)v - 2 c_\alpha c_{\alpha_3}(c_{\alpha_2}^2 M_h^2 -M_{s_2}^2 + M_{s_1}^2)s_{\alpha_3}v_2}{2 v v_2^2},\\
\lambda_6 &=  \frac{\text{tan}\beta(M_{A^0}^2 s_\theta^2 + c_\theta^2 M_{A_2}^2 s_{\theta_3}^2)v \text{ tan}\beta + 2 c_{\alpha_2}c_{\alpha_3}(M_h^2 - M_{s_1}^2) s_\alpha s_{\alpha_2}v_1)}{2v v_1^2},\\
\lambda_7 &=  \frac{\text{tan}\beta(M_{A^0}^2 s_\theta^2 + c_\theta^2 M_{A_2}^2 s_{\theta_3}^2)v \text{ tan}\beta + 2 c_{\alpha_3}s_{\alpha}(c_{\alpha_2}^2 M_h^2 - M_{s_1}^2 + M_{s_2}^2 s_{\alpha_2}^2) s_{\alpha_3}v_2)}{2v v_2^2}.
\end{aligned}
\end{equation}

\end{appendices}
\newpage
\bibliographystyle{jhep}
\bibliography{jhepv2} 

\end{document}